\documentclass[12pt]{report}
\usepackage{graphicx,amsmath,epsf,a4}               %% Preamble.

\begin{document}
\pagestyle{empty}
\title{
%Non-adiabatic 
Adiabatic and entropy perturbations
     in cosmology}

\vspace{5.0cm}

\author{{\sc Christopher Gordon} \\ \\ \\ \\ \\ \\ \\ \\ \\
        Submitted for the Degree of Doctor of Philosophy \\ \\
        Department of Computer Science and Mathematics \\ \\
        University of Portsmouth, UK.}

\date{November 2001}
\maketitle
\thispagestyle{empty}

%%%%%%%%%%%Double spacing%%%%%%%%%%%%%%%%%%%
%\begin{double}

%----------------------------------------------------------------------------
%                            ACKNOWLEDGMENTS
%----------------------------------------------------------------------------
\newpage

\vspace*{6.0cm}

%\begin{center}
\section*{Acknowledgements}
Firstly, I would like to thank Roy Maartens for supervising my PhD and
collaborating with me on part of the work in this thesis. Also, I
appreciate that he gave me the freedom to follow my own interests.

I would also like to thank my other collaborators during the PhD:
Luca Amendola, Arjun Berera, Bruce Bassett, Misao Sasaki and David
Wands. Working with them has been very stimulating and enjoyable. 

I am also grateful for financial support from the University of
Portsmouth and the Overseas Research Council. Lastly, I would like to
thank my friends and family for their support and encouragement.

%\end{center}

\newpage
\pagestyle{plain}
\pagenumbering{roman}

\begin{abstract}
This thesis presents a study of the effect and generation of
non-adiabatic perturbations in Cosmology.

We study adiabatic (curvature) and entropy (isocurvature)
perturbations produced during a period of cosmological inflation that
is driven by multiple scalar fields with an arbitrary interaction
potential.  A local rotation in field space is performed to separate
out the adiabatic and entropy modes.  The resulting field equations
show explicitly how on large scales entropy perturbations can source
adiabatic perturbations if the background solution follows a curved
trajectory in field space, and how adiabatic perturbations cannot
source entropy perturbations in the long-wavelength limit. It is the
effective mass of the entropy field that determines the amplitude of
entropy perturbations during inflation. 
%We present two applications of
%the equations. First, 
We show why one in general expects the adiabatic
and entropy perturbations to be correlated at the end of inflation,
and calculate the cross-correlation in the context of a double
inflation model with two non-interacting fields \cite{gordonadent}. 
%Second, 
Then, we consider
two-field preheating after inflation, examining conditions under which
entropy perturbations can alter the large-scale curvature perturbation
and showing how our new formalism has advantages in numerical
stability when the background solution follows a non-trivial
trajectory in field space \cite{gordonadent,gordonsting}.

Then we compare the latest cosmic microwave background data with
theoretical predictions including correlated adiabatic and CDM
isocurvature perturbations with a simple power-law dependence.  We
find that there is a degeneracy between the amplitude of correlated
isocurvature perturbations and the spectral tilt.  A negative (red)
tilt is found to be compatible with a larger isocurvature
contribution. 
%Estimates of the baryon and CDM densities are found to
%be almost independent of the isocurvature amplitude.
The main result
is that current microwave background data do not exclude a dominant
contribution from CDM isocurvature fluctuations on large scales, and
marginally favour a significant fraction \cite{gordoncmb}.

We then study perturbations in Randall-Sundrum-type brane-world
cosmologies. The density perturbations generate Weyl curvature in the
bulk, which in turn backreacts on the brane via stress-energy
perturbations. On large scales, the perturbation equations contain a
closed system on the brane, which may be solved without solving for
the bulk perturbations. Bulk effects produce a {\em non-adiabatic} mode,
even when the matter perturbations are adiabatic, and alter the
background dynamics. As a consequence, the standard evolution of
large-scale fluctuations in general relativity is modified. The metric
perturbation on large-scales is {\em not} constant during high-energy
inflation \cite{gordonbranes}. 
%It is constant during the radiation era, except at most
%during the very beginning, if the energy is high enough.

The effect of non-linear perturbations on initiating inflation is examined from
  the perspective of both spacetime embedding and scalar field
  dynamics.  
%The spacetime embedding problem is solved for arbitrary
 % initial spatial curvature $\Omega$, which generalizes previous works
%  that primarily treat the flat case $\Omega=1$.
 Scalar field dynamics
  that is consistent with the embedding constraints are examined, with
  the additional treatment of damping effects. The effects of
  inhomogeneities on the embedding problem also are considered.  A
  category of initial conditions are identified that are not acausal
 and can develop into an inflationary regime \cite{gordonic}.

Finally the work is summarised and  current and future extensions are
discussed.

\end{abstract}
%
%
%----------------------------------------------------------------------------
%                               CONTENTS
%----------------------------------------------------------------------------
\newpage
%\pagestyle{plain}
%\pagenumbering{roman}

\tableofcontents
\listoffigures                     %% Uncomment this to generate list

\newpage
%-----------------------------------------------------------------------------
%                        Symbols and terminology
%-----------------------------------------------------------------------------

\newpage
\pagenumbering{arabic}

\chapter{Introduction}

\label{intro}
In this thesis I have aimed to present the work I did during my PhD.
It is based on the publications and a preprint I collaborated on during that time
\cite{gordonadent,gordonsting,gordoncmb,gordonbranes,gordonic}. My
other publication \cite{mnf} was completed before starting my PhD and
so is not included in the thesis.

In this introduction Chapter I try to place the work done during the
thesis in context. To do this I give a summary of the current state of
cosmology both in terms of theory and experiment. My summary is
centered around the issue of adiabatic and entropy perturbations in
Cosmology and is by no means exhaustive. There are a large number of
good text books available where a more exhaustive treatment is
available, see for example \cite{KT,LiLy00}. 
%In this introduction, the
%references are also only for a few example articles and books. A more complete
%list of references can be found in \cite{KT,LiLy00}.

\section{Big bang cosmology}
The big bang model is based on the assumption that the Universe is
isotropic and homogeneous on large scales. There is much evidence for
this. In particular the cosmic microwave background (CMB) has been measured
to be isotropic to one part in $10^5$ \cite{COBE}.  Together with the
weak Copernican principle, i.e. all cosmic observers see a nearly
isotropic CMB, this implies that the universe is nearly homogeneous on
large scales \cite{SME}.  Large scale structure studies support this.

This assumption then leads to the following space-time metric:
\[
ds^2 = -dt^2 + a(t)^2\left\{ \frac{dr^2}{1-K r^2} +  r^2
d\theta^2 + \sin^2\theta + d\varphi^2  \right\}  
\]
where $a(t)$ is the scale factor, $t$ is the time coordinate, $r,
\theta, \varphi$ are the spatial polar coordinants and $K$ is the
curvature which can be negative, zero or positive.

Using the Einstein equations and assuming the matter in the universe
is a mixture of perfect fluids we get the Friedman equation:
\begin{equation}
H(t)^2 = \frac{8\pi G }{3} \rho(t) - \frac{K}{a(t)^2}
\label{feq}
\end{equation}
 where $H=\dot{a}/a$ is the Hubble parameter, $G$ is Newton's
constant, $\rho$ is the energy density.
%, $\Lambda$ is the cosmological
%constant term and $K$ is the curvature, which can be positive,
%negative or zero.

 In the 1920's Hubble observed that the red shift of
light emitted from galaxies increases with their distance and thus
showed that $H$ is positive and so the Universe is expanding. If
$\dot{a}$ is positive then $a$ must have been zero at some time in the
past. This model of a  Universe expanding from an initial singularity
is often referred to as the `big bang' model. However at very high
energy densities quantum corrections are thought to change the Friedman
equation and perhaps avoid the singularity.

%The $Lambda$ can be
%interpreted as the vacuum energy. 

%Although the presence of
%$\Lambda$ has been controversial, recent Super Nova measurements
%\cite{SN} have provided strong evidence that we may be living in a
%Universe where $Lambda$ is the dominant term in the Friedman
%equation. 

From energy momentum conservation we get the continuity equation 
\[
d(\rho a^3) = - p d(a^3)
\]
where $p$ is the pressure.
For the simple equation of state 
\[
p=w\rho,
\]
 we then get
\[
\rho \propto a^{-3(1+w)}.
\]
Some cases of interest are 
\[ \begin{array}{rrrl}
\mbox{Radiation: } &(p= \frac{1}{3} \rho) &\rightarrow &\rho \propto
a^{-4},\\
\mbox{Matter: } & (p=0) & \rightarrow &\rho \propto a^{-3},\\
\mbox{Vacuum energy: }& (p=-\rho)&\rightarrow&\rho \propto
\mbox{constant}.
\end{array}
\]
So for a Universe consisting of matter, radiation and vacuum energy
(also known as a cosmological constant) the Friedman equation becomes
\[
H(t)^2 = \frac{8\pi G}{3} \left(\frac{\rho_r(t_0) a(t_0)^4}{a(t)^4} +
\frac{\rho_m(t_0)a(t_0)^3}{a(t)^3} + \rho_v(t_0)  \right) -
\frac{K}{a^2}  
\]
where subscripts $r, m$ and $v$ refer to radiation, matter and vacuum
respectively. The constants of integration are set at some initial
time $t_0$. As $a$ is increasing with time, it can be seen from the
above equation that the Universe can go through different stages where
different components in the Friedman equation will dominate. Clearly if
$\rho_v(t_0)$ is non-zero and $K$ is non-positive, then eventually the vacuum energy will
dominate no matter how large the other initial densities or curvature
are.  If $K$ is positive then it is possible that the expansion
will be halted and reversed before the vacuum energy dominates the
expansion. Current experimental data roughly indicates the following
sequence of events: About 15 billion years ago there was the big
bang, and the Universe was radiation dominated. After several hundred
thousand years, the Universe became matter dominated. Then about three
billion years ago (red-shift one), the Universe became vacuum energy
dominated. In order for the Universe to have such a long period of
radiation and matter domination,
% it indicates that 
$K$ must be very
close to zero.

For radiation the temperature $T$ is related to the energy density by
\[
\rho \propto T^4.
\]
Today the radiation component of the Universe is still detectable even
though it only contributes a minute fraction of the current energy
density of the Universe. It has been red-shifted to microwave
frequencies and is known as the cosmic microwave background. It
has a black body spectrum due to the radiation being in thermal
equilibrium. Its detection was one of the main pieces
of evidence for the big bang theory.  Thus the temperature magnitude
and evolution of the early Universe can be inferred. This allows the
modelling of the formation of helium and the other light elements from
hydrogen during the early Universe, a process known as ``big bang
nuclear synthesis'' (BBN).  These inferred proportions of hydrogen,
helium and other light elements are in good agreement with observations
and provide further solid evidence for big bang Cosmology.

\section{Structure formation}
Although the Friedman equation provides an excellent description of
the behaviour on average, there is still the task of explaining the
formation of galaxies and galaxy clusters. Clearly gravitational
attraction will cause any inhomogeneities present in the Universe to
increase with time. So the early Universe would have to be more
homogeneous than today, but still there need to be some inhomogeneities
that can grow into the large scale structure we see today.

In the hot early Universe, the CMB radiation would have been coupled
to the baryons by Thomson scattering. The average path lengths of the
photons would have been very short as a consequence. However as the
Universe expanded and the temperature dropped to about $10^3$ K
the radiation became decoupled from the baryons and so the path length
of the photons became almost unlimited. Thus, since several hundred
thousand years after the big bang, the radiation photons would have
been free streaming. As a consequence when we observe them, we are
able to see a snap shot of what the Universe looked like at the time
of decoupling. In effect we have a picture of the spatial temperature
variation on a sphere surrounding us with a radius of about fifteen
billion light years. This allows us to infer what the density
inhomogeneities were at that time.

The first accurate detection of the inhomogeneities (or perturbations)
was made by the COBE satellite \cite{COBE}. It measured the
temperature variation on degree scales and showed it to be one part in
$10^5$. It also found that the power spectrum of the perturbations was
close to scale invariant.  This is a particularly simple perturbation
spectrum as it implies there is no characteristic length scale for the
perturbations. 

%Since then there have been a number of balloon
%experiments who have mapped the inhomogeneity on scales as low as 0.3
%degrees \cite{newcmb}. Within the next few years there will be
%satellite experiments which will map the CMB to even greater accuracy
%\cite{newcmbsat}. 

In reality the scale invariance is broken on scales smaller than COBE
was able to measure.  This is because the equations describing the
evolution of cosmological perturbations (see for example \cite{MFB})
show that a characteristic scale for the perturbations is the Hubble
length $1/H$.  Typically the adiabatic pressure does not effect the
perturbations on scales larger than the Hubble length. During the
standard cosmological evolution the Hubble radius is increasing in
size faster than the scale of the perturbations, i.e. for a
perturbation with a wave number $k/a$, the ratio $k/(aH)$ will be
increasing with time. Thus perturbations with scales initially greater
than the Hubble length will eventually have scales smaller than the
Hubble length. At this stage, pressure can counteract the tendency of
gravitational collapse of the perturbations. This can lead to
`acoustic oscillations' where the magnitude of the perturbation
oscillates with time. These oscillations lead to peaks in the CMB
power spectrum which have been observed in recent CMB experiments
\cite{net,halverson,lee}.

\section{Inflation}
 Although the big bang model is well validated by
experiment, it does seem to require rather special initial conditions.
At the time of decoupling the Hubble horizon (which is roughly equal
to the causal horizon) would only be about one square degree. Yet the
CMB is the same temperature to one part in $10^5$ in all parts of the
sky. What's more the deviations from homogeneity have the special form
of having a scale invariant power spectrum on scales larger than the
Hubble horizon. In the big bang model there is no causal way of
achieving this special setup and it just has to be put in by hand as an initial
condition. These are known as the `homogeneity' and `inhomogeneity'
problems, i.e. why is the Universe so homogeneous and why are the
deviations from homogeneity of such a special form.

Another unusual thing is how close the curvature $K$ must be to
zero in order to have such a long radiation and matter dominated
era. This is known as the flatness problem. 

A natural solution to these and various other problems can be provided
by the model known as inflation (see \cite{LiLy00} for a recent
review). From particle physics models one expects that under the
extreme conditions of the early Universe there would be one or more
scalar fields. It can be shown (see for example \cite{linde}) that
provided the potential energy of the scalar field dominates the
gradient and velocity energy in a volume larger than a Hubble volume
and the potential is sufficiently flat, then any inhomogeneities in the
field will be smoothed out. The criterion that the potential energy has
to dominate over a Hubble volume to some extent undermines the
solution to the horizon problem. If the `chaotic inflation' scenario is
considered, where the scalar field is taken to be distributed randomly
throughout the Universe at the Planck time, then it can be shown that
in some areas this criterion will be met just by chance, i.e. there is
a finite probability of the criterion being met and so if there is a
large area of randomly fluctuating scalar field, there is a large
probability that the criterion will be met somewhere. However if one
wants to consider inflation starting in some isolated patch well after
the Planck time, such as in `new inflation', then the homogeneity
criterion becomes more restrictive.
This issue is examined further 
% I present an analysis of this issue
%in more detail in
% Appendix
in  Chapter \ref{ic}.

The other criterion that the potential must be sufficiently flat is
quantified by the slow roll conditions:
\begin{eqnarray}
\epsilon(\phi) &=& \frac{1}{16\pi G} \left(\frac{V'}{V} 
 \right)^2 \ll 1 \\
|\eta(\phi)| &=& \left| \frac{1}{8\pi G} \frac{V''}{V}  \right| \ll 1
\end{eqnarray}
where $\phi$ is the scalar field value, primes denote differentiation
with respect to $\phi$ and $V$ denotes the potential of $\phi$. It can
be shown that if the slow roll conditions are satisfied then the scale
factor accelerates with time and the curvature is driven to zero and
so inflation solves the flatness problem. To be more specific 
\[
\ddot{a} > 0
\]
where dot denotes differentiation with respect to proper time. This is in
contrast to ordinary matter where $a$ deaccelerates and the curvature
moves away from zero.

For a homogeneous scalar field, the evolution is given by the Klein
Gordon equation
 \[
 \ddot{\phi} + 3H\dot{\phi} + \frac{dV}{d\phi}=0
 \]
For a homogeneous scalar field dominated Universe the Friedman equation is given
by
\[
H^2 = \frac{8\pi G}{3} \left( \frac{\dot{\phi}^2}{2} + V(\phi)
\right).
\]
Applying the slow roll conditions, the Klein Gordon and Friedman
equations can be approximated by 
\[
H^2 \approx \frac{8\pi G}{3} V(\phi)
\]
and
\[
3H \dot{\phi} \approx -V'(\phi).
\]
A common potential used is
\[
V= \frac{1}{2} m^2 \phi^2
\]
where $\sqrt{V''}=m$ is the effective mass of $\phi$.

Remarkably inflation also solves the inhomogeneity problem. On
sub-Hubble scales, quantum fluctuations in the value of $\phi$
occur. In inflation the Hubble scale is almost constant, while the
scale factor is rapidly increasing. Thus the sub-Hubble vacuum
fluctuations eventually are stretched to super-Hubble scales. At this
stage they lead to classical perturbations in the value of $\phi$ on
super-Hubble scales.

It can be shown (see for example \cite{LiLy00}) that the power spectrum
of fluctuations generated by inflation has spectral index
\[
n = 1 + 2\eta - 6\epsilon
\]
where $n=1$ represents scale invariance.  These fluctuations are
inherited by the radiation era when the scalar field decays into
radiation and other forms of matter after inflation ends. So
remarkably, the same slow roll conditions solve the homogeneity,
flatness and inhomogeneity problems.

\section{Reheating}
Inflation ends when the slow roll conditions are violated. This is the
stage when $\phi$ (the `inflaton') oscillates in the valley of the
potential $V$.  It is then thought to decay into photons ($\gamma$),
non-baryonic or `cold dark matter' matter ($c$), baryonic matter ($b$)
and neutrinos ($\nu$).  This process is known as `reheating'.

Simple aspects of this process of the transfer of energy from the
inflaton can be modelled by the addition of a scalar field $\chi$
which has a coupling term $\frac{1}{2}g^2\phi^2\chi^2$ in the
potential, where $g$ is a coupling constant. Neglecting perturbations
in the metric (a full discussion or perturbations is given in Chapter
\ref{ch:adent}), the equation of motion for the perturbation of wave
number $k$ of $\chi$ is given by:
\[
\delta \ddot{\chi}_k + 3H\delta\dot{\chi}_k +\left( \frac{k^2}{a^2} +
g^2\phi^2 \right) \delta \chi_k = 0.
\]
At the end of inflation $\phi$ is oscillating in the valley of the
potential. For $V=\frac{1}{2}m^2 \phi^2$ the behaviour of $\phi$ is
well approximated by 
\[
\phi \approx \phi_0 \sin mt
\]
where $\phi_0$ is the amplitude of the oscillations. for small wave
lengths ($k/a \gg H$) the expansion can be neglected. The
perturbation equation can then be transformed to the well known
Mathieu equation (see for example \cite{mathieu})
\[
\frac{d^2 \delta \chi_k}{dz^2} + (A - 2q\cos 2z)\delta \chi_k = 0
\]
where $A = k^2/(m^2a^2) + 2q$, $q=g^2\phi_0^2/(4m^2)$,
and $z=mt$. The behaviour of this equation can be understood with the
aid of the Mathieu chart \cite{mathieu} where there exist bands of
instability where for certain value of $A$ and $q$ there is
exponential growth of $\delta \chi_k$. Including expansion means that modes
of $\delta \chi_k$ move in and out of bands of resonance.
%As $\phi$ is oscillating at the end of inflation, this equation can
%undergo parametric resonance which can lead to exponential growth of
%$\delta \chi_k$. 
This process is known as `preheating' \cite{KLS1,KLS2}.
This represents the transfer of energy from the
homogeneous field $\phi$ into the perturbations of $\chi$. The
homogeneous field $\phi$ represents a condensate of zero momentum
particles while $\delta \chi_k$ represent bosons with momentum
$k$. Thus preheating models the exponential production of
bosons. Fermions can also be considered. Once most of the energy of
$\phi$ has been transferred into $\delta \chi$ then the $\chi$
and remaining $\phi$ particles can in turn decay into other particles
which eventually thermalize and make up the contents of the radiation
era.

The effect of having a preheating phase is to end the oscillating
phase of $\phi$ more rapidly and it can lead to a higher
temperature. There is also the possibility that the super-Hubble scale
perturbations could be amplified. This would have the undesirable
consequence of the complicated and uncertain physics of preheating having
an effect on the observed anisotropies in the CMB. This issue will be
discussed further in Chapter \ref{preheating}.

\section{Adiabatic and entropy perturbations}
Another special property of the fluctuations generated by a single
scalar field is that they are adiabatic, i.e. the perturbation in
$\phi$ can be expressed as a time shift in the background scalar
field:
\[
\delta \phi = \dot{\phi}(t) \delta t(t, {\bf x})
\]
where $\bf x$ is the spatial position vector and $\delta t$ is the
time shift of the background solution needed to produce the
perturbation.

The adiabatic nature of the perturbations is inherited by the
perturbations in the radiation era once the scalar field decays. So we
have 
\[
\delta t = \frac{\delta \rho_\gamma}{\dot{\rho_\gamma}} = \frac{\delta
  \rho_\nu}{\dot{\rho_\nu}} = \frac{\delta \rho_c}{\dot{\rho_c}} =
\frac{\delta \rho_b}{\dot{\rho_b}}.
\]
%??need to define quantities
Using the continuity equation
\[
\dot{\rho} = -3H(\rho +p)
\]
the adiabatic condition becomes
\[
 \frac{3}{4}\frac{\delta \rho_\gamma}{{\rho_\gamma}} =
\frac{3}{4}\frac{\delta \rho_\nu}{{\rho_\nu}} = 
\frac{\delta \rho_c}{{\rho_c}} = \frac{\delta \rho_b}{\rho_b} .
\]
This can also be expressed in terms of perturbations in particle
number $n$
\[
 \delta \left(\frac{ n_i}{n_j}\right) = 0
\]
where $i$ and $j$ can be $\gamma, \nu, c$ or $b$.
Although the adiabatic condition is quite restrictive, it does seem to
be in good agreement with observations (see for example \cite{net,halverson,lee})
and as mentioned above is also predicted by the simplest inflationary
models. However a priori there seems to be no reason why the primordial
perturbations have to satisfy the adiabatic condition. 

For example if there are two scalar fields ($\phi$ and $\chi$) present
during inflation then the perturbations in each field do not
necessarily have to correspond to the same time shift, i.e.
\[
\frac{\delta \phi}{\dot{\phi}} \not= \frac{\delta \chi}{\dot{\chi}}.
\]
The generation of and evolution of adiabatic perturbations during
multi-field inflation models is examined further in Chapter
\ref{ch:adent}. Another important issue that is discussed is the
possibility of there being correlations between the adiabatic and
entropy perturbations.

If there are entropy perturbations in the inflation stage, these can
lead to entropy perturbations in the radiation dominated phase.
For example we could have
\[
\frac{\delta \rho_c}{\rho_c} \not= \frac{3}{4} \frac{\delta \rho_\gamma}{\rho_\gamma}.
\]
The entropy perturbation is defined as that part of the perturbation
which does not satisfy the adiabatic condition, i.e.
\[
{\cal S}_{c \gamma} = \frac{\delta \rho_c}{\rho_c} - \frac{3}{4} \frac{\delta
  \rho_\gamma}{\rho_\gamma}
\]
where 
${\cal S}_{c \gamma}$ is the entropy perturbation between cold dark
matter and photons. 

The effect of entropy perturbations in the radiation era on the
observed CMB spectrum is discussed in Chapter \ref{ch:cmb}. Using current
data the magnitude and correlation of the entropy perturbation is
estimated.

\section{Brane world cosmology}
String/M-theory theory predicts the existence of extra spatial
dimensions.  As a simple example, our Universe could be a three plus
one hyper-surface (the `brane') embedded in a four plus one space-time
(the `bulk').

Until recently it was thought that any extra dimensions would be too
small to be detectable, since it was thought that the deviations in
gravity would contradict existing experiments. However Arkani-Hamed
{\it et. al.} \cite{ADD} presented models with large compact extra
dimensions, and then Randall and Sundrum (RS) \cite{RS1,RS2} showed that even
an infinite extra dimension is possible since gravity can be localized
near the brane by the curvature of the bulk.

% that if the bulk has a five dimensional
%anti-Desitter metric then large extra dimensions are possible.
%Current experiments looking for deviations of
%Newton's inverse square law indicate that the extra dimensions could be
%detectable at energies as low as a TeV.

In the RS model, the bulk is anti-de Sitter and the brane has positive
vacuum energy (the `tension') which cancels out the bulk negative
vacuum energy. 
%It also contains the matter of the standard model.
The particles of the standard model are confined to the brane.

There has been much interest in seeing whether such a
scenario would have any cosmological implications (see \cite{roy} for
a review). In particular the question of whether the extra dimension
would lead to an identifiable signature on the CMB is of great
interest. 

The bulk affects the perturbation equations on the brane in two
ways. It adds terms which are quadratic in the brane energy
momentum tensor. It also adds terms from the projected Weyl tensor of
the bulk. These can be viewed as an additional fluid on the brane and
so lead to the possibility of additional entropy perturbations. This
and other related issues are discussed in Chapter \ref{ch:branes}.

% A formalism for decomposing a multi-field fluctuation into an adiabatic
% and entropy part is proposed. This formalism is then used to study
% under what circumstances multi-field models can lead to non-adiabatic
% perturbations in the radiation era.

% In Chapter \ref{ch:cmb} the effect of correlated non-adiabatic perturbations on
% the CMB is investigated. Current data is used to constrain the
% quantity and nature of the non-adiabatic perturbations. 
% %The chapter is
% %based on the work in \cite{gordon_cmb}.

% In Chapter \ref{ch:branes}, the effects of an extra spatial dimension
% on the perturbation equations is investigated.  %This work is based on
% %\cite{gordon_branes}.  The M-theory inspired ``brane world'' models
% Brane world models \cite{ransun} are studied, where matter is confined
% to a 3 + 1 dimensional membrane and gravity exists in 4 + 1
% dimensions. It is shown that the presence of the extra dimension can
% introduce non-adiabatic perturbations.  %This chapter uses the
% %covariant approach to %perturbations while the other chapters use the
% %metric based approach.

% The results of the thesis are then summarised in Chapter
% \ref{ch:conc} and current and future extensions to the work are
% discussed. 

\chapter{Adiabatic and entropy perturbations from inflation}

\label{ch:adent}

\newcommand\eq[1]{Eq.~(\ref{#1})}
\newcommand\eqs[2]{Eqs.~(\ref{#1}) and (\ref{#2})}
\newcommand\eqss[3]{Eqs.~(\ref{#1}), (\ref{#2}) and (\ref{#3})}
\newcommand\eqsss[4]{Eqs.~(\ref{#1}), (\ref{#2}), (\ref{#3})
and (\ref{#4})}
\newcommand\eqssss[5]{Eqs.~(\ref{#1}), (\ref{#2}), (\ref{#3}),
(\ref{#4}) and (\ref{#5})}
\newcommand\eqst[2]{Eqs.~(\ref{#1})--(\ref{#2})}

\newcommand\ee{\end{equation}}
\newcommand\be{\begin{equation}}
\newcommand\eea{\end{eqnarray}}
\newcommand\bea{\begin{eqnarray}}
\renewcommand{\topfraction}{0.99}

\section{Introduction}

%A period of accelerated expansion -- 
As discussed in the introduction, inflation 
%-- 
in the early
universe has become the standard model for the origin of structure.
%in the universe. 
Inhomogeneities in the present matter
distribution can be traced back to quantum fluctuations in the
fields driving inflation which are stretched beyond the Hubble
scale during inflation. In the simplest models of inflation driven by a
single scalar field, these fluctuations produce a primordial
adiabatic spectrum
whose amplitude can be characterized by the comoving curvature
perturbation ${\cal R}$, which remains constant on super-Hubble
scales until the perturbation comes back within the Hubble scale
long after inflation has ended.

As soon as one considers more than one scalar field, one must also
consider the role of non-adiabatic fluctuations. This can have
important consequences, both in affecting the evolution of the
curvature perturbation (often referred to as the `adiabatic
perturbation'), but also in the possibility of seeding isocurvature
(or `entropy') perturbations after inflation.

Previous studies have demonstrated that non-adiabatic pressure
perturbations can alter the curvature perturbation on super-Hubble
scales either during inflation~\cite{GBW,Shinji} or
after~\cite{HK,Hu,WMLL,Bassett:1999mt,BKM1,Finelli:1999bu}.  A general formalism to evaluate the
curvature perturbation at the end of inflation in multiple field
models was developed in Ref.~\cite{SS}.  In the presence of
non-adiabatic fluctuations, one must follow the evolution of perturbed
fields on super-Hubble scales, in particular tracking the perturbation
in the integrated expansion~\cite{Star,SS,Salopek,SasTan,LR,WMLL}, in
order to 
evaluate the large-scale curvature perturbation at late
times~\cite{Polarski:1992dq,PolSta94,SY,GBW,SS,Salopek,SasTan,MukSte,julien}.

However no similar formalism has been developed so far to evaluate
the isocurvature perturbation in the general case. Instead,
isocurvature perturbations have been studied in a number of
particular models of inflation~\cite{many,Polarski:1992dq,PolSta94}.
These fluctuations typically arise as baryon modes
(e.g.~\cite{baryon}) or cold dark matter modes~\cite{P}, but
neutrino isocurvature modes have also been considered~\cite{BMT}.
Recently, it has been pointed out~\cite{Langlois,BMT2} that it is
rather natural to expect the curvature and isocurvature
perturbations to be correlated, which yields distinctive
observational results~\cite{Langlois2}, in contrast to the
pure or uncorrelated isocurvature perturbations usually tested against
observations~\cite{isocmb,enqvistBM}.

In this Chapter we will develop a general formalism to study the
evolution of both curvature and isocurvature perturbations in a wide
class of multi-field inflation models by decomposing field
perturbations into perturbations along the background trajectory in
field space (the adiabatic field perturbation), and orthogonal to the
background trajectory (the entropy field).  We allow an arbitrary
interaction potential for the fields, and, although we concentrate
upon the case of two scalar fields, the general approach can be easily
extended to $N$ fields, where there will be $N-1$ entropy fields
orthogonal to the background trajectory. This was done for a specific
assisted inflation model in Ref.~\cite{Karim}. We will work in the
metric based approach of Bardeen~\cite{Bardeen} in order to define
gauge-invariant cosmological perturbations, but our formalism can also
be applied to the study of multiple scalar fields in other
approaches~\cite{Hwang,Ellis,Durrer}.

We begin by reviewing the standard results obtained in single field
models, emphasizing the suppression of non-adiabatic fluctuations on
large-scales. We then extend our analysis to general two-field models,
defining an adiabatic field and an entropy field, whose fluctuations,
though uncorrelated on small scales, may develop correlations through
the subsequent evolution.
%Thesis:
% We present two specific models of two-field
%inflation, one with non-interacting fields, the other a model of
%interacting fields which undergo preheating after inflation.

\section{Perturbation equations for multiple scalar fields}

We consider $N$ scalar fields  with  Lagrangian density:
\begin{gather}
{\cal L} = -V(\varphi_1,\cdots,\varphi_N)
 -\frac{1}{2} \sum_{I=1}^{N} g^{\mu\nu} \varphi_{I,\mu}\varphi_{I,\nu}
 \,,
\label{lag}
\end{gather}
and minimal coupling to gravity. 
In order to study the evolution of linear perturbations in the scalar
fields, we make the standard splitting $\varphi_I(t,{\bf x})
\to\varphi_I(t)+\delta\varphi_I(t,{\bf x})$. 
The field equations, derived from Eq.~(\ref{lag}) for the background
homogeneous fields, are
\begin{equation}
\label{eq:KG} 
\ddot{\varphi}_I + 3H\dot{\varphi}_I + V_{\varphi_I} = 0\,,
\end{equation}
where $V_{x} = {\partial V}/{\partial x}$, and the Hubble rate, $H$,
in a spatially flat Friedmann-Robertson-Walker (FRW) universe, is
determined by the Friedman equation:
\begin{gather}
H^2 = \left(\frac{\dot{a}}{a}\right)^2 =\frac{8\pi G}{3} \left[
V(\varphi_I) + \frac{1}{2}  \sum_I \dot\varphi_I^{~2}
 \right]\,, \label{eq:hubble}
\end{gather}
with $a(t)$ the FRW scale factor.

Consistent study of the linear field fluctuations $\delta\varphi_I$
requires that we also consider linear scalar perturbations of the
metric, corresponding to the line element\footnote{
We follow the notation of Ref.~\cite{MFB}, apart from our use of
$A$ rather than $\phi$ as the perturbation in the lapse function.
}
\begin{eqnarray}
ds^2 &=& - (1+2A)dt^2 + 2aB_{,i}dx^idt \nonumber\\ &&~{} +
a^2\left[ (1-2\psi)\delta_{ij} + 2E_{,ij}\right] dx^idx^j \,,
\end{eqnarray}
where we have not at this stage specified any particular choice of
gauge~\cite{MFB,Bardeen,KS}.

Scalar field perturbations, with comoving wavenumber $k=2\pi
a/\lambda$ for a mode with physical wavelength $\lambda$, then
obey the perturbation equations
\begin{eqnarray}
&& \ddot{\delta\varphi}_I + 3H\dot{\delta\varphi}_I
 + \frac{k^2}{a^2} \delta\varphi_I + \sum_J V_{\varphi_I\varphi_J}
\delta\varphi_J
  \nonumber\\ &&~~{}=
-2V_{\varphi_I}A + \dot\varphi_I \left[ \dot{A} + 3\dot{\psi} +
\frac{k^2}{a^2} (a^2\dot{E}-aB) \right] \,. \label{eq:perturbation}
\end{eqnarray}
The metric terms on the right-hand-side, induced  by the scalar
field perturbations, obey the energy and momentum constraints
\begin{eqnarray}
3H\left(\dot\psi+HA\right) + 
\frac{k^2}{a^2}\left[\psi+H(a^2\dot{E}-aB)\right] &=& -4\pi G \delta\rho \,,
\label{eq:densitycon}
\\
\dot\psi + HA &=& -4\pi G \delta q \,.
\label{eq:mtmcon}
\end{eqnarray}
The total energy and momentum perturbations are given in terms of
the scalar field perturbations by
\begin{eqnarray}
\delta\rho &=& \sum_I\left[
 \dot\varphi_I \left( \dot{\delta\varphi}_I -\dot\varphi_I A \right)
 + V_{\varphi_I}\delta\varphi_I \right]
\label{eq:density} \\ 
\delta q_{,i} &=& - \sum_I \dot{\varphi}_I
\delta\varphi_{I,i} \,. 
\label{eq:mtm}
\end{eqnarray}
These two equations can be combined to construct a gauge-invariant
quantity, the comoving density perturbation~\cite{Bardeen}
\begin{eqnarray}
\label{def:epsilonm} \epsilon_m &\equiv& \delta\rho -3H\delta q
\nonumber\\ &=& \sum_I \left[ \dot\varphi_I \left(
\dot{\delta\varphi}_I - \dot\varphi_I A \right) - \ddot\varphi_I
\delta\varphi_I \right]\,,
\end{eqnarray}
which is sometimes used to represent the total matter perturbation.

Because the anisotropic stress vanishes to linear order for scalar
fields minimally coupled to gravity, we have a further constraint on
the metric perturbations:
\begin{equation}
\label{eq:aniso} \left( a^2\dot{E}-aB
\right)^{\displaystyle{\cdot}} + H \left( a^2\dot{E}-aB \right) +
\psi - A = 0 \,.
\end{equation}

The coupled perturbation
equations~(\ref{eq:perturbation})--(\ref{eq:mtm}) and
(\ref{eq:aniso}) are probably most often solved in the zero-shear
(or longitudinal or conformal Newtonian) gauge, in which
$a^2\dot{E}_\ell-aB_\ell=0$~\cite{MFB}. The two remaining metric
perturbation variables which appear in the scalar field
perturbation equation, $A_\ell\equiv\Phi$ and $\psi_\ell\equiv\Psi$, are
then equal in the absence of any anisotropic stress by
Eq.~(\ref{eq:aniso}).

Another useful choice is the spatially flat gauge, in which
$\psi_Q=0$~\cite{KS,Hwang}.  The scalar field perturbations in this
gauge are sometimes referred to as the Sasaki or Mukhanov
variables~\cite{SM}, which have the gauge-invariant definition
\begin{equation}
\label{def:QI}
Q_{I} \equiv \delta\varphi_I + \frac{\dot\varphi_I}{H}\psi \,.
\end{equation}
The shear perturbation in the spatially flat gauge is simply
related the curvature perturbation, $\Psi$, in the zero-shear
gauge:
\begin{equation}
a^2\dot{E}_Q - aB_Q = a^2\dot{E} - aB + \frac{1}{H}\psi = \frac{1}{H}\Psi \,.
\end{equation}
The energy and momentum constraints, Eqs.~(\ref{eq:densitycon})
and~(\ref{eq:mtmcon}), in the spatially flat gauge thus yield
\begin{eqnarray}
\label{eq:Psi}
\frac{k^2}{a^2} \Psi
 &=& - {4\pi G} \epsilon_m
 \,,\\
\label{eq:AQ} HA_Q &=& - 4\pi G \delta q_Q \,,
\end{eqnarray}
where $\epsilon_m$ is given in Eq.~(\ref{def:epsilonm}), and
from Eq.~(\ref{eq:mtm}) we have $\delta q_Q = -\sum_I\dot\varphi_I
Q_I$. 

The equations of motion, Eq.~(\ref{eq:perturbation}), rewritten in
terms of the Sasaki-Mukhanov variables, and using
Eqs.~(\ref{eq:Psi}) and (\ref{eq:AQ}) to eliminate the metric
perturbation terms in the spatially flat gauge, become~\cite{TN}:
\begin{eqnarray}
&&\ddot Q_I + 3 H \dot Q_I + \frac{k^2}{a^2} Q_I  \nonumber\\
&&~~{}
 + \sum_{J} \left[ V_{\varphi_I\varphi_J} - \frac{8 \pi G}{a^3}
\left( \frac{a^3}{H} \dot \varphi_I \dot \varphi_J
\right)^{\displaystyle{\cdot}} \right] Q_J = 0 \,. \label{eq:multi
Q}
\end{eqnarray}

\subsection{Curvature and  entropy perturbations}

The comoving curvature perturbation~\cite{Lukash,Lyth85} is given by
\begin{eqnarray}
\label{def:calR} {\cal R} &\equiv& \psi - \frac{H}{\rho+p} \delta q
 \nonumber\\
&=&  \sum_I \left(\frac{\dot\varphi_I}{\sum_J \dot\varphi_J^2}
\right) Q_{I} \,.
\end{eqnarray}
This can also be given in terms of the metric perturbations in the
longitudinal gauge as~\cite{MFB}
\begin{equation}
{\cal R} = \Psi - {H\over\dot{H}} \left( \dot\Psi+H\Phi \right) \,.
\end{equation}

For comparison we give the curvature perturbation on
uniform-density hypersurfaces,
\begin{equation}
{}-\zeta \equiv \psi + H {\delta\rho\over \dot\rho} \,,
\end{equation}
first introduced by Bardeen, Steinhardt and Turner~\cite{BST} as a
conserved quantity for adiabatic perturbations on large
scales~\cite{MarSch,WMLL}. It is related to the comoving curvature
perturbation in Eq.~(\ref{def:calR}) by a gauge transformation
\begin{equation}
{}-\zeta = {\cal R} + {2\rho\over3(\rho+p)} \left( {k\over aH}
\right)^2 \Psi \,,
\end{equation}
where we have used the constraint equation~(\ref{eq:Psi}) to
eliminate the comoving density perturbation, $\epsilon_m$. Note
that ${\cal R}$ and $-\zeta$ thus coincide in the limit $k\to0$.

Both ${\cal R}$ and $-\zeta$ are commonly used to characterise the
amplitude of adiabatic perturbations as both remain constant for
purely adiabatic perturbations on sufficiently large scales as a
direct consequence of local energy-momentum conservation~\cite{WMLL},
allowing one to relate the perturbation spectrum on large scales to
quantities at the Hubble scale crossing during inflation in the simplest
inflation models~\cite{BST,LL93}.

A dimensionless definition of the total entropy perturbation
(which is automatically gauge-invariant) is given by
\begin{equation}
\label{S_total}
{\cal S} = H \left( {\delta p \over \dot{p}} - {\delta\rho \over
\dot\rho} \right) \,,
\end{equation}
which can be extended to define a generalised entropy perturbation
between any two matter quantities
$x$ and $y$:
\begin{equation}
{\cal S}_{xy} = H \left( {\delta{x}\over \dot{x}} - {\delta{y}
\over \dot{y}} \right) \,.
\end{equation}
The total entropy perturbation in Eq.~(\ref{S_total}) for $N$ scalar
fields is given by 
\begin{equation}
\label{defSN} 
{\cal S} = \frac {
2\left( \dot{V} + 3H
\sum_J \dot\varphi_J^2 \right) \delta V 
 + 
2\dot{V}\sum_I \dot\varphi_I (
\dot{\delta\varphi}_I - \dot\varphi_I A )
} {3
\left(2\dot{V}+3H\sum_J \dot\varphi_J^2\right) \sum_I
\dot\varphi_I^2} \,,
\end{equation}
where the perturbation in the total potential energy is given by
$\delta V = \sum_I V_{\varphi_I}\delta\varphi_I$.

The change in ${\cal R}$ on large scales (i.e., neglecting spatial
gradient terms) can be directly related to the non-adiabatic part of
the pressure perturbation~\cite{GBW,WMLL,FB}
\begin{equation}
\dot{\cal R} \approx -3H {\dot{p}\over\dot\rho} {\cal S} \,.
\end{equation}
We will  now consider the evolution of the adiabatic and entropy
perturbations in both one- and two-field models of inflation.

\subsection{Single field}

Perturbations in a single self-interacting scalar field obey the
gauge-dependent equation of motion
\begin{eqnarray}
\label{eq:varphi} && \ddot{\delta\varphi} + 3H\dot{\delta\varphi}
+ \left( \frac{k^2}{a^2}+V_{\varphi\varphi} \right) \delta\varphi
   \nonumber\\ &&~~{}=
-2V_{\varphi}A + \dot\varphi \left[ \dot{A} + 3\dot{\psi} +
\frac{k^2}{a^2} (a^2\dot{E}-aB) \right] \,,
\end{eqnarray}
subject to the energy and momentum constraint equations given in
Eqs.~(\ref{eq:densitycon}--\ref{eq:mtm}).
%, where the energy
%and momentum perturbations for a single scalar field are given by
%\begin{eqnarray}
%\delta\rho &=& \dot\varphi \left( \dot{\delta\varphi} -\dot\varphi
%A \right) + V_{\varphi}\delta\varphi\,, \label{eq:density1} \\
%\delta q_{,i} &=& - \dot{\varphi}\delta\varphi_{,i} \,.
%\label{eq:mtm1}
%\end{eqnarray}

The scalar field perturbation in the spatially flat gauge has the
gauge-invariant definition, Eq.~(\ref{def:QI}),
\begin{equation}
Q_{\varphi} \equiv \delta\varphi +
{\dot\varphi\over H}\psi\,.
\end{equation}
For a single field this is directly related to the curvature
perturbation in the comoving gauge, where the momentum,
$\delta q=-\dot\varphi\delta\varphi$,
vanishes
%and hence, from Eq.~(\ref{eq:mtm1}), $\delta\varphi_m=0$,
\begin{equation}
\label{eq:zeta1}
{\cal R} = \psi + {H\over \dot\varphi} \delta\varphi = {H\over
\dot\varphi} Q_\varphi \,.
\end{equation}

%{}From Eq.~(\ref{def:epsilonm}) we have the comoving density perturbation
%\begin{equation}
%\label{eq:epsilonm} \epsilon_m = \dot\varphi \left(
%\dot{\delta\varphi} - \dot\varphi A \right) -
%\ddot\varphi\delta\varphi \,,
%\end{equation}
%and hence, from Eqs.~(\ref{eq:density1}) and~(\ref{eq:mtm1}), we
%obtain the constraints
%\begin{eqnarray}
%{k^2 \over a^2} \Psi &=& -4\pi G \left[ \dot\varphi \left(
%\dot{\delta\varphi} - \dot\varphi A \right)
% - \ddot\varphi\delta\varphi \right] \,,\label{a}\\
%A_Q &=& 4\pi G {\dot\varphi \over H} Q_\varphi \,.
%\end{eqnarray}
%Substituting these expressions for the metric perturbations into
%Eq.~(\ref{eq:varphi}), yields the decoupled equation of motion for
%the scalar field perturbation,
%\begin{equation}
%\ddot Q_\varphi + 3 H \dot Q_\varphi + \left[ \frac{k^2}{a^2} +
%V_{\varphi\varphi} - \frac{8 \pi G}{a^3} \left(\frac{a^3}{H}
%\dot\varphi^2 \right)^{\displaystyle{\cdot}} \right] Q_\varphi = 0
%\,. \label{eom:Q}
%\end{equation}

It is not obvious that the intrinsic entropy perturbation for a single
scalar field, obtained from Eq.~(\ref{defSN}),
\begin{equation}
\label{eq:intS}
{\cal S}
 = {2V_\varphi \over 3\dot\varphi^2(3H\dot\varphi+2V_\varphi)}
 \left[ \dot\varphi \left( \dot{\delta\varphi} - \dot\varphi A \right) -
\ddot\varphi \delta\varphi\right] \,,
\end{equation}
should vanish on large scales. Because the scalar field obeys a
second-order equation of motion, its general solution contains two
arbitrary constants of integration, which can describe both
adiabatic and entropy perturbations.  However ${\cal S}$ for a
single scalar field is proportional to the comoving density
perturbation given in Eq.~(\ref{def:epsilonm}), and this in turn is
related to the metric perturbation, $\Psi$, via
Eq.~(\ref{eq:Psi}), so that~\cite{Bassett:1999mt}
\begin{equation}
\label{eq:S1}
{\cal S}
 = - {V_{\varphi} \over 6\pi G\dot\varphi^2[3H\dot\varphi+2V_\varphi]}
 \left( \frac{k^2}{a^2} \Psi \right) \,.
\end{equation}
In the absence of anisotropic stresses, $\Psi$ must be of order $A_Q$,
by Eq.~(\ref{eq:aniso}), and hence the non-adiabatic pressure becomes
small on large scales~\cite{SS,Bassett:1999mt,LR}. The amplitude of
the asymptotic solution for the scalar field at late times (and hence
large scales) during inflation thus determines the amplitude of an
adiabatic perturbation.

The change in the comoving curvature perturbation is given by
\begin{equation}\label{zetadot2}
\dot{\cal R} = {H\over \dot{H}} {k^2 \over a^2} \Psi
 \,,
\end{equation}
and hence the rate of change of the curvature perturbation, given
by $d\ln{\cal R}/d\ln a\sim (k/aH)^2$, becomes negligible on large
scales during single-field inflation.

\subsection{Two fields}

In this section we will consider two interacting scalar fields,
$\phi\equiv\varphi_1$ and $\chi\equiv\varphi_2$. The analysis
developed here should be straightforward to extend to include
additional scalar fields, but we do not expect to see any
qualitatively new features in this case, so for clarity we restrict
our discussion here to two fields.

In order to clarify the role of adiabatic and entropy
perturbations, their evolution and their inter-relation, we define
new adiabatic and entropy fields
by a rotation in field space. The ``adiabatic field'', $\sigma$,
represents the path length along the classical trajectory, such that
\begin{equation}
 \label{eq:sigma}
\dot\sigma = (\cos\theta) \dot\phi + (\sin\theta) \dot\chi \,,
\end{equation}
where
\begin{equation}
  \label{eq:cos sin}
\cos\theta = \frac{\dot{\phi}}{\sqrt{\dot{\phi}^2 +
\dot{\chi}^2}}, \quad \sin\theta =
\frac{\dot{\chi}}{\sqrt{\dot{\phi}^2 + \dot{\chi}^2}}\,.
\end{equation}
This definition, plus the original equations of motion for $\phi$ and
$\chi$, give
\begin{equation}
 \label{eq:sigma_dot_dot}
 \ddot{\sigma} + 3H\dot{\sigma} + V_\sigma = 0\,,
\end{equation}
where
\begin{eqnarray}
\label{eq:V_sigma}
V_\sigma &=& (\cos \theta) V_\phi + (\sin\theta) V_\chi\,.
\end{eqnarray}
As illustrated in Fig.~\ref{fig:decomposition}, $\delta \sigma$ is
the component of the two-field perturbation vector along the
direction of the background fields' evolution.
\begin{figure}[htbp]
  \begin{center}
\begin{picture}(0,0)%
\epsffile{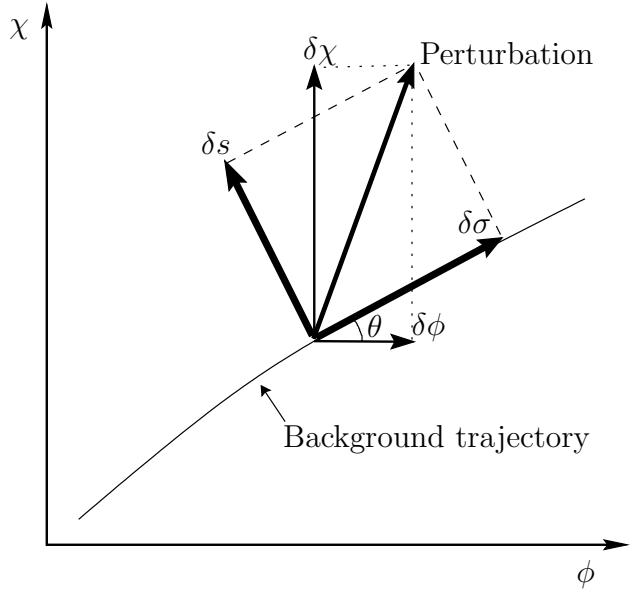}%
\end{picture}%
\setlength{\unitlength}{3947sp}%
\begingroup\makeatletter\ifx\SetFigFont\undefined%
\gdef\SetFigFont#1#2#3#4#5{%
  \reset@font\fontsize{#1}{#2pt}%
  \fontfamily{#3}\fontseries{#4}\fontshape{#5}%
  \selectfont}%
\fi\endgroup%
\begin{picture}(3897,3705)(1271,-3424)
\put(4085,-1171){\makebox(0,0)[lb]{\smash{\SetFigFont{12}{14.4}
{\rmdefault}{\mddefault}{\updefault}$\delta \sigma$}}}
\put(2990,-2517){\makebox(0,0)[lb]{\smash{\SetFigFont{12}{14.4}
{\rmdefault}{\mddefault}{\updefault}Background trajectory}}}
\put(3847,-113){\makebox(0,0)[lb]{\smash{\SetFigFont{12}{14.4}
{\rmdefault}{\mddefault}{\updefault}Perturbation}}}
\put(3116,-76){\makebox(0,0)[lb]{\smash{\SetFigFont{12}{14.4}
{\rmdefault}{\mddefault}{\updefault}$\delta \chi$}}}
\put(2476,-676){\makebox(0,0)[lb]{\smash{\SetFigFont{12}{14.4}
{\rmdefault}{\mddefault}{\updefault}$\delta s$}}}
\put(3798,-1813){\makebox(0,0)[lb]{\smash{\SetFigFont{12}{14.4}
{\rmdefault}{\mddefault}{\updefault}$\delta \phi$}}}
\put(3517,-1825){\makebox(0,0)[lb]{\smash{\SetFigFont{12}{14.4}
{\rmdefault}{\mddefault}{\updefault}$\theta$}}} \put(1271,
84){\makebox(0,0)[lb]{\smash{\SetFigFont{12}{14.4}{\rmdefault}
{\mddefault}{\updefault}$\chi$}}}
\put(4831,-3366){\makebox(0,0)[lb]{\smash{\SetFigFont{12}{14.4}
{\rmdefault}{\mddefault}{\updefault}$\phi$}}}
\end{picture}
\caption[An illustration of the decomposition of an arbitrary perturbation
into an adiabatic ($\delta \sigma$) and entropy ($\delta s$)
component.]{
An illustration of the decomposition of an arbitrary perturbation
into an adiabatic ($\delta \sigma$) and entropy ($\delta s$)
component. The angle of the tangent to the background trajectory
is denoted by $\theta$. The usual perturbation decomposition,
along the $\phi$ and $\chi$ axes, is also shown.
}
    \label{fig:decomposition}
  \end{center}
\end{figure}
%The parameter $\sigma$ acts as an affine parameter for the
%background solution. 
Conversely, fluctuations orthogonal to the
background classical trajectory represent non-adiabatic
perturbations, and we define the ``entropy field'', $s$, such that
\begin{gather}
\label{eq:s}
\delta s = (\cos\theta) \delta\chi - (\sin\theta) \delta\phi\,.
\end{gather}
From this definition, it follows that $s=$constant along the
classical trajectory, and hence entropy perturbations are
automatically gauge-invariant~\cite{StewartWalker}. Perturbations
in $\delta\sigma$, with $\delta s=0$, describe adiabatic
field perturbations, and this is why we refer to $\sigma$ as the
``adiabatic field''.

The total momentum of the two-field system, given by
Eq.~(\ref{eq:mtm}), is then
\begin{equation}
\delta q_{,i} = - \dot\phi\delta\phi_{,i} - \dot\chi\delta\chi_{,i}
= - \dot\sigma\delta\sigma_{,i} \,,
\end{equation}
and the comoving curvature perturbation in Eq.~(\ref{def:calR}) is
given by
\begin{eqnarray}
\label{eq:zeta}
{\cal R}
 &=& \psi + H\left( \frac{\dot\phi\delta\phi
              + \dot\chi\delta\chi}{\dot\phi^2 + \dot\chi^2}\right) 
\nonumber \,,\\
 &=& \psi + \frac{H}{\dot{\sigma}} \delta\sigma \,.
\end{eqnarray}
This expression, written in terms of the adiabatic field, $\sigma$,
is identical to that given in
% Eqs.~(\ref{eq:mtm1}) and
Eq.~(\ref{eq:zeta1}) for a single field.

We can also write Eq.~(\ref{eq:zeta}) as
\begin{equation}
{\cal R}=(\cos^2\theta){\cal R}_\phi+(\sin^2\theta) {\cal
R}_\chi\,,
\end{equation}
where we define the comoving curvature perturbation for each of the
original fields as
\begin{equation}
{\cal R}_I \equiv \psi + {H\over\dot\varphi_I}\delta\varphi_I =
{H\over\dot\varphi_I}Q_I \,.
\end{equation}
However, even fields with no explicit interaction will in general have
non-zero intrinsic entropy perturbations on large scales in a
multi-field system due to their gravitational interaction, so that ${\cal
R}_{I}$ for each field is not conserved. Although the intrinsic
entropy perturbation for each field is still of the form given by
Eq.~(\ref{eq:intS}), it is no longer constrained by Eq.~(\ref{eq:Psi})
to vanish as $k\to0$.  This is in contrast to the case of
non-interacting perfect fluids, where it is possible to define a
constant curvature perturbation for each fluid on large
scales~\cite{WMLL}.

The comoving matter perturbation in Eq.~(\ref{def:epsilonm}) can
be written as
\begin{equation}
\label{epsilonm2} \epsilon_m = \dot\sigma \left(
\dot{\delta\sigma} - \dot\sigma A \right) -
\ddot\sigma\delta\sigma + 2 V_s \delta s \,,
\end{equation}
which acquires an additional term, compared with the single-field
case, due to the dependence of the potential upon $s$, where
\begin{equation}
V_s = (\cos\theta) V_\chi-(\sin\theta) V_\phi\,.
\end{equation}
The perturbed kinetic energy of $s$ has no contribution to
first-order as in the background solution $\dot{s}=0$, by
definition.

The total entropy perturbation, Eq.~(\ref{defSN}), for the two
fields can be written as
\begin{eqnarray}
&&{\cal S}
 = {2 \over 3\dot\sigma^2(3H\dot\sigma+2V_\sigma)}\times\nonumber\\
&&~~~{}\times \left\{ V_\sigma \left[ \dot\sigma \left(
\dot{\delta\sigma} - \dot\sigma A \right) - \ddot\sigma
\delta\sigma \right] + 3H\dot\sigma^2 \dot\theta\delta s \right\}
\,. \label{eq:S2}
\end{eqnarray}
Combining Eqs.~(\ref{eq:Psi}), (\ref{epsilonm2}) and~(\ref{eq:S2}), we
can write
\begin{equation}
{\cal S}
 = - {V_{\sigma} \over 6\pi G\dot\sigma^2[3H\dot\sigma+2V_\sigma]}
 \left( \frac{k^2}{a^2} \Psi \right)
- {2V_s \over3\dot\sigma^2} \delta s
\,.
\end{equation}
Comparing this with the single-field result given in
Eq.~(\ref{eq:S1}), we see that the entropy perturbation on large
scales is due solely to the relative entropy perturbation between
the two fields, described by the entropy field $\delta s$.

The change in the comoving curvature perturbation is given by~\cite{GBW,FB}
%Eq.~(\ref{eq:zeta_dot_old}) 
%rate of change is given by~\cite{GBW,FB}
\begin{equation}
\label{eq:zeta_dot_old}
  \dot{{\cal R}} = \frac{H}{\dot{H}}\frac{k^2}{a^2} \Psi
+{1\over2}H \left({\delta\phi\over\dot{\phi}}-{\delta
\chi\over\dot{\chi}}\right)
\frac{d}{dt}\left({\dot{\phi}^2-\dot{\chi}^2\over\dot{\phi}^2+
\dot{\chi}^2}\right) \, ,
\end{equation}
which can be expressed neatly in terms of the new variables:
\begin{equation}
  \label{eq:zeta_dot_new}
\dot{{\cal R}} = \frac{H}{\dot{H}}\frac{k^2}{a^2} \Psi +
\frac{2H}{\dot{\sigma}} \dot{\theta} \delta s \,,
\end{equation}
where
\begin{equation}
\dot{\theta} = -\frac{V_{s}}{\dot{\sigma}} \,.
\end{equation}
The new source term on the right-hand-side of this equation,
compared with the single-field case, Eq.~(\ref{zetadot2}), is
proportional to the relative entropy perturbation between the two
fields, $\delta s$.
Clearly, there can be significant changes to ${\cal R}$ on
large scales if the entropy perturbation is not
suppressed and if the background solution follows a curved
trajectory, i.e., $\dot\theta\neq0$, in field space~\cite{LR}.
This can then produce a change in the comoving curvature on
arbitrarily large scales (i.e., even in the limit
$k\to0$)~\cite{GBW,Bassett:1999mt}.

Equations of motion for the adiabatic and entropy field perturbations
can be derived from the perturbed scalar field
equations~(\ref{eq:perturbation}), to give
\begin{eqnarray} \label{eq:adiabatic1}
&& \delta \ddot{\sigma} + 3 H \delta \dot{\sigma} + \left(
\frac{k^2}{a^2} + V_{\sigma\sigma} - \dot{\theta}^2 \right)
 \delta \sigma \nonumber \\&&~~{}=
 -2V_{\sigma} A  + \dot{\sigma} \left[ \dot{A} +
3\dot\psi + \frac{k^2}{a^2} (a^2\dot{E}-aB) \right] \nonumber\\
&&~~~~~~{} +2(\dot\theta\delta s)^{\displaystyle{\cdot}} -
2{V_\sigma \over \dot\sigma} \dot\theta\delta s \,,
\end{eqnarray}
and
\begin{eqnarray}\label{eq:entropy1}
 &&\ddot{\delta s} + 3H\dot{\delta s} +
\left(\frac{k^2}{a^2}
  + V_{ss} - \dot{\theta}^2 \right) \delta s
\nonumber \\&&~~{}= -2{\dot\theta \over \dot\sigma} \left[
\dot\sigma ( \dot{\delta\sigma} - \dot\sigma A ) - \ddot\sigma
\delta \sigma \right] \,,
\end{eqnarray}
where 
\begin{eqnarray}
  \label{eq:V_sigma_sigma}
V_{\sigma\sigma}& = &(\sin^2\theta)V_{\chi\chi} +
(\sin2\theta)V_{\phi\chi} + (\cos^2\theta)V_{\phi\phi} \,,\\
  \label{eq:V_ss}
V_{ss} &=& (\sin^2\theta) V_{\phi\phi} - (\sin2\theta)V_{\phi\chi}
+ (\cos^2\theta)V_{\chi\chi}\,.
\end{eqnarray}
When $\dot\theta=0$, the adiabatic and entropy perturbations
decouple.
%\footnote
%{
If we employ the slow-roll approximation for the background
fields, $\dot\phi\simeq -V_\phi/3H$ and $\dot\chi\simeq
-V_\chi/3H$, we obtain $\dot\theta\simeq0$. This reflects the fact
that the rate of change of $\theta$ is slow -- instantaneously it
moves in an approximately straight line in field space. But the
integrated change in $\theta$ cannot in general be neglected.
Even working within the slow-roll approximation, fields do not in
general follow a straight line trajectory in field space.
%}.
The equation of motion for $\delta\sigma$ then reduces to that for
a single scalar field in a perturbed FRW spacetime, as given in
Eq.~(\ref{eq:varphi}), while the equation for $\delta s$ is that
for a scalar field perturbation in an {\em unperturbed} FRW
spacetime.

The only source term on the right-hand-side 
in Eq.~(\ref{eq:entropy1})
for the entropy perturbation comes from the intrinsic entropy
perturbation in the $\sigma$-field (see Eq.~\ref{eq:intS}). From Eqs.~(\ref{eq:Psi})
and~(\ref{epsilonm2}) we have
\begin{eqnarray}
\dot\sigma ( \dot{\delta\sigma} - \dot\sigma A ) - \ddot\sigma
\delta \sigma = 2\dot\sigma \dot\theta \delta s - {k^2\over 4\pi G
a^2} \Psi \,,\label{eq:int_entropy}
\end{eqnarray}
and hence we can rewrite the evolution
equation for the entropy perturbation as
\begin{equation}
\label{eq:entropy}
\ddot{\delta s} + 3H\dot{\delta s} + \left(\frac{k^2}{a^2}
  + V_{ss} + 3\dot{\theta}^2 \right) \delta s =
{\dot\theta\over\dot\sigma} {k^2 \over 2\pi G a^2} \Psi
\,.
\end{equation}
Note that this evolution equation is automatically gauge-invariant
and holds in any gauge. On large scales the inhomogeneous source
term becomes negligible, and we have a homogeneous second-order
equation of motion for the entropy perturbation, decoupled from
the adiabatic field and metric perturbations. If the initial
entropy perturbation is zero on large scales, it will remain so.

By contrast, we cannot neglect the metric back-reaction for the
adiabatic field fluctuations, or the source terms due to the
entropy perturbations. Working in the spatially flat gauge,
defining
\begin{equation}
Q_\sigma = \delta\sigma_Q = \delta\sigma + {\dot\sigma \over H}
\psi \,,
\end{equation}
and using
\begin{equation}
A_Q = 4\pi G {\dot\sigma \over H} Q_\sigma \,, \label{eq:A_Q}
\end{equation}
we can rewrite the equation of motion for the adiabatic field
perturbation as
\begin{eqnarray}
 && \ddot{Q}_\sigma + 3 H \dot{Q}_\sigma +
\left[ \frac{k^2}{a^2} + V_{\sigma\sigma} - \dot{\theta}^2 - {8\pi
G\over a^3} \left( {a^3\dot\sigma^2\over H}
\right)^{\displaystyle{\cdot}} \right] Q_\sigma \nonumber \\
&&~~~{}= 2(\dot\theta\delta s)^{\displaystyle{\cdot}} - 2\left(
{V_\sigma \over \dot\sigma} + {\dot{H}\over H} \right)
\dot\theta\delta s \,.\label{eq:adiabatic}
\end{eqnarray}
When $\dot\theta=0$, this reduces to the single-field equation of motion,
%~(\ref{eom:Q}), 
but for a curved trajectory in field
space, the entropy perturbation acts as an additional source term
in the equation of motion for the adiabatic field perturbation,
even on large scales.

In order for small-scale quantum fluctuations to produce
large-scale (super-Hubble) perturbations during inflation, a field
must be ``light'' (i.e., overdamped). The effective mass for the
entropy field in Eq.~(\ref{eq:entropy}) is
$\mu^2_s=V_{ss}+3\dot\theta^2$. For $\mu_s^2>{3\over2}H^2$, the
fluctuations remain in the vacuum state and fluctuations on large
scales are strongly suppressed. The existence of large-scale
entropy perturbations therefore requires
\begin{equation}\label{mus}
\mu^2_s\equiv V_{ss}+3\dot\theta^2 < {3\over2}H^2 \,.
\end{equation}

\section{Application to entropy/adiabatic correlations from inflation}

Equations~(\ref{eq:entropy}) and~(\ref{eq:adiabatic}) are the key
equations which govern the evolution of the adiabatic and entropy
perturbations in a two field system. Together with constraint
equations~(\ref{eq:int_entropy}) and~(\ref{eq:A_Q}) for the metric
perturbations, they form a closed set of equations. They allow one to
follow the effect on the adiabatic curvature perturbation due to the
presence of entropy perturbations, absent in the single field
model. This in turn will allow us to study the resulting correlations
between the spectra of adiabatic and entropy perturbations produced on
large-scales due to quantum fluctuations of the fields on small-scales
during inflation.

A useful approximation commonly made when studying field
perturbations during inflation, is to split the evolution of a
given mode into a sub-Hubble regime ($k>aH$), in which the Hubble
expansion is neglected, and a super-Hubble regime ($k<aH$), in
which gradient terms are dropped.

If we assume that both fields $\phi$ and $\chi$ are light (i.e.,
overdamped) during inflation, then we can take the field
fluctuations to be in their Minkowski vacuum state on sub-Hubble
scales. This gives their amplitudes at Hubble crossing ($k=aH$) as
\begin{equation}
  \label{eq:ics_old}
Q_I\big|_{k=aH} = \frac{H_k}{\sqrt{2k^3}}\,e_I(k)\,,
\end{equation}
where $I=\phi,\chi$, $H_k$ is the Hubble parameter when the mode
crosses the Hubble radius (i.e., $H_k=k/a$), and $e_\phi$ and
$e_\chi$ are independent Gaussian random variables satisfying
\begin{equation}
\label{eq:Gaussian} \langle e_I(k)\rangle = 0 \,, \quad \langle
e_I(k)e_J^*(k')\rangle = \delta_{IJ}\, \delta(k-k')\,,
\end{equation}
with the angled brackets denoting ensemble averages. It follows
from our definitions of the entropy and adiabatic perturbations in
Eqs.~(\ref{eq:sigma}) and~(\ref{eq:s}) that their distributions at
Hubble crossing have the same form:
\begin{equation}
  \label{eq:ics_new}
 Q_\sigma\big|_{k=aH} = \frac{H_k}{\sqrt{2k^3}}e_\sigma(k)\,,
 \quad
 \delta s\big|_{k=aH} = \frac{H_k}{\sqrt{2k^3}}e_s(k)\,,
\end{equation}
where $e_\sigma$ and $e_s$ are Gaussian random variables obeying
the same relations given in Eq.~(\ref{eq:Gaussian}), with
$I,J=\sigma,s$.

Super-Hubble modes are assumed to obey the equations of motion
given in Eqs.~(\ref{eq:adiabatic}) and~(\ref{eq:entropy}), which
we will write schematically as
\begin{eqnarray}
\label{eq:hatOsigma} \hat{O}^\sigma (Q_\sigma) &=& \hat{S}^\sigma
(\delta s)\,, \\ \hat{O}^s (\delta s) &=& 0\,,
\end{eqnarray}
where $\hat{O}^\sigma (Q_\sigma)$ and $\hat{O}^s (\delta s)$ are
obtained by setting $k=0$ on the left-hand side of
Eqs.~(\ref{eq:adiabatic}) and~(\ref{eq:entropy}) respectively, and
$\hat{S}^\sigma (\delta s)$ is given by the right-hand side of
Eq.~(\ref{eq:adiabatic}). As remarked before, there is no source
term for $\delta s$ appearing on the right-hand side of
Eq.~(\ref{eq:entropy}) once we neglect gradient terms. The general
super-Hubble solution can thus be written as
\begin{eqnarray}
Q_\sigma &=& A_+ f_+(t) + A_- f_-(t) + P(t) \,,\\ \delta s &=& B_+
g_+(t) + B_- g_-(t)\,,
\end{eqnarray}
where the real functions $f_{\pm}$ and $g_\pm$ are the
growing/decaying modes of the homogeneous equations,
$\hat{O}^\sigma(f_{\pm})=0$ and $\hat{O}^s(g_{\pm})=0$, and $P(t)$
is a particular integral of the full inhomogeneous
equation~(\ref{eq:hatOsigma}). 
Note that the slow-roll growing-mode solution $f_+\propto \dot\sigma/H$.

Henceforth we shall consider only slow-roll inflation where the
evolution can be approximated by 
first-order equations [dropping $\ddot{\delta s}$ and
$\ddot{Q}_\sigma$ in Eqs.~(\ref{eq:entropy}) and~(\ref{eq:adiabatic})],
%During slow-roll evolution, the decaying modes can be neglected, 
so that we have\footnote
{We note that in non-slow-roll scenarios the decaying modes
  may not be negligible on super-Hubble scales, which could 
  affect the correlations between adiabatic and entropy perturbations.}
\begin{eqnarray}
\label{SRsols}
Q_\sigma &\simeq& A f(t) + P(t) \,,\\
\delta s &\simeq& B g(t)\,.
\end{eqnarray}
We can, without loss of generality, take $f=1=g$ and $P=0$
when $k=aH$, so that the amplitudes of the growing modes at
Hubble-crossing are given by Eqs.~(\ref{eq:ics_new}) as
\begin{equation}
A(k) = \frac{H_k}{\sqrt{2k^3}}e_\sigma(k)\,,
 \quad
B(k) = \frac{H_k}{\sqrt{2k^3}}e_s(k)\,.
\end{equation}
From Eq.~(\ref{eq:hatOsigma}), we see that the amplitude of the
particular integral $P(t)$ at later times will be
correlated with the amplitude of the entropy perturbation, $B$,
and we can write $P(t)=B\tilde{P}(t)$, where $\tilde{P}(t)$ is a
real function
independent of the random variables $e_\sigma, e_s$.

In order to quantify the correlation, we define
\begin{equation}
\langle x(k) y^*(k') \rangle \equiv {2\pi^2\over k^3}\, {\cal
C}_{xy} \, \delta(k-k') \,.
\end{equation}
The adiabatic and entropy power spectra are given by
\begin{eqnarray}
{\cal P}_{Q_\sigma} \equiv
 {\cal C}_{Q_\sigma Q_\sigma}
 &\simeq& \left( {H_k\over2\pi} \right)^2
\left[ |f^2| + |\tilde{P}^2|\right] \,,\\ 
{\cal P}_{\delta s} \equiv 
 {\cal C}_{\delta s\delta s}
 &\simeq& \left( {H_k\over2\pi} \right)^2 |g^2| \,,
\end{eqnarray}
while the dimensionless cross-correlation is given by
\begin{equation}
\frac{{\cal C}_{Q_\sigma\delta s}} 
     {\sqrt{{\cal P}_{Q_\sigma}}\sqrt{{\cal P}_{\delta s}}}
 \simeq {g\tilde{P} \over \sqrt{g^2}\sqrt{|f^2|+|\tilde{P}^2|}} \,.
\end{equation}
Note that the adiabatic power spectrum at late times is always
enhanced if it is coupled to entropy perturbations [i.e., $P(t)\neq0$,
in Eq.~(\ref{SRsols})], as the entropy field fluctuations at
Hubble-crossing provide an uncorrelated extra source.

As an illustration, we consider the correlations in the adiabatic
and entropy perturbations at the start of the radiation era,
produced after double inflation, as studied in
Ref.~\cite{Langlois}. The double-inflation potential for two
non-interacting but massive scalar fields is:
\begin{equation}
  \label{eq:double potential}
V = \frac{1}{2}m_\phi^2 \phi^2 + \frac{1}{2}m_\chi^2 \chi^2\,.
\end{equation}
Following \cite{Polarski:1992dq}, it is possible to parametrise
the background scalar field trajectory in polar coordinates when
both fields are slow-rolling:
\begin{equation}
  \label{eq:di background}
\chi \simeq \sqrt{\frac{N}{2\pi G}}\sin\alpha\,, \quad \phi \simeq
\sqrt{\frac{N}{2\pi G}}\cos\alpha\,,
\end{equation}
where $N = -\ln(a/a_{\rm end})$ is the number of e-folds until the
end of inflation. The background trajectory can then be expressed
as:
\begin{equation}
  \label{eq:di background soln}
N \simeq N_0 \frac{(\sin\alpha)^{2/(R^2 - 1)}}
{(\cos\alpha)^{2R^2/(R^2 - 1)}}\,,
\end{equation}
where $R = {m_\chi}/{m_\phi}$. The scalar field position angle,
$\alpha$, can be related to the scalar field velocity angle,
$\theta$, which we used to define the adiabatic and entropy
perturbations:
\begin{equation}
  \label{eq:sin theta sin Theta}
\tan\theta \simeq -\frac{m_\chi^2}{3H\dot{\sigma}}
\sqrt{\frac{N}{2\pi G}}
%\sin
\tan\alpha\,.
\end{equation}

The scalar field $\chi$ is assumed to decay into cold dark matter
while the scalar field $\phi$ decays into radiation. The
entropy/isocurvature at the start of the radiation-dominated era
is described by
\begin{equation}
  \label{eq:rad isoc}
S_{\rm rad} \equiv \frac{\delta \rho_c}{\rho_c} -
 {3\over4} \frac{\delta\rho_\gamma}{\rho_\gamma}\,.
\end{equation}
In Ref.~\cite{Langlois}, it is shown how the super-Hubble
perturbations in the radiation era can be determined in terms of
the perturbations during the inflationary era.  The fluctuations
in both $\phi$ and $\chi$ fields can contribute to both the
adiabatic and entropy perturbations. The adiabatic component comes
directly from the comoving curvature perturbation, ${\cal R}$, at
the end of inflation, and is given by
\begin{equation}
{\cal R}_{{\rm rad}} \simeq  -\sqrt{4\pi G} \sqrt{{N_k\over k^3}}
H_k\left[(\sin\alpha_k) e_\chi(k)  +(\cos\alpha_k)
e_\phi(k)\right]\,. \label{eq:Phi rad}
\end{equation}
The isocurvature perturbation at the start of the
radiation-dominated era is related to the entropy perturbation
between the two fields at the end of inflation~\cite{PolSta94}
\begin{equation}
S_{{\rm rad}} \simeq - {2\over3} m_\chi^2 {1\over H} \left(
{\delta\chi\over\dot\chi} - {\delta\phi\over\dot\phi} \right) \,,
\end{equation}
which yields
\begin{equation}
    \label{eq:entropy_entropy}
S_{{\rm rad}} \simeq -\sqrt{4\pi G}\sqrt{{N_k\over k^3}}H_k\left[ R^4
{\rm sec}\,\alpha_k + {\rm cosec}\,\alpha_k \right]e_s(k)\,,
\end{equation}
and
\begin{eqnarray}
\nonumber &&{\cal R}_{{\rm rad}} \simeq \sqrt{4\pi G}\sqrt{{N_k\over
k^3}} H_k
\frac{R^2\tan\alpha_k\sin\alpha_k}{\sqrt{R^2\tan^2\alpha_k +
1}}\times
\\&&~~{} \times \left\{\left[
\frac{1}{R^2\tan^2\alpha_k} + 1 \right]e_\sigma(k) +
\left[\frac{1-R^2}{R^2\tan\alpha_k}\right]e_s(k)  \right\}.
\label{eq:adiabatic_ent_adi}
\end{eqnarray}
The entropy perturbation during the radiation era only depends on
the entropy perturbation at Hubble-crossing during the
inflationary era, while the adiabatic perturbation during the
radiation era depends on both the adiabatic and entropy
perturbations at Hubble-crossing. This is consistent with
equations (\ref{eq:entropy}) and (\ref{eq:adiabatic1}), showing
that the entropy perturbation sources the adiabatic perturbation
on super-Hubble scales, but not vice versa.

As both equations (\ref{eq:entropy_entropy}) and
(\ref{eq:adiabatic_ent_adi}) depend on the random variable $e_s$,
the adiabatic and entropy perturbations will be correlated, and we
find
\begin{equation}
\frac{{\cal C}_{{\cal R}_{\rm rad}S_{\rm rad}}} 
     {\sqrt{{\cal P}_{{\cal R}_{\rm rad}}} \sqrt{{\cal P}_{S_{\rm rad}}}}
 \simeq \frac{(R^2-1)\sin
2\alpha_k}{2\sqrt{R^4\sin^2\alpha_k + \cos^2\alpha_k)}}\,.
\end{equation}
This correlation is investigated fully in \cite{Langlois} in terms
of the usual scalar field perturbation variables. An interesting
point that can easily be seen from
Eq.~(\ref{eq:adiabatic_ent_adi}) is that ${\cal R}_{{\rm rad}}$
will depend only on $e_\sigma$ if $R\equiv m_\chi/m_\phi=1$. Thus,
there will be no correlation if $R=1$. As can be seen from
Eq.~(\ref{eq:di background soln}), $\alpha$ will be constant for
$R=1$ and thus so will $\theta$; a straight-line background
trajectory will be obtained for $R=1$. This is consistent with
Eq.~(\ref{eq:adiabatic1}), where it can be seen that the entropy
component only sources the adiabatic component on large scales if
$\dot{\theta}\not=0$.

\section{Conclusions}

We have introduced a new formalism in which to follow the evolution of
adiabatic and entropy perturbations during inflation with multiple
scalar fields.  We decompose arbitrary field perturbations into a
component parallel to the background solution in field space, termed
the {\em adiabatic\/} perturbation, and a component orthogonal to the
trajectory, termed the {\em entropy\/} perturbation. We have rederived
the field equations in terms of these rotated fields in
Eqs.~(\ref{eq:entropy}) and~(\ref{eq:adiabatic}). These show that the
adiabatic perturbation on large scales can be driven by the entropy
perturbation, while the entropy perturbation itself obeys a
homogeneous second-order equation on super-Hubble scales.  There can
only be significant change in the large-scale comoving curvature
perturbation if there is a non-negligible entropy perturbation, {\em and}
if the background trajectory in field space is curved.

Our formalism can be applied to evaluate the correlation between the
adiabatic and entropy perturbations at the end of inflation.  As an
example we considered the example of two non-interacting fields in
double inflation, calculating the cross-correlation between the
adiabatic and entropy perturbations.

%%%%%%%%%%%%%%%%%%%%%%%%%%%%%%%%%%%%%%%%%%%%%%%%%%%%%%%%%%%%%%%%%%%%%%%%

% {\em Note added:} After completing this work we became aware of
% related work by Hwang and Noh~\cite{HwangNoh} who also study entropy
% perturbations in multiple field inflation. They find that the
% adiabatic and entropy modes decouple on super-horizon scales when
% the effect of curvature of the trajectory in field
% space is neglected, but we have shown that this cannot in general be
% assumed, even in models of slow-roll inflation.

\chapter{Preheating}

\label{preheating}
\newcommand{\ssec}{\subsection}
\newcommand{\beq}{\begin{equation}}
\newcommand{\beqn}{\begin{eqnarray}}
\newcommand{\eeq}{\end{equation}}
\newcommand{\eeqn}{\end{eqnarray}}
\newcommand{\vp}{\varphi}
\newcommand{\dvp}{\delta\phi}
\newcommand{\ts}{\textstyle}
\newcommand{\rd}{\displaystyle{\cdot}}
\def\lb{\label}
\def\sec{\section}

%%%%%%%%%%%%%%%%%%%%%%%%%%%%%%%%%%%%%%%%%%%%%%%%%%%%%%%%%%%%%%%%%%%%%

\sec{Introduction}

Standard inflationary models must end with a phase of reheating
during which the inflaton, $\phi$, transfers its energy to other
fields. Reheating itself may begin with a violently nonequilibrium
``preheating" era, when coherent inflaton oscillations lead to
resonant particle production (see \cite{KLS2} and refs. therein).
Until recently, preheating studies implicitly assumed that
preheating proceeds without affecting the spacetime metric. In
particular, causality was thought to be a ``silver bullet,"
ensuring that on cosmologically relevant scales, the non-adiabatic
effects of preheating could be ignored.

%In fact, exciting, super-Hubble scale perterbations is possible during
During preheating  metric perturbations may be resonantly amplified
on all length scales \cite{BKM1,Bassett:1999mt,mm,BV}. Causality is not violated
precisely because of the huge coherence scale of the inflaton
immediately after inflation \cite{BKM1,Bassett:1999mt} (see also \cite{Finelli:1999bu}).
 Strong preheating (with resonance parameter $q \gg 1$; see
 Chapter \ref{intro} for overviews and notation) typically leads to
 resonant amplification of scalar metric perturbation modes
 $\Phi_k$, possibly even on super-Hubble scales (i.e., $k/aH \ll
 1$).
%, where $a$ is the scale factor and $H$ the Hubble rate).
% One of
% our aims is to answer the question ``how typical is {\em
% typical}?"

Understanding when super-Hubble scales are amplified
%The answer 
is crucial since preheating can lead to e
anisotropies in the CMB. Observational
limits rule out those models that produce unbridled nonlinear
growth, but models which pass the metric preheating test on { COBE}
scales may nevertheless leave a non-adiabatic signature of
preheating in the { CMB}. 

\section{Entropy suppression}

In this section we use the entropy/adiabatic decomposition of the
perturbation equations developed in Chapter \ref{ch:adent} to
investigate the dynamics of super-Hubble perturbations during a period
of preheating at the end of inflation. We consider three models,
encompassed by the general effective potential
\begin{equation}
  \label{eq:mod_pot}
V = \frac{1}{2} m^2\phi^2 + \frac{\lambda}{4}\phi^4 +
\frac{1}{2}g^2\phi^2\chi^2 + \tilde{g}^2 \phi^3\chi\,.
\end{equation}
The essence of preheating lies in the parametric amplification of
field perturbations due to the time-dependence of their effective
mass, e.g., $m_{\chi}^2\equiv V_{\chi\chi} = g^2\phi^2$. In the
simplest cases, the inflaton $\phi$ simply oscillates at the end of
inflation.

Preheating typically amplifies long-wavelength modes
preferentially. As discussed in~\cite{BKM1,Finelli:1999bu,Bassett:1999mt},
amplification of super-Hubble modes does not lead to a violation
of causality, due to the super-Hubble coherence of the inflaton
oscillations set up by the prior inflationary phase. If ${\cal R}$
is amplified on super-Hubble scales, this will alter the resulting
imprint on the anisotropies of the 
%cosmic microwave background
 CMB, and break the simple link between CMB observations and
inflationary models.

We consider first the case where the inflaton is massive
($m\neq0$) and neglect its self-interaction ($\lambda=0$).
The traditional resonance parameter for the strength of preheating at
the end of inflation is
\begin{equation}
  \label{eq:q}
  q = \frac{g^2\phi_0^2}{4m^2}\, , 
\end{equation}
where $\phi_0$ is the initial value of $\phi$ at the beginning of
preheating.
In the massive case, where modes move through the resonance bands
of the Mathieu chart, and for inflation at high energies where the
expansion of the universe is very vigorous, $q$  needs to be much
larger than one if the parametric resonance is to be
efficient~\cite{KLS2}.
It is possible to have large $q$ even for small coupling,
$g^2\ll1$, as $m\ll\phi_0\sim M_{\rm Pl}$.
We can write the effective mass of the $\chi$ during inflation as
\begin{equation}
  \label{eq:chi_mass}
\frac{m_\chi^2}{H^2} \approx \frac{3q}{\pi}\frac{M_{\rm
    Pl}^2}{\phi_0^2}\, .
\end{equation}
 It then follows from Eq.~(\ref{eq:chi_mass}) that $\chi$
must be heavy during inflation for this simple potential if efficient
preheating is to be obtained.

Any change in the curvature perturbation ${\cal R}$ on very large
scales must be due to the presence of non-adiabatic perturbations.
In
\cite{Jedamzik:2000um,LLMW}, it was shown how, if $m_{\chi}^2 \gg
m_{\phi}^2$ during inflation
with $\lambda=0=\tilde{g}$, then the $\chi$ field and hence any
non-adiabatic perturbations on large scales are exponentially
suppressed during inflation, and no change to ${\cal R}$ occurs before
backreaction ends the resonance.

However, when $\tilde{g} \neq 0$, the $\chi$ field will have a nonzero
vacuum expectation value (vev) during inflation {\em even along the
valley of the potential}. 
In the slow-roll limit for $\phi$, this vev is determined by
$V_\chi=0$, which gives (see Eq.~\ref{eq:KG})
\begin{equation}
  \label{eq:mod_traj}
  \chi \approx -\frac{\tilde{g}^2}{g^2}\phi\,.
\end{equation}
The $\tilde{g}$ coupling has the effect of rotating the valley of
the potential -- which the attractor trajectory approximately
follows -- from $\chi=0$, through an angle
\begin{equation}
  \label{eq:rotation}
  \theta \approx - \frac{\tilde{g}^2}{g^2}\,,
\end{equation}
where, to ensure that the chaotic inflation scenario is not drastically
altered, we assume~\cite{gordonsting} 
\begin{equation}
  \label{eq:small_g_tilde}
  \frac{\tilde{g}}{g} \ll 1\,.
\end{equation}

The effect of $\tilde{g}$ is to change the attractor for both $\chi$
and $\delta\chi$ during inflation, since the $\chi$ and $\delta\chi$
equations of motion (Eqs.~\ref{eq:KG} and \ref{eq:perturbation}) gain
inhomogeneous driving terms proportional to $\tilde{g}^2\phi^3$.  This
does not necessarily imply that ${\cal R}$ will be amplified by
preheating at the end of inflation as purely adiabatic perturbations
along the slow-roll attractor now have a component along $\chi$ as
well as $\phi$.  In order to determine whether or not the evolution of
the comoving curvature perturbation, ${\cal R}$, on super-Hubble
scales is affected, we need to follow the evolution of the entropy
field perturbation\footnote{ {}From Eq.~(\ref{eq:zeta_dot_new}) we see
  that $\dot\theta\delta s$ must be non-zero to change ${\cal R}$ on
  large scales.  Because $\dot\theta\approx0$, from
  Eq.~(\ref{eq:rotation}), the entropy remains decoupled from the
  adiabatic perturbation during slow-roll inflation in this model. But
  at the end of inflation, during preheating, $\dot\theta\neq0$.  },
defined by Eq.~(\ref{eq:s}), which gives
\begin{equation}
\delta s \approx \delta\chi + \frac{\tilde{g}^2}{g^2} \delta\phi \,.
\end{equation}
In the limit $\tilde{g}/g\to0$ we recover $\delta s\to\delta\chi$.
Crucially, the evolution equation~(\ref{eq:entropy}) for the entropy
perturbation has {\em no} inhomogeneous terms in the long-wavelength
($k \rightarrow 0$) limit, even for $\tilde{g}\neq0$, and entropy
perturbations will only be non-negligible on super-Hubble scales if
the entropy field is light during inflation.

In the slow-roll limit and on large scales, the evolution
equation~(\ref{eq:entropy}) for the entropy perturbation has the
approximate solution~\cite{BD}
\begin{equation}
\delta s \propto a^{-3/2} \left(\frac{k}{aH}\right)^{-\nu}\,,
\end{equation}
where 
\begin{equation}
\nu^2 = \frac{9}{4} - \frac{\mu_s^2}{H^2}\,, \label{nus}
\end{equation}
and the effective mass of the entropy field, $\mu_{s}$ is defined in
Eq.~(\ref{mus}).
The power spectrum of entropy perturbations is
\begin{equation}
{\cal P}_{\delta s} \propto H^3 \left(\frac{k}{aH}\right)^{3-2
{\rm Re}(\nu)} \,. \label{powerspec2}
\end{equation}
The real part of $\nu$ vanishes for $\mu_s^2/H^2 > 9/4$,
leaving a steep $k^3$ blue spectrum, which is exponentially
suppressed with time.

Using Eqs.~(\ref{eq:V_ss}), (\ref{eq:mod_pot}), (\ref{eq:rotation}),
and (\ref{eq:small_g_tilde}), one finds that 
\begin{equation}
  \label{eq:mod_entropy_mass}
\frac{\mu_s^2}{H^2} \approx \left[1 - 4q\left({\tilde{g}\over
g}\right)^4\left({\phi\over\phi_0}\right)^2\right]^{-1}
\frac{3qM_{\rm Pl}^2}{\pi\phi_0^2}\,,
\end{equation}
$\mu_s^2/H^2$ has a local minimum for $\tilde{g}=0$. Thus the
additional $\tilde{g}$ term in Eq.~(\ref{eq:mod_pot}) serves to {\em
increase} the entropy mass relative to the Hubble parameter, and so
does not avoid the suppression of the entropy perturbation. The
$\tilde{g}$ term therefore does not significantly alter the spectral
index of the spectrum of entropy perturbations, which remains steep if
$q \gg 1$.  The strongly blue spectrum implies that non-linear
backreaction is dominated by small-scale modes, which go nonlinear
long before the cosmological modes, implying that resonance ends
before ${\cal R}$ changes~\cite{KLS2,Jedamzik:2000um}.

We have also integrated the field equations numerically to avoid
relying on any slow-roll-type approximations.
To numerically evaluate the entropy perturbation, one could
simulate the original perturbation variables $\delta \phi$ and
$\delta \chi$, using Eq.~(\ref{eq:perturbation}), and then work
out $\delta s$ algebraically via Eq.~(\ref{eq:s}). However, this
approach is prone to numerical instability when the entropy
perturbation is suppressed.
To illustrate this, we take
$\tilde{g}=8\times 10^{-3}g$ and $q = 3.8\times10^5$ 
After about 60 e-folds of inflation, one can see analytically that
$\delta s \sim 10^{-40}$. Numerically, $\delta\chi\cos\theta \sim
\delta \phi\sin \theta\sim10^{-8}$ during inflation. So in order
to obtain a high enough accuracy to model the suppression of
$\delta s$, we require that $\delta\chi\cos\theta $ and $ \delta
\phi\sin \theta$ have to be simulated to a relative accuracy of
$\sim10^{-8}/10^{-40} = 10^{-32}$. This means approximately 32
significant figures are needed, which is beyond the capability of
standard numerical ordinary differential equation integration
routines.

If instead we use the new adiabatic and entropy field perturbations
and integrate Eqs.~(\ref{eq:entropy}) and (\ref{eq:adiabatic}), then
this numerical instability does {\em not} occur, since one no longer
needs to find the difference between two nearly equal
quantities.
Simulation results using these equations are compared
with the results using the old field perturbation
equations~(\ref{eq:perturbation}) in Fig.~\ref{fig:old_new}.
\begin{figure}[h]
%  \begin{center}
%\input figure2.pstex_t
\begin{picture}(0,0)%
\epsffile{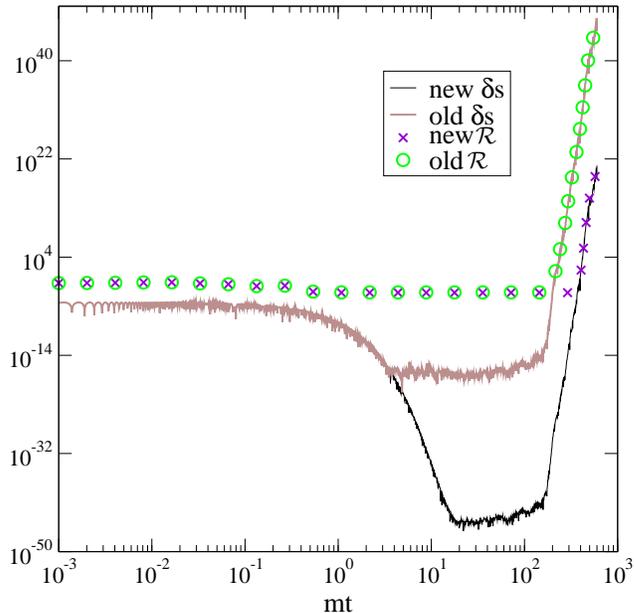}%
\end{picture}%
\setlength{\unitlength}{3947sp}%
\begingroup\makeatletter\ifx\SetFigFont\undefined%
\gdef\SetFigFont#1#2#3#4#5{%
  \reset@font\fontsize{#1}{#2pt}%
  \fontfamily{#3}\fontseries{#4}\fontshape{#5}%
  \selectfont}%
\fi\endgroup%
\begin{picture}(4107,4107)(3364,-6781)
\put(6376,-3741){\makebox(0,0)[lb]{\smash{\SetFigFont{9}{10.8}{\rmdefault}{\mddefault}{\updefault}$\cal R$}}}
\put(6321,-3896){\makebox(0,0)[lb]{\smash{\SetFigFont{9}{10.8}{\rmdefault}{\mddefault}{\updefault}$\cal R$}}}
\end{picture}
\caption[ Numerical simulations of the entropy and comoving curvature
 perturbations during inflation and preheating.]{
 Numerical simulations of the entropy and comoving curvature
 perturbations during inflation and preheating, with $\lambda=0$,
 $g=2\times10^{-3}$, $\tilde{g}=8\times10^{-3}g$ and $m=10^{-6}M_{\rm
   pl}$.  The
 `new' prefix indicates that the field perturbations were evaluated by
 numerically integrating Eqs.~(\ref{eq:entropy}), and
 (\ref{eq:adiabatic}), while the `old' prefix indicates that the
 perturbations were evaluated by integrating the original field
 equations~(\ref{eq:perturbation}).  We have not included any
 higher-order corrections such as backreaction from small-scale
 perturbations which would shut down the resonant amplification of
 $\delta s$ at some point.
 }
     \label{fig:old_new}
 \end{figure}
 The
simulations show that the growth in $\cal R$ is driven by $\delta s$,
in concordance with Eq.~(\ref{eq:zeta_dot_new}). As can be seen, the
numerical result using the field perturbation equations fails to track
the exponential decay of the entropy during inflation and thus
underestimates the delay in the growth of $\cal R$.

In practice, we find a similar instability if we
try to construct the gauge-invariant metric perturbation, $\Psi$,
required in Eq.~(\ref{eq:entropy}) in terms of the constraint
Eq.~(\ref{eq:int_entropy}). This includes the intrinsic entropy
perturbation in the $\sigma$ field, which does become small at late
times/large scales, but results from the diminishing difference between
finite terms. It is more stable numerically to follow the value of
$\Psi$ at late times using the evolution equation
\begin{equation}
\dot\Psi + \left( H - {\dot{H}\over H} \right) \Psi = 4\pi G\dot\sigma
Q_\sigma \,,
\end{equation}
which can be obtained from the definition of $\Psi$ given in
Eq.~(\ref{eq:Psi}) and the metric constraint
equations~(\ref{eq:mtmcon}) and~(\ref{eq:aniso}).

Note that the adiabatic/entropy decomposition becomes
ill-defined if $\dot\sigma=0$, i.e. both fields stop rolling, and this can
cause numerical instability during preheating if the trajectory is
confined to a narrow valley. 
This can occur, for instance, when $\tilde{g}=0$ and only the $\phi$
field oscillates.
The original field perturbations $\delta\phi$ and $\delta\chi$ remain
well-defined, although the comoving curvature perturbation ${\cal R}$,
defined in  Eq.~(\ref{eq:zeta}) becomes singular when
$\dot\sigma=0$~\cite{HK}. 
This does not happen for the simulation results shown in
Fig.~\ref{fig:old_new} with $\tilde{g}\neq0$ where the fields
oscillate in a two-dimensional potential well.

The massive inflaton potential ($m\neq0$) safeguards the conservation
of ${\cal R}$ by a bootstrap effect: if preheating is strong, $q \gg
1$, then the entropy perturbation is heavy during inflation; on the
other hand, if the entropy is light during inflation, then $q \leq 1$
and preheating is very weak. This is not altered by a rotation of the
trajectory in field space ($\tilde{g}\neq0$) as can most quickly be
seen by noting, from Eqs.~(\ref{eq:V_sigma_sigma})
and~(\ref{eq:V_ss}), that
\begin{equation}
\label{sumrule}
V_{\sigma\sigma} + V_{ss} = V_{\phi\phi} + V_{\chi\chi} \,.
\end{equation}
Thus if the $\chi$ field is very massive ($V_{\chi\chi}\gg H^2$), we
must have $V_{\sigma\sigma} + V_{ss}\gg H^2$. For slow-roll inflation
we require $V_{\sigma\sigma}\ll H^2$ and hence $V_{ss}\gg H^2$.

\section{Unsuppressed entropy}
The suppression does not necessarily occur in massless ($m=0$)
self-interacting ($\lambda\neq0$) inflation models~\cite{BV,Finelli:1999bu,ZBS}.
This latter class of models is almost conformally invariant, allowing
analytical results from Floquet theory to be applied.  The Floquet
index, $\mu_k$, which determines the rate of exponential growth, can
reach its maximum as $k/aH \to 0$, when $g^2/\lambda = 2 n^2$ for
integer $n$, thereby implying maximum growth for the
longest-wavelength perturbations.  
Assuming slow-roll inflation driven by $V\approx \lambda\phi^4/4$, we
see from Eq.~(\ref{sumrule}) that $V_{\sigma\sigma} +
V_{ss}>V_{\chi\chi}=g^2\phi^2$ and thus that the entropy field is
massive ($V_{ss}>9H^2/4$) whenever
\begin{equation}
\frac{g^2}{\lambda} > 8\pi \frac{\phi^2}{M_{\rm Pl}^2} \,.
\end{equation}
However, we can have resonance at large scales for $n=1$ and
$g^2/\lambda = 2$, when the entropy field need not be heavy during
inflation and no exponential suppression takes place, so that the
subsequent growth of ${\cal R}$ is explosive~\cite{BV}. The growth of
${\cal R}$ occurs before backreaction can shut off the resonant growth
of the entropy perturbations $\delta s$~\cite{BV,Finelli:1999bu,TBV,ZBS}.
Although the region of parameter space around $g^2/\lambda = 2$ is
thus ruled out, the same does not hold for $g^2/\lambda \gg 1$, since
the entropy field is then heavy during inflation and $\delta s$ is
again suppressed.

Models where the entropy effective mass is simply very small during
inflation but then becomes large at preheating can also effect large
scales.

\section{New cosmological effects}

% Our eventual goal  must  be to calculate physical quantities such
% as the power spectrum of $\Phi_k$. Since $P_{\Phi} = k^3
% |\Phi_k|^2/2\pi^2$, one might be concerned that these strong
% preheating effects at $k \rightarrow 0$ would be made irrelevant
% by the $k^3$ phase space factor. Perhaps the easiest way to see
% that this is not so is to look at the evolution of $\zeta_k$.
% Since $\zeta_k$ is {\em not} conserved for small  $k$ (see Fig.
% \ref{fig1}), the standard normalization of the {\sc cmb} spectrum
% is increased. This can only take place if the power spectrum of
% metric fluctuations is strongly affected as $k \rightarrow 0$.
% This is understandable since preheating acts only as a non-trivial
% transfer function $T(k)$.

%^
Beyond the effects discussed in \cite{BKM1,Bassett:1999mt},
metric preheating can lead to a host of interesting new effects.\\
$\bullet$ The growth of
$\zeta_k$ implies amplification of isocurvature modes in unison
with adiabatic scalar modes on super-Hubble scales. Preheating
thus yields the possibility of inducing a post-inflationary
universe with both isocurvature and adiabatic modes on large
scales. The effect of having mixtures of isocurvature and adiabatic
perturbations on the CMB is discussed in Chapter \ref{ch:cmb}.
%If these are uncorrelated and of roughly equal strength,
%the corresponding Doppler peaks will tend to cancel
% \cite{doppler}. (This mechanism is independent of the one
% discussed in \cite{BKM1,Bassett:1999mt}, which requires nonlinearity to persist
% until decoupling.) However, if the adiabatic and isocurvature
% modes are strongly correlated, this would create the possibility
% of a ``smoking gun" finger-print of preheating. 
%The challenge
%remains to distinguish such correlations from those induced in
%hybrid inflation.
\\ 
$\bullet$ Because the metric perturbations can
go nonlinear, whether on sub- or super-Hubble scales, the
corresponding $\chi$ density perturbations $\delta$ typically have
non-Gaussian statistics. This is simply a reflection of the fact
that $-1\leq\delta <\infty$, so that the distribution of necessity
becomes skewed and non-Gaussian. Further,
% in Class II models,
when $\langle\chi\rangle = 0$ during inflation, $\chi$
perturbations in the energy density will necessarily be
non-Gaussian (chi-squared distributed), even if $\delta\chi_k$ is
Gaussian distributed, since stress-energy components are quadratic
in the fluctuations (see e.g. \cite{nongauss}).  
%Non-Gaussian
%effects are therefore an intrinsic part of many metric preheating
%models 
%(particularly those in Class II),
% and open up a potential
%signal for detection in future experiments.\\ 
\\$\bullet$ Another
new feature we can identify is the breaking of conformal
invariance. Once metric perturbations become large on some scale,
the metric on that scale cannot be thought of as taking the simple
Friedmann-Robertson-Walker ({FRW}) form, and
conformal invariance is lost. This is particularly important for
the production of  primordial magnetic fields, which are usually
strongly suppressed due to the conformal invariance of the Maxwell
equations in a {FRW} background. The coherent oscillations of
the inflaton during preheating further provide a natural cradle
for producing a primordial seed for the observed large-scale
magnetic fields. A charged inflaton field, with kinetic term
$D_{\mu} \phi (D^{\mu}\phi)^*$, will  couple to electromagnetism
through the gauge covariant derivative $D_{\mu} = \nabla_{\mu} -
ie A_{\mu}$. This  will naturally lead to parametric resonant
amplification of the existing magnetic field, which
%^will
could produce
large-scale coherent seed fields on the required super-Hubble
scales without fine-tuning
\cite{pbvm}. (Note that a tiny seed field {\em must} exist
during inflation due to the conformal trace anomaly and one-loop
QED corrections in curved spacetime \cite{mag}.)\\

\section{Conclusion}
The effect of preheating on the large-scale curvature perturbation has
been examined using the formalism developed in Chapter \ref{ch:adent}.
The mass of the entropy field
during inflation is a crucial quantity. If the entropy field is heavy,
then any fluctuations on large scales are suppressed to negligible
values at the beginning of preheating. This squeezing of the entropy
perturbation is most accurately modelled numerically using our
evolution equation for the entropy perturbation. If it is estimated
from the usual field equations, it may contain large numerical errors
when there is a non-trivial background trajectory in field space.

For models where efficient preheating can occur with a light entropy
during inflation, large scale perturbations are effected by
preheating.  A range of cosmological implications for such models was
discussed.

% In conclusion, the suppression discussed in \cite{Jedamzik:2000um,I} is highly
% sensitive to the form of the particle interactions considered;
% when couplings are considered which are found in most realistic
% particle physics models, the effects of \cite{Jedamzik:2000um,I} recede.
% Instead, in models from either of the two general classes
% highlighted here, preheating
% %^will generically
% can produce a strong amplification of metric perturbations on
% cosmologically significant scales. 

%We can rule out potentials that do lead to strong amplification of
%super-Hubble scale perterbations.
%Metric preheating thus allows
% us to rule out models in which backreaction effects fail to
% prevent
% %^
% super-Hubble nonlinear growth, and shows that in the surviving
% models, there will typically be some signature of preheating
% imprinted on the power spectrum. The robustness of the
% amplification further demonstrates the need to move towards more
% realistic models of preheating in order to develop a realistic
% understanding of the predictions of inflation for observational
% cosmology.

\chapter{Correlated adiabatic and entropy perturbations and the Cosmic Microwave Background}

\label{ch:cmb}

%
% Correlated perturbations from inflation and the cosmic microwave background
%
%  Luca Amendola, Christopher Gordon, David Wands and Misao Sasaki
%
%   astro-ph/0107089
%   submitted 5th July 2001
%
%\documentstyle[aps,prl,floats,graphicx]{revtex}
%\tighten 

%\newcommand\eq[1]{Eq.~(\ref{#1})}
%\newcommand\eqs[2]{Eqs.~(\ref{#1}) and (\ref{#2})}
%\newcommand\eqss[3]{Eqs.~(\ref{#1}), (\ref{#2}) and (\ref{#3})}
%\newcommand\eqsss[4]{Eqs.~(\ref{#1}), (\ref{#2}), (\ref{#3})
%and (\ref{#4})}
%\newcommand\eqssss[5]{Eqs.~(\ref{#1}), (\ref{#2}), (\ref{#3}),
%(\ref{#4}) and (\ref{#5})}
%\newcommand\eqst[2]{Eqs.~(\ref{#1})--(\ref{#2})}
\newcommand\<{\left[}
\renewcommand\>{\right]}

\def\R{\cal R}
\def\S{\cal S}

\section{Introduction}

Increasingly accurate measurements of temperature anisotropies in the
cosmic microwave background sky offer the prospect of precise
determinations of both cosmological parameters and the nature of the
primordial perturbation spectra.
The recent Boomerang \cite{net}, DASI \cite{halverson} and
Maxima \cite{lee} data have shown evidence for 
three peaks in the CMB
temperature anisotropy power spectrum as expected 
in inflationary scenarios.
In this context the CMB data support the current `concordance' model
based on a spatially flat Friedmann-Robertson-Walker universe
dominated by cold dark matter and a cosmological constant \cite{tegmark}.
In addition, the CMB data no longer shows any signs of
being in conflict with the big bang nucleosynthesis data
\cite{adiablikli}. 

In the studies which have estimated the cosmological and primordial
parameters with these new data sets, only the case of purely adiabatic
perturbations has been considered so far,  i.e. the perturbation
in the relative number densities, $\delta n/ n$, of different particle
species is taken to be zero.  Although this assumption is justified
for perturbations originating from single field inflationary models,
it does not necessarily follow when there is more than one field
present during inflation (see for example
\cite{GBW,Langlois,gordonadent,HwangNoh,bartolo}). Other possible
primordial modes are isocurvature \cite{BMT,BMT2} (also referred to as
``entropy'') modes in which the particle ratios are perturbed but the
total energy density is unperturbed in the comoving gauge.

Most previous studies have examined the extent to which a
statistically independent isocurvature contribution to the primordial
perturbations may be constrained by CMB and large-scale structure data
\cite{isocmb,enqvistBM}.  
It has recently been shown that multi-field inflationary models in
general produce correlated adiabatic and isocurvature perturbations
\cite{Langlois,gordonadent,HwangNoh,bartolo}.  These correlations can
dramatically change the observational effect of adding isocurvature
perturbations \cite{Langlois2,BMT,BMT2}.
Up until now, only the case of scale-invariant correlated adiabatic
and entropy perturbations has been considered. Trotta {\em et al.\/}
\cite{trotta} found (with an earlier CMB dataset) 
that in this case the cold dark matter (CDM) isocurvature mode
was likely to be very small if not entirely absent, though they did
find that a neutrino isocurvature mode contribution \cite{BMT,BMT2} was
not ruled out.
In this Chapter we examine whether a correlated CDM isocurvature mode
is better favoured by the recent CMB data when a tilted power law
spectrum is allowed.

\section{Theory}

Non-adiabatic perturbations are produced during a period of slow-roll
inflation in the presence of two or more light scalar fields, whose
effective masses are less than the Hubble rate. On sub-horizon scales,
fluctuations remain in their vacuum state so that when fluctuations 
reach the horizon scale their amplitude is given by
%\begin{equation}
\( \label{Hover2pi}
\hat{\delta\phi}_{i*} \simeq \left( {H_* / 2\pi} \right) \hat{a}_i
\)
%\end{equation}
where the subscript $*$ denotes horizon-crossing and $\hat{a}_i$ are
independent normalised Gaussian random variables, obeying $\langle
\hat{a}_i \hat{a}_j \rangle=\delta_{ij}$. 
The total comoving curvature and entropy perturbation at any time
during two-field inflation can quite generally be given in terms of
the field perturbations, along and orthogonal to the background
trajectory, (see Chapter \ref{ch:adent})
% DW: LEFT THESE EQUATIONS IN FOR NOW. BUT WE COULD TAKE THEM OUT
% LATER IF WE NEED TO FIND SPACE, AS THE PRECEDING DESCRIPTION IN
% WORDS COULD SUFFICE. 
\begin{eqnarray}
\label{RandS}
\hat{\cal R} & \propto & \cos\theta\,  \hat{\delta\phi}_1
 + \sin\theta\,  \hat{\delta\phi}_2
\,,\\
\hat{\cal S} & \propto & -\sin\theta\,  \hat{\delta\phi}_1
 + \cos\theta\,  \hat{\delta\phi}_2
\,,
\end{eqnarray}
where $\theta$ is the angle of the inflaton trajectory in field
space.
Although the curvature and entropy perturbations are uncorrelated at
horizon-crossing, any change in the angle of the trajectory,
$\theta$, will begin to introduce correlations.
%~\cite{gordonadent}.
%
Further correlations may be introduced by the model dependent dynamics
when inflation ends and the fields' energy is transformed into
radiation and/or dark matter.  The comoving curvature perturbation,
${\cal R}_{\rm rad}$,
on large-scales during the radiation-dominated era is
related to the conformal Newtonian metric perturbation, $\Phi$, by
${\cal R}_{\rm rad}=3\Phi/2$.  The isocurvature perturbation is 
\[
{\cal
  S}_{\rm rad}=\frac{\delta\rho_{\rm cdm}}{\rho_{\rm
  cdm}}-\frac{3}{4}\frac{\delta\rho_\gamma}{\rho_\gamma}
\]
 and remains constant on
large scales until it re-enters the horizon.
On large scales the CMB temperature perturbation can be expressed in
terms of the primordial perturbations \cite{Langlois}
\begin{equation}
  \label{eq:largescales}
  \frac{\hat{\delta T}}{T} \approx \frac{1}{5} \left( \hat{{\cal
  R}}_{\rm rad} - 2\hat{{\cal S}}_{\rm rad} \right) \,. 
\end{equation}

The general transformation of linear curvature and entropy
perturbations from horizon-crossing during inflation to the beginning
of the radiation era will be of the form
\begin{equation}
\label{transform}
\left( 
\begin{array}{c}
\hat{\cal R}_{{\rm rad}} \\ 
\hat{\cal S}_{{\rm rad}}
\end{array}
\right) = 
\left( 
\begin{array}{cc}
1 & T_{{\cal R}{\cal S}} \\ 
0 & T_{{\cal S}{\cal S}}
\end{array}
\right) 
\left( 
\begin{array}{c}
\hat{\cal R}_* \\ 
\hat{\cal S}_*
\end{array}
\right) \,,
\end{equation}
Two of the matrix coefficients, $T_{{\cal R}{\cal R}}=1$ and $T_{{\cal
    S}{\cal R}}=0$, are determined by the physical requirement that
the curvature perturbation is conserved for purely adiabatic
perturbations and that adiabatic perturbations cannot source entropy
perturbations on large scales~\cite{WMLL}.
The remaining terms will be model dependent.
If the fields and their decay products completely thermalize after
inflation then $T_{{\cal S}{\cal S}}=0$ and there can be no entropy
perturbation if all species are in thermal equilibrium characterised
by a single temperature, $T$. This means that it is unlikely that a
neutrino isocurvature perturbation could be produced by inflation
unless the reheat temperature is close to that at neutrino
decoupling shortly before primordial nucleosynthesis takes place. On
the other hand, a cold dark matter species could remain decoupled at
temperatures close to, or above, the supersymmetry breaking scale
yielding $T_{{\cal S}{\cal S}}$.
The simplest assumption being that one of the fields can itself be
identified with the cold dark matter~\cite{Langlois}.

The slow evolution (relative to the Hubble rate) of light fields after
horizon-crossing translates into a weak scale dependence of both the
initial amplitude of the perturbations at horizon crossing, and the
transfer coefficients $T_{{\cal R}{\cal S}}$ and $T_{{\cal S}{\cal
    S}}$.  Parameterising each of these by simple power-laws over the
scales of interest, requires three power-laws to
describe the scale-dependence in the most general adiabatic and
isocurvature perturbations,
\begin{eqnarray}
\hat{\cal R}_{\rm rad} &=& A_r k^{n_1} \hat{a}_r + A_s k^{n_3} \hat{a}_s
 \,, \label{R} \\
\hat{\cal S}_{\rm rad} &=& B k^{n_2} \hat{a}_s \,. \label{S}    
\end{eqnarray}
The generic power-law spectrum of adiabatic perturbations from single
field inflation can be described by two parameters, the amplitude and
tilt, $A$ and $n$. Uncorrelated isocurvature perturbations require a
further two parameters, whereas we now have in general six parameters.
The dimensionless cross-correlation
 \begin{equation}
 \cos\Delta = 
 \frac{\langle{\cal R}_{\rm rad}{\cal S}_{\rm rad}\rangle}
 {(\langle{\cal R}_{\rm rad}^2\rangle \langle{\cal S}_{\rm
     rad}^2\rangle)^{1/2}}
 =  \frac{ {\rm sign}(B)\ A_s k^{n_3}}
           {\sqrt{A_r^2 k^{2n_1} + A_s^2 k^{2n_3}} } 
% \,,
\end{equation}
is in general scale-dependent.

\section{Likelihood analysis}

We will investigate in this Chapter the restricted case where all the
spectra share the same spectral index and hence 
$\Delta$ is scale-independent.  This might naturally arise in the case of
almost massless fields where the scale-dependence of the field
perturbations is primarily due to the decrease of the Hubble rate
during inflation,  which is common to
both perturbations and yields $n_i<0$.
In the following analysis
we also allow $n_i>0$, but we shall see that blue power spectra of
this type are not favoured by the data.

We then have four parameters, $A=\sqrt{A_r^2+A_s^2}$, $B$, $\Delta$,
and $n$ describing the effect of correlated perturbations, 
where $n=1+2n_i$ is defined to coincide with the standard definition
of the spectral index for adiabatic perturbations.  
We leave an investigation of the full six parameters for future
work.

By defining the entropy-to-adiabatic
ratio \( B^{*}=B/A \) the parameter \( A \) becomes an overall amplitude
that can be marginalized  analytically (see below). In the following,
to simplify notation, we write \( A=1 \) and drop the star from \(
B^{*} \).
We limit the analysis to \( B>0 \) and \( 0<\Delta <\pi  \), since there
is complete symmetry under \( \Delta \rightarrow -\Delta  \) and
under \( (B\rightarrow -B,\Delta \rightarrow \pi -\Delta ) \). Further,
we allow three background cosmological parameters to vary, 
\(
\omega _{b}\equiv \Omega _{b}h^{2}\, \, ,\quad \omega _{c}\equiv
\Omega _{cdm}h^{2}\, \, ,\) and \( \Omega _{\Lambda } 
\)
where $\Omega_{b,cdm,\Lambda}$ is the density parameter for baryons, CDM and
the cosmological constant, respectively. Since we assume spatial
flatness, the Hubble constant
is \[
h^2={\frac{\omega _{c}+\omega _{b}}{1-\Omega _{\Lambda }}}
.\]
Our aim is therefore to constrain the six parameters 
\[
\alpha_i \equiv \left\{
B,\Delta ,n, \omega _{b},\omega _{c},\Omega _{\Lambda } \right\} \,,
\]
by comparison with CMB observations.  We consider the COBE data
analysed in \cite{bon}, and the recent high-resolution Boomerang
\cite{net} and Maxima data \cite{lee}.  In order to concentrate on the
role of the primordial spectra (and limit the numerical computation
required) we will fix the reionisation history (no reionisation),
neutrino masses (zero) and spatial curvature (zero). We will also
neglect any contribution from tensor (gravitational wave)
perturbations.

We will use a CMBFAST code \cite{sel} modified in order to allow
correlated perturbations to calculate the expected CMB angular power
spectrum, $C_l$, for all parameter values. (Our $C_l$ is defined as
$C_l=l(l+1)C^*_l/(2\pi)$ 
where $C^*_l$ is the square of the multipole amplitude). The
computations required 
can be considerably reduced by expressing the spectrum for a generic
value of \( B \) and \( \Delta \) as a function of the spectra for
other values. Let us denote the purely adiabatic and isocurvature
spectra when \( 
B=1 \) as \( [ C_{l}]_{\rm ad} \) and \( [C_{l}]_{\rm iso} \)
respectively, and the correlation term
for totally correlated perturbations \( B=1,\Delta=0 \) as \(
[C_{l}]_{\rm corr} \) . Then we can write the generic spectrum
for arbitrary $B$ and $\Delta$ as
\begin{equation}
C_{l} = [C_{l}]_{\rm ad} + 
B^{2}[C_{l}]_{\rm iso} + 2B\cos\Delta [C_{l}]_{\rm corr} 
\label{total}
\end{equation}
We can obtain $[C_{l}]_{\rm corr}$ from \eq{total} and using
any $B\cos\Delta\neq0$.  The library spectra $[C_{l}]_{\rm ad}$ and
$[C_{l}]_{\rm iso}$ and $[C_{l}]_{\rm corr}$ can then be used to
evaluate $C_{l}$ for any $B$ and $\Delta$.
A different set of library spectra will be needed for each set of cosmological
parameters.
When $n_1\not=n_3$ then $\Delta$ is not generally scale independent
and so it would be necessary to evaluate the shape of the
cross-correlation spectra $[C_{l}]_{\rm corr}$ for each
form of $\Delta(k)$, but one can always perform the scaling with
respect to $B$ analytically. 

The remaining input parameters requested by the CMBFAST code are set
as follows: \( T_{cmb}=2.726K, \) \( Y_{He}=0.24,N_{\nu
  }=3.04,\tau_{c}=0 \). In the analysis of
 \cite{net} \( \tau_{c} \), the optical
depth to Thomson
scattering, was also included in the
general likelihood and, in the flat case, was found to be compatible
with zero at slightly more than 1\( \sigma \).  Therefore here, to further reduce the
parameter space, we assume \( \tau _{c} \) vanishes. 
We did not include the cross-correlation between band powers because it
is not available, but it should be less than 10\% according to \cite{net}.
 An offset log-normal approximation to the
band-power likelihood has been advocated by \cite{bon} and adopted by
\cite{net,lee}, but the quantities necessary for its evaluation are
not available.  Since the offset log-normal reduces to a log-normal in
the limit of small noise we evaluated the log-normal likelihood
\begin{equation}
-2\log L(\alpha _{j})=\sum _{i}\frac{\left[ Z_{\ell ,t}(\ell
 _{i};\alpha _{j})-Z_{\ell ,d}(\ell _{i})\right] ^{2}}{\sigma _{\ell
 }^{2}} 
\end{equation}
where \( Z_{\ell }\equiv \log \hat{C}_{\ell } \), the subscripts
\( t \) and \( d \) refer to the theoretical quantity and to the
real data, \( \hat{C}_{\ell } \) are the spectra binned over some
interval of multipoles centered on \( \ell _{i} \), \( \sigma _{\ell } \)
are the experimental errors on \( Z_{\ell ,d} \), and the parameters
are denoted collectively as \( \alpha _{j} \).

The overall amplitude parameter \( A \) can be integrated out analytically
using a logarithmic measure \( d\log A \) in the likelihood.
Analogously, an analytic integration can get rid of the calibration
uncertainty of the Boomerang and Maxima data (see \cite{net,lee}),
to obtain the final likelihood function that we discuss in the following.
We neglected beam and pointing errors, but we checked that the results
do not change significantly even increasing the calibration errors by 50\%.
We assume a linear integration measure for all the other parameters.

In order to compare with the Boomerang and Maxima analyses we assume
uniform priors as in \cite{net}, with the parameters confined in the
range 
\[
\begin{array}{l}
 B\in (0,3), \quad \Delta \in (0,\pi ),\quad n\in
(0.6,1.4)\\  \omega_{b}\in (0.0025,0.08), \quad
\omega _{c}\in (0.05,0.4), \quad \Omega _{\Lambda }\in (0,0.9).
\end{array}
\]
 As
extra priors, the value of \( h \) is confined in the range \(
(0.45,0.9) \) and the universe age is limited to \( >10 \) Gyr as in
\cite{net}. A grid of \( \sim 10,000 \) multipole CMB spectra is used
as a database over which we interpolate to produce the likelihood
function.

Figure~1 shows the one of the best cases in our database,
corresponding to \( (B,n ,\omega _{b},\omega _{c},\Omega _{\Lambda
  })=(0.3,0.8,0.02,0.1,0.7)\) and $\Delta=0$. The case $\Delta=\pi/4$
provided an equally good fit.
\begin{figure}[ht]
{\centering \includegraphics[width=12cm]{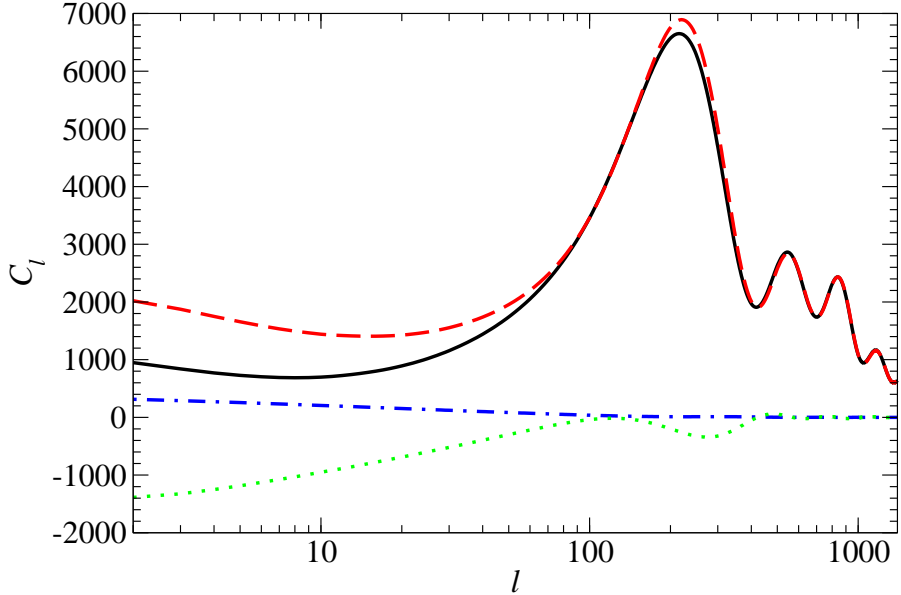}}
\caption[A decomposition of the best fit CMB spectra.]{
A decomposition of the best fit spectra (solid line) which has
  $\Delta=0$. The adiabatic (dashes), entropy (dash-dot) and
  correlation (dots) contributions are also plotted. }
\end{figure}
In the figure the adiabatic  $ ([ C_{l}]_{\rm ad} )$, entropy
($B^2[C_l]_{\rm iso}$) and correlated $(2B\cos\Delta[C_{l}]_{\rm corr})$
components are shown.
As can be seen, the effect of adding a positively
correlated component is to reduce the height of the low-$l$ plateau
relative to the acoustic peaks \cite{Langlois2}. This is in contrast
to the uncorrelated case where the addition of entropy perturbations
reduces the peak height relative to the plateau.

We found a near-degeneracy between \( B \) and \( n \) when \( \Delta
=0 \): the effect of adding maximally correlated isocurvature
perturbations mimics an increase in the primordial slope. This makes
clear the importance of varying \( n \) when studying correlated
isocurvature perturbations: a lower \( n \) allows a larger \( B \) to
be consistent with the CMB data.

In Fig. 2 we plot a series of two-dimensional likelihood functions;
all the other parameters have been marginalized in turn.
The contour lines of the cosmological parameters \( \omega_b \) and
\( \omega_c \) are almost parallel
to \( B \) for \( B<1 \). This means that the isocurvature
perturbations do not alter significantly the best estimates for these
cosmological parameters. 
On the other hand, increasing \( B \) moves the region of confidence for \(
\Omega _{\Lambda } \) and of \( n \) toward smaller values.
\begin{figure}[ht]
%%%CG fixed bounding box 
{\centering \includegraphics[width=12cm,bb=101 447 419
  755]{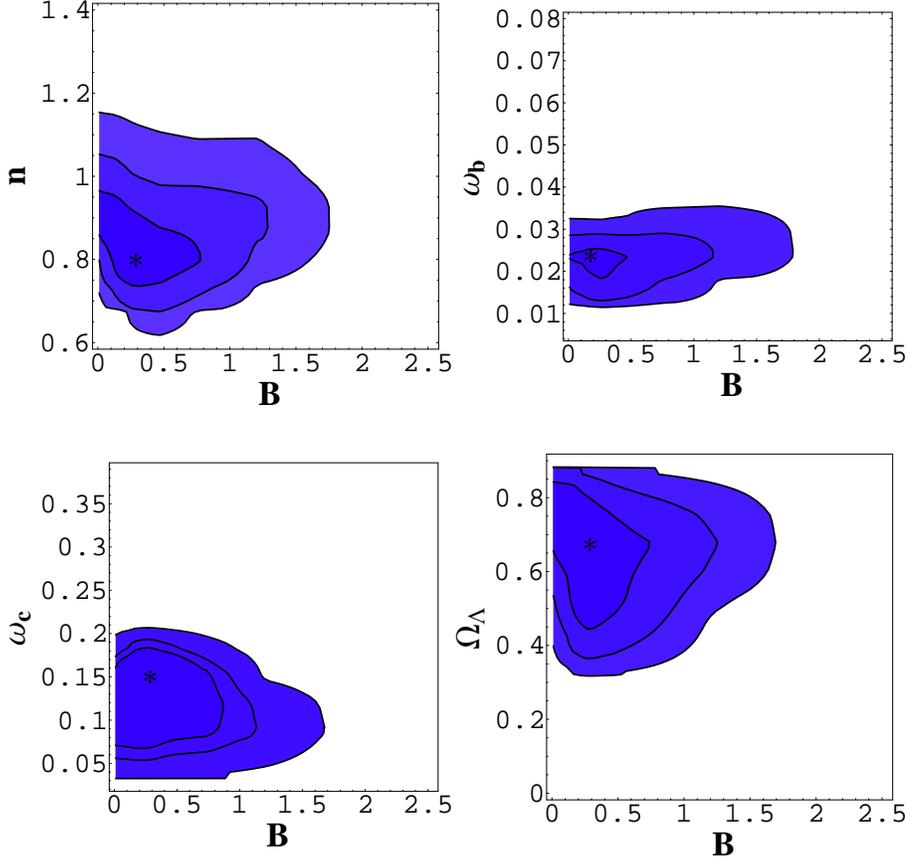} \par} 
\caption[Two-dimensional likelihood contour plots.]{
Two-dimensional likelihood contour plots. The contours enclose
40\%, 86\% and 99\% of the likelihood and the stars mark the peaks.
See text for explanation.}
\end{figure}

In Fig. 3 we plot the one-dimensional likelihood functions obtained by
marginalizing all the remaining parameters. 
Panel \emph{a} shows that the contribution of
isocurvature perturbations can be as large as the adiabatic
perturbations, or even larger: we find that \( B<1.5 \) to 95\% c.l..
In contrast, if the isocurvature perturbations are uncorrelated, their
fraction cannot exceed 70\% (\( B<0.7 \)) to the same c.l.. It is
intriguing to observe that the likelihood of \( B \) peaks around 0.3:
that is, a non-zero contribution of isocurvature perturbations is {\em
  more} likely than a vanishing contribution.  In the same panel we
show as a long dashed line the likelihood assuming experimental errors
reduced to one third, a precision within reach of the forthcoming
satellite experiments: the curve shows that this level of precision
would allow the detection of a finite isocurvature contribution.
Equally interesting, in panel \emph{b} we see that the likelihood of
the correlation \({\rm cos} \Delta \) peaks near unity (maximal
correlation), but has a not negligible probability everywhere in its
domain. The likelihood functions for \( n \) and \( \Omega _{\Lambda }
\) move toward smaller values, as anticipated, while the CDM and the
baryon density estimates remain largely unaffected. The average values
are 
\[
\begin{array}{l}
 n=0.87\pm 0.1,\quad \omega_b=0.023\pm 0.005, \\
\omega_c=0.12\pm 0.04,\quad  \Omega_\Lambda=0.63\pm 0.13 .
\end{array}
\]
\begin{figure}[ht]
%%%CG fixed bounding box 
{\centering \includegraphics[width=12cm, bb=99 405 473
  763]{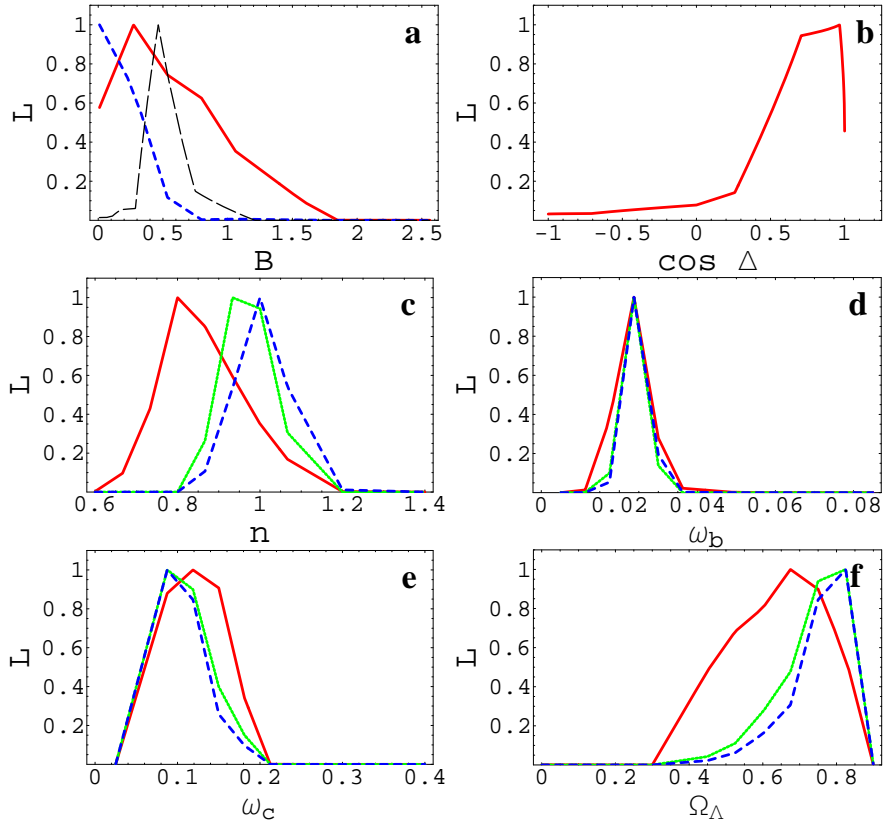} \par} 
\caption[One-dimensional likelihood functions.]{
One-dimensional likelihood functions in arbitrary units. 
  Green (light) solid lines for the purely adiabatic models ( \mbox{$ B=0 $});
  blue short-dashed lines for uncorrelated fluctuations ($ \cos\Delta
  =0 $); red (dark) solid lines for correlated fluctuations. See
  text for further explanation.} 
\end{figure}

\section{Conclusions}

By contrast, Enqvist {\em et al} \cite{enqvistBM} found that
a large uncorrelated isocurvature contribution is only consistent with
blue tilted slopes. The reason for this difference is that
correlations can cause the acoustic peak height to increase relative
to the Sachs Wolfe plateau (see Fig.\ 1) unlike the case of
independent perturbations where the relative height always decreases.
Trotta {\em et al} \cite{trotta} found that the CMB data was not
consistent with a significant CDM isocurvature contribution because
they restricted the primordial slope, $n$, to be unity. As can be seen
from Fig.\ 2 our $n=1$ likelihood contours also indicate a very low
isocurvature contribution. But when $n$ is allowed to be less than one
the isocurvature contribution can be even larger than the adiabatic
contribution.

As can be seen from Figs.\ 2 and 3 our estimates of $\omega_b$ and
$\omega_c$ are virtually unaffected by the addition of correlated CDM
isocurvature perturbations. Thus, in our model, the nature of the
isocurvature component can be investigated almost independently of the
composition of the matter component.

The main conclusion of this Chapter is that Boomerang and Maxima
are consistent with a large correlated CDM isocurvature perturbation
contribution when the spectral slopes are tilted to the red ($n<1$).
The higher precision of future satellite data has the potential to
detect the isocurvature contribution, if any, thereby showing that
inflation was not a single-field process.

\chapter{Non-adiabatic perturbations from brane world effects}

\label{ch:branes}

\def \D {\mbox{D}}
\def \curl {\mbox{curl}\,}
\def \ep {\varepsilon}

\def \gtrsim {\stackrel{>}{\sim}}

\section{Introduction}

According to string and M-theory, gravity is a higher-dimensional
theory, reducing to Einstein's four-dimensional theory of general
relativity at low enough energies. In the brane-world scenario,
the standard model matter fields are confined to a 3-brane in
$1+3+d$ dimensions, while the gravitational field can propagate in
the bulk, i.e., also in the $d$ extra dimensions, being localized
at the brane at low energies. Recent developments show that the
$d$ extra space dimensions need not be small, or even compact,
thus allowing the intriguing possibility that corrections could
occur even at TeV scales.

These exciting theoretical developments may offer a promising
route towards a quantum gravity theory. However, as well as
theoretical elegance, they must also pass the increasingly
stringent tests provided by cosmological observations. Primarily,
this involves developing higher-dimensional perturbation theory
and then applying it to analyze the generation and evolution of
density and tensor perturbations on the brane, leading to a
prediction of the CMB  anisotropies
and galaxy distribution.

This is an ambitious and difficult programme, but initial steps have
already been taken, at least in the case of a particular class of
models that generalize the Randall-Sundrum models~\cite{RS1}.
Large-scale adiabatic density perturbations from inflation on the
brane have been computed~\cite{mwbh} (see also~\cite{cll}), using
the conservation of the curvature perturbation on uniform-density
hypersurfaces. This conservation follows from adiabaticity and the
conservation of energy-momentum on the brane, and is independent
of the form of the field equations~\cite{WMLL}. In~\cite{mwbh},
the backreaction effect of metric fluctuations in the fifth
dimension was neglected. In the general case, i.e., incorporating
also the fluctuations in the nonlocal quantities that carry the
bulk influence onto the brane, it has been shown that { 
large-scale density perturbations contain a closed system on the
brane}---and thus can in principle be evaluated purely from
initial conditions on the brane, without knowledge of bulk
dynamics~\cite{m}.
Note that not all large-scale scalar perturbations can be computed
intrinsically -- the relation between the metric potentials $A$ and
$\Psi$ is mediated by anisotropic stress imprinted on the brane by the
bulk Weyl tensor and this stress cannot be evaluated without solving
the bulk equations \cite{LMSW}.
 In this Chapter, we solve the closed system described
in~\cite{m} to find the evolution of large-scale density
perturbations on the brane. We show that extra-dimensional effects
introduce a non-adiabatic mode on the brane.

A general perturbation formalism has been developed~\cite{pert,bd},
encompassing equations on the brane and in the bulk, and in principle
able to describe all scales.  However, the general equations are
extremely complicated, in particular since the mode equations are
partial differential equations. A first application of the equations
has been made to large-scale tensor perturbations from inflation on
the brane~\cite{lmw}. Unlike the scalar case, large-scale tensor
perturbations cannot be evaluated without the bulk perturbation
equations.

We develop the outline argument presented first in~\cite{m}, and
analyze large-scale density perturbations and their evolution, from
after Hubble-crossing in inflation through the radiation era. This
provides part of the information needed for predicting the large-angle scalar
anisotropies generated in CMB temperature, and seeing how the bulk
effects modify general relativistic predictions.
However, the Sachs-Wolfe (SW) effect cannot be computed without knowledge
of the Weyl anisotropic stress \cite{LMSW}. It is possible to estimate
the SW effect by making assumptions about the Weyl anisotropic stress
\cite{BM}, but a complete solution requires solving the full 5D
perturbation problem.

 We show that in
general, the perturbation $\Phi$ (a covariant analog of the
Bardeen metric perturbation) is {  no longer constant during
high-energy inflation, but grows.} However, $\Phi$ is constant during the
radiation era, as in general relativity, except at most in the
early radiation era, if the energy density is still high relative
to the brane tension. 
% This means that 
% %{
% % 
%   the Sachs-Wolfe plateau in the CMB anisotropies is likely to be
%   preserved, but the limits which COBE measurements place on
%   inflationary potentials will be changed, and could also become
%   sensitive to the form of the potential.
% %}
% %\footnote{
% %
% The Sachs-Wolfe formula has not yet been calculated in brane-world
% cosmologies, but we expect that the corrections to the general
% relativistic result will be very small, since the energy scale at
% last scattering is much less than the brane tension.
% %
% %}

\section{Brane dynamics}

We follow the 1+3 covariant approach and notation of~\cite{m} which is
different from (but ultimately equivalent to) the metric-based
approach to perturbations described in Chapter \ref{ch:adent}. The
5-dimensional (bulk) field equations are
\begin{equation}
\widetilde{G}_{AB} =
\widetilde{\kappa}^2\left[-\widetilde{\Lambda}\widetilde{g}_{AB}
+\delta(\chi)\left\{ -\lambda g_{AB}+T_{AB}\right\}\right]\,,
\label{1}
\end{equation}
where tildes denote the bulk generalization of standard general
relativity quantities, and $\widetilde{\kappa}^2=
8\pi/\widetilde{M}_{\rm p}^3$, where $\widetilde{M}_{\rm p}$ is
the fundamental 5-dimensional Planck mass, which is typically much
less than the effective Planck mass on the brane, $M_{\rm
p}=1.2\times 10^{19}$ GeV. The brane is given by $\chi=0$, so that
a natural choice of coordinates is $x^A=(x^\mu,\chi)$, where
$x^\mu=(t,x^i)$ are spacetime coordinates on the brane and the brane a
moving hypersurface. The brane
tension is $\lambda$, and $g_{AB}=\widetilde{g}_{AB}-n_An_B$ is
the induced metric on the brane, with $n_A$ the spacelike unit
normal to the brane. Standard-model matter fields confined to the
brane make up the brane energy-momentum tensor $T_{AB}$ (with
$T_{AB}n^B=0$). The bulk cosmological constant
$\widetilde{\Lambda}$ is negative, and is the only 5-dimensional
stress energy. (See~\cite{mw} for the modification of this
approach in the case where there is also a scalar field in the
bulk.)

The most general background bulk with homogeneous and isotropic
induced metric on the brane is the 5-dimensional
Schwarzschild-anti-de Sitter metric~\cite{msm}, with the black
hole mass leading to an effective radiation-like correction to the
Friedmann equation on the brane~\cite{bdel}.

The field equations induced on the brane are derived via an
elegant geometric approach in~\cite{sms}, leading to new terms
that carry bulk effects onto the brane:
\begin{equation}
G_{\mu\nu}=-\Lambda g_{\mu\nu}+\kappa^2
T_{\mu\nu}+\widetilde{\kappa}^4S_{\mu\nu} - {\cal E}_{\mu\nu}\,,
\label{2}
\end{equation}
where $\kappa^2=8\pi/M_{\rm p}^2$. The various energy scales are
related to each other via
\begin{equation}
\lambda=6{\kappa^2\over\widetilde\kappa^4} \,, ~~ \Lambda =
{\textstyle{1\over2}}\widetilde\kappa^2\left(\widetilde{\Lambda}+
{\textstyle{1\over6}}\widetilde\kappa^2\lambda^2\right)\,,
\label{3}
\end{equation}
and the high-energy regime is $\rho \gtrsim \lambda$. Bulk
corrections to the Einstein equations on the brane are of two
forms: firstly, {  the matter fields contribute local quadratic
energy-momentum corrections} via the tensor $S_{\mu\nu}$, and
secondly, there are {  nonlocal effects from the free
gravitational field in the bulk,} transmitted via the projection
${\cal E}_{\mu\nu}$ of the bulk Weyl tensor. The matter
corrections are given by
\begin{eqnarray}
S_{\mu\nu}&=&{\textstyle{1\over12}}T_\alpha{}^\alpha T_{\mu\nu}
-{\textstyle{1\over4}}T_{\mu\alpha}T^\alpha{}_\nu\nonumber\\
&&~~{}+ {\textstyle{1\over24}}g_{\mu\nu} \left[3 T_{\alpha\beta}
T^{\alpha\beta}-\left(T_\alpha{}^\alpha\right)^2 \right]\,.
\label{3'}
\end{eqnarray}
The projection of the bulk Weyl tensor is
\begin{equation}
{\cal E}_{AB}=\widetilde{C}_{ACBD}n^C n^D\,,
\label{4}
\end{equation}
which is symmetric and traceless and without components orthogonal
to the brane, so that ${\cal E}_{AB}n^B=0$ and ${\cal E}_{AB}\to
{\cal E}_{\mu\nu}g_A{}^\mu g_B{}^\nu$ as $\chi\to 0$.

The Weyl tensor $\widetilde{C}_{ABCD}$ represents the free,
nonlocal gravitational field in the bulk, i.e., the part of the
field that is not directly determined at each point by the
energy-momentum tensor at that point. The local part of the bulk
gravitational field is the Einstein tensor $\widetilde{G}_{AB}$,
which is determined locally via the bulk field equations
(\ref{1}). Thus ${\cal E}_{\mu\nu}$ transmits nonlocal
gravitational degrees of freedom from the bulk to the brane,
including tidal (or Coulomb), gravito-magnetic and transverse
traceless (gravitational wave) effects~\cite{m}.

If $u^\mu$ is the 4-velocity comoving with matter (which we assume
is a perfect fluid or minimally-coupled scalar field), we can
decompose the nonlocal term as
\begin{equation}
{\cal E}_{\mu\nu}={-6\over\kappa^2\lambda}\left[{\cal
U}\left(u_\mu u_\nu+{\textstyle {1\over3}} h_{\mu\nu}\right)+{\cal
P}_{\mu\nu}+{\cal Q}_{\mu}u_{\nu}+{\cal Q}_{\nu}u_{\mu}\right]\,,
\label{6}
\end{equation}
where $h_{\mu\nu}=g_{\mu\nu}+u_\mu u_\nu$ projects into the
comoving rest-space. Here
\[
{\cal U}=-{\textstyle{1\over6}}\kappa^2 \lambda\, {\cal
E}_{\mu\nu}u^\mu u^\nu
\]
is {  an effective nonlocal energy density} on the brane (which
need not be positive), arising from the free gravitational field
in the bulk. It carries Coulomb-type effects from the bulk onto
the brane. There is {  an effective nonlocal anisotropic stress}
\[
{\cal P}_{\mu\nu}=-{\textstyle{1\over6}}\kappa^2 \lambda\left[
h_\mu{}^\alpha h_\nu{}^\beta-{\textstyle{1\over3}}h^{\alpha\beta}
h_{\mu\nu}\right] {\cal E}_{\alpha\beta}
\]
on the brane, which carries Coulomb, gravito-magnetic and
gravitational wave effects of the free gravitational field in the
bulk. The {  effective nonlocal energy flux} on the brane,
\[
{\cal Q}_\mu ={\textstyle{1\over6}}\kappa^2 \lambda\,
h_\mu{}^\alpha {\cal E}_{\alpha\beta}u^\beta\,,
\]
carries Coulomb and gravito-magnetic effects from the free
gravitational field in the bulk. (Note that there is no energy
flux in the bulk, and thus no transfer of energy between bulk and
brane; this situation changes if bulk scalar fields are
present~\cite{bd,mw}.)

\section{Local and nonlocal conservation equations}

The local and nonlocal bulk modifications may be consolidated into
an effective total energy-momentum tensor:
\begin{equation}
G_{\mu\nu}=-\Lambda g_{\mu\nu}+\kappa^2 T^{\rm tot}_{\mu\nu}\,,
\label{6'}
\end{equation}
where
\begin{equation}
T^{\rm tot}_{\mu\nu}= T_{\mu\nu}+{6\over \lambda}S_{\mu\nu}-
{1\over\kappa^2}{\cal E}_{\mu\nu}\,. \label{6''}
\end{equation}
The effective total energy density, pressure, anisotropic stress
and energy flux are
\begin{eqnarray}
\rho^{\rm tot} &=& \rho\left(1+{\rho\over2\lambda}\right)+{6 {\cal
U}\over\kappa^4\lambda}\,, \label{a}\\ p^{\rm tot} &=& p+
{\rho\over2\lambda}(\rho+2p) +{2{\cal U}\over\kappa^4\lambda}\,,
\label{b}\\ \pi^{\rm tot}_{\mu\nu} &=&{6\over
\kappa^4\lambda}{\cal P}_{\mu\nu}\,, \label{c}\\ q^{\rm tot}_\mu
&=& {6\over \kappa^4\lambda}{\cal Q}_\mu \,.\label{d}
\end{eqnarray}

The brane energy-momentum tensor separately satisfies the
conservation equations, $\nabla^\nu T_{\mu\nu}=0 $, giving
\begin{eqnarray}
&&\dot{\rho}+\Theta(\rho+p)=0\,,\label{pc1}\\ && \D_\mu
p+(\rho+p)A_\mu =0\,,\label{pc2}
\end{eqnarray}
where a dot denotes $u^\nu\nabla_\nu$, $\Theta=\D^\mu u_\mu$ is
the volume expansion rate of the $u^\mu$ congruence,
$A_\mu=\dot{u}_\mu$ is its 4-acceleration, and $\D_\mu$ is the
projected covariant spatial derivative. The Bianchi identities on
the brane imply that the projected Weyl tensor obeys the
constraint
\begin{equation}
\nabla^\mu{\cal E}_{\mu\nu}={6\kappa^2\over\lambda}\nabla^\mu
S_{\mu\nu}\,. \label{5}
\end{equation}
This shows how nonlocal bulk effects are sourced by local bulk
effects, which include spatial gradients and time derivatives:
{  evolution and inhomogeneity in the matter fields can generate
nonlocal gravitational effects in the bulk, which backreact on the
brane.} The brane energy-momentum tensor and the consolidated
effective energy-momentum tensor are {  both} conserved
separately. Conservation of $T^{\rm tot}_{\mu\nu}$ gives, upon
using Eqs. (\ref{a})--(\ref{pc2}), propagation equations for the
nonlocal energy density ${\cal U}$ and energy flux ${\cal Q}_\mu$.
In linearized form, these are
\begin{eqnarray}
&& \dot{\cal U}+{\textstyle{4\over3}}\Theta{\cal U}+\D^\mu{\cal
Q}_\mu  =0 \,, \label{lc1'}\\&& \dot{\cal Q}_{\mu}+4H{\cal Q}_\mu
+{\textstyle{1\over3}}\D_\mu{\cal U}+{\textstyle{4\over3}}{\cal
U}A_\mu \nonumber\\&&~~~{}+\D^\nu{\cal P}_{\mu\nu}
=-{\textstyle{1\over6}} \kappa^4(\rho+p) \D_\mu \rho
\,,\label{lc2'}
\end{eqnarray}
where $H=\dot{a}/a$ ($={1\over3}\Theta$) is the Hubble rate in the
background. The nonlocal tensor mode,
which is the part of ${\cal P}_{\mu\nu}$ that 
 satisfies $\D^\nu{\cal
P}_{\mu\nu}=0 \neq {\cal P}_{\mu\nu}$, does not enter the nonlocal
conservation equations. Furthermore, there is no evolution
equation at all for ${\cal P}_{\mu\nu}$, reflecting the fact that in
general the equations do not close on the brane, and one needs
bulk equations to determine brane dynamics. There are bulk degrees
of freedom whose impact on the brane cannot be predicted by brane
observers, for example, incoming gravitational radiation from the bulk. 
%{ 
 The evolution of the nonlocal energy density and
flux, which carry scalar and vector modes of the bulk
gravitational field, is determined on the brane, while the
evolution of the nonlocal anisotropic stress, which carries
scalar, vector and tensor modes of the bulk field, is not.
%}

The generalized Raychaudhuri equation on the brane in linearized
form is
\begin{eqnarray}
&&\dot{\Theta}+{\textstyle{1\over3}}\Theta^2 -{\rm D}^\mu
A_\mu+{\textstyle{1\over2}}\kappa^2(\rho + 3p) -\Lambda
\nonumber\\&&~~{}= -{\textstyle{1\over2}} \kappa^2
(2\rho+3p){\rho\over\lambda} -{6 {\cal U}\over\kappa^2\lambda}\,,
\label{prl}
\end{eqnarray}
where the general relativistic case is recovered when the
right-hand side is set to zero. In the background, this gives
\begin{eqnarray}
\dot{H}&=&-H^2-{\kappa^2\over6}\left[\rho+3p+{\rho\over 2\lambda}
(2\rho+3p)\right]\nonumber\\&&~~{}+{1\over3}\Lambda- {2\over
\kappa^2\lambda }{\cal U}_o\left({a_o\over
a}\right)^4\,,\label{bray}
\end{eqnarray}
where the solution for ${\cal U}$ follows from Eq.~(\ref{lc1'}),
$a_o$ is the initial scale factor and ${\cal U}_o={\cal U}(a_o)$.
The first integral of this equation is the generalized Friedmann
equation on the brane:
\begin{equation}\label{f}
H^2={\kappa^2\over3} \rho\left(1 +{\rho\over2\lambda}\right) +
{1\over3}\Lambda -{K\over a^2}+ {2\over \kappa^2\lambda } {\cal
U}_o\left({a_o\over a}\right)^4\,,
\end{equation}
where $K=0,\pm1$. Local bulk effects modify the background
dynamics. In particular, inflation at high energies
($\rho\gtrsim\lambda$) proceeds at a higher rate than the
corresponding rate in general relativity. This introduces
important changes to the dynamics of the early
universe~\cite{mwbh,cll,eu}, and accounts for an increase in the
amplitude of scalar and tensor fluctuations at
Hubble-crossing~\cite{mwbh,lmw}.

Using Eqs.~\ref{f} and \ref{bray} the condition for inflation
becomes~\cite{mwbh}
\begin{equation}\label{inf}
w<-{1\over3}\left({2\rho+\lambda\over \rho+\lambda} \right)\,,
\end{equation}
where $w=p/\rho$. As $\rho/\lambda\to\infty$, we have
$w<-{2\over3}$, while the general relativity condition
$w<-{1\over3}$ is recovered as $\rho/\lambda\to 0$.

If ${\cal U}_o=0$, i.e., if the background bulk is conformally
flat, then Eqs.~(\ref{a}) and (\ref{b}) show that the effective
equation of state index for the total energy-momentum tensor is
\begin{equation}\label{w}
w^{\rm tot}\equiv {p^{\rm tot}\over \rho^{\rm tot}}
={w+(1+2w)\rho/2\lambda \over 1+\rho/2\lambda}\approx 1+2w\,,
\end{equation}
where the last equality holds at very high energies
($\rho\gg\lambda$). Thus for slow-roll inflation, $w^{\rm tot}$
and $w$ are both close to $-1$. The high-energy inflation
condition $w<-{2\over3}$ is $w^{\rm tot}<-{1\over3}$. During
high-energy reheating with $w\approx 0$ on average, we have
$w^{\rm tot}\approx 1$, so that the effective equation of state is
stiff, while high-energy radiation-domination ($w={1\over3}$) has
$w^{\rm tot}\approx{5\over3}$, i.e., an ultra-stiff effective
equation of state. The effective sound speed at very high energies
is also altered:
\begin{equation}\label{cs}
(c_{\rm s}^2)^{\rm tot}\equiv {\dot{p}^{\rm tot} \over
\dot{\rho}^{\rm tot}} \approx c_{\rm s}^2+w+1\,,
\end{equation}
where $c_{\rm s}^2=\dot{p}/\dot{\rho}$.

\section{Scalar perturbations on the brane}

When the vorticity of the fluid vanishes, scalar perturbations are
covariantly (as well as gauge-invariantly and locally) characterized
as the case when all perturbed quantities are expressible as spatial
gradients of scalars. In particular, the nonlocal perturbed bulk
effects are described by $\D_\mu{\cal U}$ and~\cite{m}
\begin{equation}\label{scal}
{\cal Q}_\mu=\D_\mu{\cal Q}\,,~~{\cal P}_{\mu\nu}=\left[
h_\mu{}^\alpha h_\nu{}^\beta-{\textstyle{1\over3}}h^{\alpha\beta}
h_{\mu\nu}\right] \D_{\alpha} \D_{\beta}{\cal P}\,.
\end{equation}
Note that 
%there is no transverse traceless mode from the bulk,
%since 
the nonlocal traceless mode has 
%nonzero 
spatial
divergence~\cite{m}:
\[
\D^\nu{\cal P}_{\mu\nu}={\textstyle{2\over3}}\D^2(\D_\mu{\cal
P})\,.
\]
The bulk gravitational field affects scalar perturbations via scalar
Coulomb modes, given by the spatial gradients of the `potentials'
${\cal U}$, ${\cal Q}$ and ${\cal P}$.

For adiabatic matter perturbations the 4-acceleration is
\begin{equation}\label{acc}
A_\mu=-{c_{\rm s}^2\over \rho(1+w)}\,\D_\mu\rho\,.
\end{equation}
The gradients
\begin{equation}\label{s3}
\Delta_\mu={a\over\rho}\D_\mu\rho\,,~~Z_\mu=a\D_\mu\Theta\,,
\end{equation}
describe inhomogeneities in the matter and expansion~\cite{ehb},
and the dimensionless gradients describing inhomogeneity in the
nonlocal quantities are~\cite{m}
\begin{equation}\label{s4}
U_\mu={a\over\rho}\D_\mu{\cal U}\,,~Q_\mu={1\over\rho} \D_\mu
{\cal Q}\,,~P_\mu={1\over a\rho}\D_\mu{\cal P}\,.
\end{equation}

The spatial gradient of the conservation equations (\ref{pc1}),
(\ref{lc1'}) and (\ref{lc2'}), and the generalized Raychaudhuri
equation (\ref{prl}), leads to a system of equations for these
gradient quantities~\cite{m}. The gradients define scalars via
their comoving divergences:
\begin{equation}\label{F}
F\equiv a\D^\mu F_\mu\,,~\mbox{ with }~ F=\Delta, Z, U, Q, P\,,
\end{equation}
where $\Delta$ is a covariant analog of the Bardeen density
perturbation $\epsilon_{\rm m}$ 
%(see~\cite{ehb}).
(see Eq.~\ref{def:epsilonm}).
 Then the system
of equations governing scalar perturbations on the brane follows
from the gradient system given in~\cite{m} as
\begin{eqnarray}
&&\dot{\Delta} =3wH\Delta-(1+w)Z\,,\label{s5}\\ &&\dot{Z}
=-2HZ-\left({c_{\rm s}^2\over 1+w}\right)
\D^2\Delta-\left({6\rho\over\kappa^2\lambda}\right) U\nonumber\\
&&~~{}-{\textstyle{1\over2}}\kappa^2 \rho\left[1+
(4+3w){\rho\over\lambda}- \left({2c_{\rm s}^2\over
1+w}\right){6{\cal U}\over\kappa^4\lambda\rho}\right]
\Delta\,,\label{s6}\\ &&\dot{U} =(3w-1)HU - \left({4c_{\rm
s}^2\over 1+w}\right)\left({{\cal U}\over\rho}\right) H\Delta
\nonumber\\&&~~{} -\left({4{\cal U}\over3\rho}\right)
Z-a\D^2Q\,,\label{s7}\\ &&\dot{Q}
=(1-3w)HQ-{1\over3a}U-{\textstyle{2\over3}} a\D^2P
\nonumber\\&&~~{} +{1\over6a}\left[ \left({8c_{\rm s}^2\over
1+w}\right){{\cal U}\over\rho}-\kappa^4
\rho(1+w)\right]\Delta\,.\label{s8}
\end{eqnarray}

In general relativity, only the first two equations apply, with
$\lambda^{-1}$ set to zero in Eq. (\ref{s6}). In this case we can
decouple the density perturbations via a second-order equation for
$\Delta$, whose independent solutions are adiabatic growing and
decaying modes. Local bulk effects modify the background dynamics,
while nonlocal bulk effects introduce new fluctuations. This leads
to fundamental changes to the simple general relativity picture.
There is no equation for $\dot{P}$, so that in general, scalar
perturbations on the brane cannot be predicted by brane observers
without additional information from the unobservable bulk. Thus in
general, one must solve also the scalar perturbations in the bulk
in order to determine the perturbation evolution on the brane.

However, there is a crucially important exception to this, arising
from the fact that $P$ only occurs in Eqs.~(\ref{s5})--(\ref{s8})
via the Laplacian term $\D^2P$.
We can use the shear propagation equation on the brane~\cite{m} to
provide an order-of-magnitude comparison of $P$ with $\Delta$:
\[
\dot{\sigma}_{\mu\nu}+2H\sigma_{\mu\nu}+E_{\mu\nu}+\D_{\langle\mu}
A_{\nu\rangle}={3\over\kappa^2\lambda}{\cal P}_{\mu\nu}\,,
\]
where $E_{\mu\nu}$ is the electric Weyl tensor on the brane. This
equation, together with Eqs.~(\ref{scal}) and (\ref{acc}) shows that
\[
{1\over\kappa^2\lambda}|\D^2{\cal P}| \sim {1\over\rho}|\D^2\rho|\,,
\]
and then Eqs.~(\ref{s3})--(\ref{F}) imply
\[
|P|\sim {\kappa^2\lambda\over a^2\rho}|\Delta|\,.
\]
Then
\[
|a\D^2P|\sim{k^2\over a^2H^2}\,{\kappa^4\rho\over a}|\Delta|\,,
\]
on using the high-energy Friedmann equation
($H^2\sim\kappa^2\rho^2/\lambda$). Thus, for $k\ll aH$, i.e. on
large scales, well beyond the Hubble horizon, we can neglect the
$\D^2P$ term in Eq.~(\ref{s8}) relative to the  $\Delta$ term.
Thus 
%{ 
 on large scales, the
system closes on the brane, and brane observers can predict scalar
perturbations from initial conditions intrinsic to the brane,
%}
without the need to solve the bulk perturbation equations.
Note that the $\D^2Q$ term in Eq.~(\ref{s7}) may also be neglected
relative to the $U$ term. This follows from the shear constraint
equation~\cite{m}
\[
\D^\nu\sigma_{\mu\nu}-{\textstyle{2\over3}}\D_\mu\Theta
=-{6\over\kappa^2\lambda}{\cal Q}_\mu\,,
\]
which gives, on taking the divergence,
\[
|Q|\sim{\kappa^2\lambda\over a\rho}|Z|\sim
{1\over aH}|U|\,,
\]
where the last relation follows from Eq.~(\ref{s6}). 
Thus
\[
|aD^2Q| \sim \frac{k^2}{a^2H^2}(H|{ U}|)
\]
and so we may neglect the $D^2Q$ term in Eq.~\ref{s7} relative to the
$U$ term.
The
system Eqs.~(\ref{s5})--(\ref{s8})
then reduces to 3 coupled equations in $\Delta$, $Z$ and
$U$, plus a decoupled equation for $Q$, which determines $Q$ once
the other 3 quantities are solved for. Thus there are in general 3
modes of large-scale density perturbations: {  a non-adiabatic
mode is introduced by bulk effects.} This mode is carried by
fluctuations $U$ in the nonlocal energy density ${\cal U}$, which
are present even if ${\cal U}$ vanishes in the background. The
fluctuations $Q$ and $P$ in the nonlocal energy flux and
anisotropic stress do not affect the density perturbations on very
large scales.

In qualitative terms, we can interpret these general results as
follows. Bulk effects lead to an effective total energy-momentum
tensor that is non-adiabatic. From Eqs.~(\ref{a}) and (\ref{b}),
we find a measure of the total effective non-adiabatic pressure
perturbation, which is the covariant analog of $\delta p^{\rm
tot}-(c_{\rm s}^2)^{\rm tot}\delta\rho^{\rm tot}$:
\begin{eqnarray}\label{nonad}
&&a\D_\mu p^{\rm tot}-(c_{\rm s}^2)^{\rm tot}a\D_\mu \rho^{\rm
tot}={6H\rho\over \kappa^4\dot{\rho}^{\rm tot}\lambda
}\left(1+{\rho\over\lambda}\right) \times
\nonumber\\&&~{}\times\left[{1\over3}-c_{\rm
s}^2-{\rho+p\over\rho+\lambda}\right] \left\{4{\cal U}\,\Delta_\mu
- 3(\rho+p) U_\mu \right\}\,.
\end{eqnarray}
In addition, Eqs.~(\ref{c}) and (\ref{d}) show that non-adiabatic
stresses and fluxes are also present, due to nonlocal bulk
effects. From this viewpoint, it is not surprising that
non-adiabatic modes arise in the density perturbations.
%Furthermore, anisotropic stresses do not affect large-scale
%density perturbations,
%which can explain how the scalar perturbation equations close on
%the brane in this case, since the generalized conservation
%equations~(\ref{pc1})--(\ref{lc2'}) form a closed system in the
%absence of ${\cal P}_{\mu\nu}$.
An analysis of the non-adiabatic large scale mode in terms of the
curvature perturbation is given in \cite{LMSW}.

Note that an alternative interpretation is also possible. We can
regard the nonlocal bulk effects as constituting a radiative
``Weyl" fluid, with energy-momentum tensor $-\kappa^{-2}{\cal
E}_{\mu\nu}$. This Weyl fluid has non-adiabatic stresses ${\cal
P}_{\mu\nu}$, and is in motion relative to the matter fluid, with
relative velocity parallel to ${\cal Q}_\mu$. Although the matter
fluid obeys energy-momentum conservation, the Weyl fluid does not;
the momentum balance equation~(\ref{lc2'}) shows that density
inhomogeneity in the matter fluid sources a momentum transfer to
the Weyl fluid.

Whatever intuitive interpretation we adopt, the concrete result is
that 
%{ 
 large-scale density perturbations on the brane may be
determined without knowledge of the bulk, and acquire a
non-adiabatic mode due to effects from the free gravitational
field in the bulk.
%}

\section{Large-scale density perturbations}

We rewrite the coupled system for large-scale perturbations by
introducing two useful new quantities. We define,
following~\cite{ehb},
\begin{equation}\label{phi}
\Phi=\kappa^2\rho a^2\Delta\,,
\end{equation}
which is a covariant analog of the Bardeen metric potential $\Phi_H$
(see Chapter~\ref{ch:adent}), and the covariant local curvature
perturbation
\begin{equation}\label{c'}
C=a\D^\mu C_\mu\,, ~C_\mu=a^3\D_\mu{R^\perp}\,,
\end{equation}
where ${R^\perp}$ is the Ricci curvature of the surfaces orthogonal
to $u^\mu$. (Note that these surfaces are in general shearing, and
non-uniform in $\rho$, $\Theta$, ${\cal U}$ and ${R^\perp}$.)

Then the coupled system for density perturbations,
Eqs.~(\ref{s5})--(\ref{s7}), can be rewritten on large scales as
\begin{eqnarray}
\dot{\Phi}&=& -H\left[1+{(1+w)\kappa^2\rho\over 2H^2}\left(1+
{\rho\over \lambda}\right)\right]\Phi \nonumber\\
&&~~{}+\left[{(1+w)\kappa^2 \rho\over 4H}\right]C -
\left[{3(1+w)a^2\rho^2\over \lambda H}\right]U\,, \label{p1}\\
\dot{C} &=& -\left[{72c_{\rm s}^2 H {\cal U}\over (1+w)\lambda
\kappa^2\rho}\right]\Phi\,, \label{p2}\\ \dot{U} &=&
H\left[3w-1-{4{\cal U}\over \kappa^2\lambda
H^2}\right]U+\left({{\cal U}\over 3a^2H\rho}\right) C \nonumber\\
&&~~{}-{2{\cal U}\over 3a^2 H\rho}\left[1+{\rho\over\lambda} +{6
c_{\rm s}^2H^2\over (1+w)\kappa^2\rho}\right]\Phi\,. \label{p3}
\end{eqnarray}

The general relativistic case is recovered when we set
$\lambda^{-1}$, ${\cal U}$ and $U$ to zero; in this case,
Eq.~(\ref{p3}) falls away, and Eq.~(\ref{p2}) reduces to
\begin{equation}\label{dotc}
C=C_o\,,~~\dot{C}_o=0\,,
\end{equation}
which expresses conservation of the covariant curvature
perturbation along each fundamental world-line. The value of $C_o$
will in general vary from world-line to word-line, so that its
conservation is local, and is {  not} an indicator of purely
adiabatic perturbations. (In general relativity, $\dot{C}=0$ on
large scales for a flat background even when there are
non-adiabatic perturbations~\cite{ehb}.) Bulk effects destroy the
local conservation of $C$ in general, by Eq.~(\ref{p2}).

However, there is an important special case when local
conservation is regained: when the nonlocal energy density ${\cal
U}$ vanishes in the background. This does not mean that
fluctuations in the nonlocal energy density are zero, i.e., we
still have $U\neq 0$ in general. It can be argued that vanishing
${\cal U}$ in the background is more natural, if one believes that
the bulk background should be conformally flat, and thus strictly
anti-de Sitter. (Quantum effects may nucleate a black hole in the
bulk~\cite{gs}, in which case the Schwarzschild-anti de Sitter
bulk, with ${\cal U}_o$ proportional to the black hole mass, would
be a natural background.) From now on, we will assume a
conformally flat background bulk and a spatially flat brane
background; thus ${\cal U}_o=0=K$ in the background generalized
Friedmann equation~(\ref{f}). Equation~(\ref{nonad}) shows that
the non-adiabatic total pressure perturbation is then proportional
to $(\rho/\lambda)U$, which will be enhanced at high energies and
suppressed at low energies.

When ${\cal U}=0$ in the background, Eq.~(\ref{dotc}) holds, and
Eq.~(\ref{p3}) gives
\begin{equation}\label{u}
U=U_of\,,~\dot{U}_o=0\,,~f=\exp \int_{a_o}^a(3w-1)\,d\ln a\,.
\end{equation}
This shows that $U$ rapidly redshifts away during inflation, so
that non-adiabatic effects from nonlocal bulk influence are small.
By contrast, the modifications to the background dynamics from
local bulk effects are strong during inflation at high energy.

The key equation~(\ref{p1}) becomes
\begin{eqnarray}
&&{d\Phi\over dN}+\left[1+{(1+w)\kappa^2\rho\over 2H^2}\left(1+
{\rho\over \lambda}\right)\right]\Phi =\nonumber\\
&&~{}\left[{(1+w)\kappa^2 \rho\over 4H^2}\right]C_o -
\left[{3(1+w)a_o^2\rho^2\over \lambda H^2}\right]e^{2N}fU_o\,,
\label{p1'}
\end{eqnarray}
where $N=\ln(a/a_o)$ is the number of e-folds. We have thus reduced the
coupled system to one simple inhomogeneous linear equation, which
may be integrated along the fundamental world-lines. Along each
world-line, the constancy of $C_o$ and $U_o$ allows us to track
the change in $\Phi$ as $w$ changes, from inflationary behavior
through to radiation- and matter-domination.

We can perform a qualitative analysis of the evolution of $\Phi$
as follows. 
%Corrections
Substituting $H^2$ from the generalized Friedman Eq.~(\ref{f}) 
into Eq.~(\ref{p1'})
and
setting ${\cal U}_0$, $K$ and $\Lambda$ to zero gives
\begin{equation}
\begin{array}{l}
{\frac {d\Phi}{dN}}+\left (1+\frac{3}{2}\,\left (1+w\right )\left (1+{
\frac {\rho}{\lambda}}\right )\left (1+\frac{1}{2}\,{\frac {\rho}{\lambda}}
\right )^{-1}\right )\Phi \\
=\frac{3}{4}\,\left (1+w\right )C_{{0}}\left (1+\frac{1}
{2}\,{\frac {\rho}{\lambda}}\right )^{-1}  \\
-9\,\left (1+w\right ){{ a_0}
}^{2}\rho\,{e^{N\left (1+3\,w\right )}}U_{{0}}{\lambda}^{-1}{\kappa}^{
-2}\left (1+\frac{1}{2}\,{\frac {\rho}{\lambda}}\right )^{-1}
\label{p1'1}
\end{array}
\end{equation}
For high-energy inflation $\rho\gg\lambda$ 
and so we can
approximate Eq.~(\ref{p1'1}) by 
\begin{eqnarray}
{\frac {d\Phi}{dN}}+\left (4+3\,w\right )\Phi=\frac{3}{2}\,{\frac {
\left (1+w\right )C_{{0}}\lambda}{\rho}}-18\,{\frac {\left (1+w\right 
){{ a_0}}^{2}{e^{N\left (1+3\,w\right )}}U_{{0}}}{{\kappa}^{2}}}
\label{p1'high}
\end{eqnarray}
As it is slow-roll inflation, we can assume $w$ and $\rho$ are
constant when solving this equation. This gives
\begin{equation}
\Phi \approx \left (1+w\right )\left (\frac{3}{2}\,{\frac {\lambda\,C_{{0}}}{\left (4+3\,w
\right )\rho}}-18\,{\frac {U_{{0}}{{ a_0}}^{2}{e^{N\left (1+3\,w
\right )}}}{\left (5+6\,w\right ){\kappa}^{2}}}\right )+{e^{\left (-4-
3\,w\right )N}}{ C_1}
\label{p1'highsoln}
\end{equation}
where $C_1$ is an integration constant. Using
$w\approx -1$ and dropping the exponentially decaying terms this gives
\begin{equation}
\mbox{high-energy inflation:}~~~\Phi \approx
{3\over2}(1+w)C_o\,{\lambda\over \rho}\,. \label{sol1}
\end{equation}
For the general relativity limit we use $\lambda\gg\rho$ and then
following the same procedure as for the high energy case we get
\begin{equation}\label{sol1'}
\Phi_{\rm gr}\approx {3\over4}(1+w)C_o\,.
\end{equation}
In general relativity, $\Phi$ remains constant on large scales
during slow-roll inflation, independent of the form of the
inflaton potential. In the brane-world, $\Phi$ is {  slowly
increasing during high-energy slow-roll inflation,} since
$\Phi\sim \rho^{-1}$ and $\rho$ is slowly decreasing. This
qualitative analysis is confirmed by the numerical integration of
a simple phenomenological model shown in Fig.~1. 
 We have modelled a smooth transition from inflation
to radiation by $w={1\over3}[(2-\alpha)\tanh(N-50)-(1-\alpha)]$,
where $\alpha$ is a small positive parameter (chosen as
$\alpha=0.1$ in the plot).
For more
realistic models, i.e., where $V(\varphi)$ is specified, the
evolution of $\Phi$ may be more complicated than shown in Fig.~1.
 For $\rho_o/\lambda\gg1$,
Eq.~(\ref{inf}) shows that inflation ends at
$N=50-2\ln[(1-2\alpha)/3]\approx 47.4$, and at $N=50$ in general
relativity. Only the lowest curve still has $\rho/\lambda\gg1$ at
the start of radiation-domination ($N$ greater than about 53), and
one can see that $\Phi$ is still growing, as confirmed by
Eq.~(\ref{sol3}).

During reheating, in periods when $w$ is approximately constant on
average (for example, $w\approx0$ for $V={1\over2}m^2\varphi^2$),
Eqs.~(\ref{pc1}) and (\ref{p1'}) imply
\begin{eqnarray}
&&\mbox{high-energy $w\approx$ constant reheating:}\nonumber\\
&&~~~~~\Phi \approx {3(1+w)\over 2(7+6w) }{\lambda\over
\rho_0}C_o\,e^{3(1+w)N}+\,{\rm const}\,. \label{sol2}
\end{eqnarray}
Thus {  high-energy $w\approx$ constant reheating on the brane
produces amplification of} $\Phi$, unlike general relativity,
where $\Phi$ remains constant on large scales during $w\approx$
constant reheating:
\begin{equation}\label{sol2'}
\Phi_{\rm gr}\approx {3(1+w)\over 2(5+3w)}C_o\,.
\end{equation}

\begin{figure} \begin{center} \includegraphics[width=12cm]{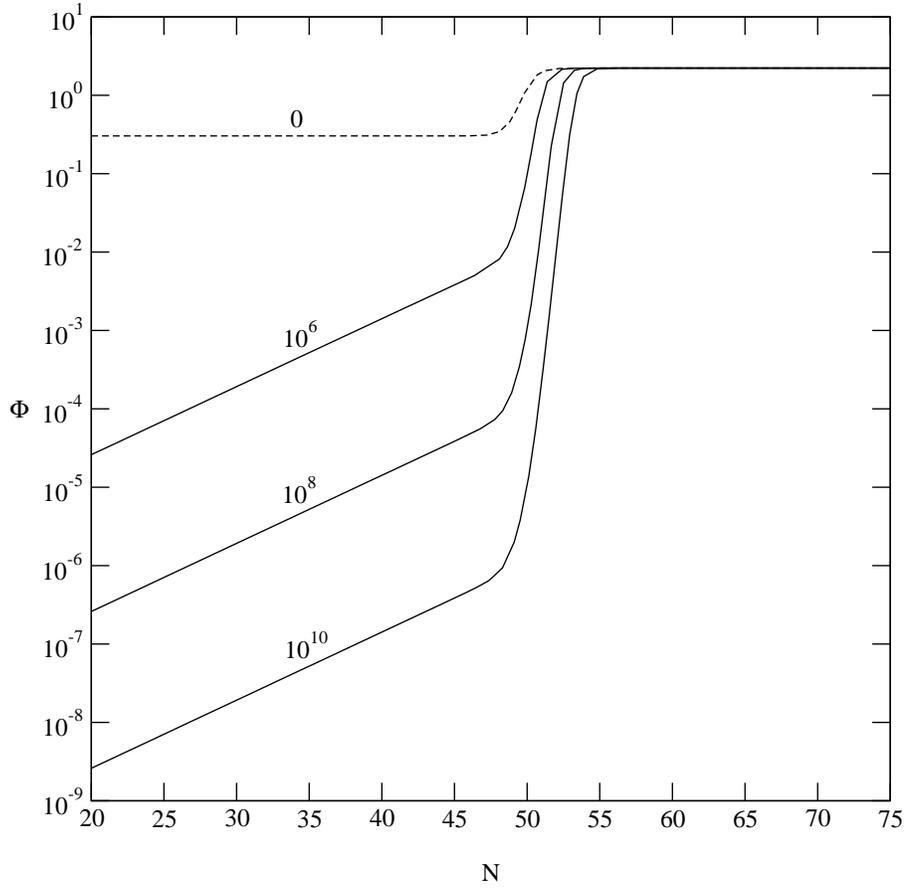} \end{center}
\caption[The evolution of $\Phi$ along a fundamental world-line.]{
The evolution of $\Phi$ along a fundamental world-line for a mode
that is well beyond the Hubble horizon at $N=0$, about 50 e-folds
before inflation ends, and remains super-Hubble through the
radiation era. Labels on the curves indicate the value
of $\rho_o/\lambda$, so that the general relativistic solution is
the dashed curve ($\rho_o/\lambda=0$).
}
\end{figure}

In the radiation era, the energy density redshifts rapidly, so
that $\rho$ quickly falls below the brane tension $\lambda$. If
the energy density at the end of reheating is high enough, then at
the start of radiation-domination we have $\rho\gg\lambda$, and we
find that $\Phi$ {  is amplified during high-energy radiation
domination:}
\begin{equation}
\mbox{high-energy radiation:}~~~\Phi \approx
{2\over9}{\lambda\over \rho_0}\,C_oe^{4N}+\,{\rm const}\,.
\label{sol3}
\end{equation}
At low energies on the brane, or in general relativity, we find
that $\Phi$ is constant:
\begin{equation}\label{sol3'}
\mbox{low-energy radiation:}~~~\Phi\approx\Phi_{\rm gr}\approx
{1\over3}C_o\,.
\end{equation}
This qualitative result is confirmed in Fig.~1. After the
radiation era, the energy scale has fallen well below the brane
tension, so that in the matter era, we recover the general relativity
result:
\begin{equation}\label{sol4}
\mbox{matter era:}~~~\Phi\approx\Phi_{\rm gr}\approx
{3\over10}C_o\,.
\end{equation}

In general relativity, the constancy of $\Phi$ during slow-roll
inflation and radiation- and matter-domination allows one to estimate
the amplification in $\Phi$. CMB large-angle anisotropies as measured
by COBE place limits on the amplified $\Phi$, and this in turn places
constraints on the inflationary potential, since the potential
determines the initial value of $\Phi$. This simple picture is
complicated by high-energy and non-local effects in the brane-world.
% We have given a rough estimate for the slow-roll inflationary
% evolution in Eq.~(\ref{sol1}). In particular, $\Phi$ grows during
% inflation, so that placing limits on inflationary parameters is more
% complicated.  The evolution of $\Phi$ is also sensitive to the form of
% the potential $V(\varphi)$, although slow-roll conditions will reduce
% this sensitivity.
Further discussion of large-angle CMB anisotropies on the brane is
given in \cite{LMSW}.

\section{Conclusions}

Using the 1+3 covariant formalism developed in~\cite{m}, we have
analyzed the evolution of large-scale density perturbations on the
brane. Density inhomogeneity on the brane generates Weyl curvature
in the bulk, which in turn backreacts on the brane, in the form of
a nonlocal energy-momentum tensor. Fluctuations in the nonlocal
energy density induce a non-adiabatic mode in large-scale density
perturbations. The fluctuations in the nonlocal energy flux are
decoupled from the density perturbations, while the nonlocal
anisotropic stress plays no role on large scales. This latter
feature is what closes the system of brane density perturbation
equations, allowing brane observers to evaluate the perturbations
on the brane without solving for the bulk perturbations.

We showed that the local and nonlocal bulk effects arising during
high-energy inflation, and any high-energy start to the radiation
era, modify the simple picture of general relativity. The local
covariant version of the metric perturbation, i.e., $\Phi$, is no
longer constant on large scales during these regimes. 
%Computing
%the constraints on inflationary potentials that are imposed by CMB
%large-angle anisotropies is therefore more complicated, and more
%model-dependent. 
We gave a rough estimate for slow-roll inflation
in Eq.~(\ref{sol1}).
%  together with Eqs.~(\ref{sol3'}) and
% (\ref{sol4}), this indicates how COBE limits can be used to impose
% constraints on the inflationary potential. A more accurate
% determination of the observational constraints on brane-world
% inflation may require numerical integration for the specified form
%of $V(\varphi)$, even for large scales. 
Numerical integration will
%also 
be required for the more complicated case, not investigated
here, when the background bulk is not conformally flat, i.e., when
the nonlocal energy density ${\cal U}$ does not vanish in the
background. In this case, the coupled system of
equations~(\ref{p1})--(\ref{p3}) can no longer be reduced to one
equation, since $C$ and $U$ are no longer locally conserved.

The formalism we have used is restricted to large scales. When a mode
approaches the Hubble radius, the gradient terms can no longer be
neglected, and the presence of these terms means that the system of
equations no longer closes on the brane. A fuller investigation
requires a formalism that can handle all scales, and which necessarily
involves the evolution of perturbations in the bulk. A covariant
formalism for bulk cosmological perturbations has not been developed,
but a metric-based formalism has been developed~\cite{pert,bd}. The
equations of this formalism are very complicated, and considerable
work remains to be done before smaller scale structure can be
predicted and compared with observations of the acoustic peaks in the
CMB anisotropies. Our results provide a useful initial step for
further developments by showing what happens on very large scales.

% Although it should be possible to reproduce the Sachs-Wolfe
% plateau and satisfy COBE limits by suitable restrictions on
% inflationary parameters, it remains to be seen whether this can be
% done consistently with the observed small-scale features of CMB
% anisotropies and with the observed matter distribution. This is
% the basis on which to confront brane-world theories with
% cosmological observations.

%\begin{appendix}
\chapter{The effect of non-linear inhomogeneities on inflation}

\label{ic}

\newcommand\xmas{\mbox{$x_{\rm mas}$}}
\newcommand\rmas{\mbox{$r_{\rm mas}$}}
\newcommand\Xmas{\mbox{$X_{\rm mas}$}}
\newcommand\Hinf{\mbox{${\rm H}_{\rm inf}$}}
\newcommand\Omegainf{\mbox{$\Omega_{\rm inf}$}}
\newcommand\Omegabg{\mbox{$\Omega_{\rm bg}$}}
\newcommand\Hbg{\mbox{${\rm H}_{\rm bg}$}}
\newcommand\rhobg{\mbox{$\rho_{\rm bg}$}}
\newcommand\rhorbg{\mbox{$\rho^{\rm bg}_{\rm r}$}}
\newcommand\rhoinf{\mbox{$\rho_{\rm inf}$}}
\newcommand\kinf{\mbox{$k_{\rm inf}$}}
\newcommand\kbg{\mbox{$k_{\rm bg}$}}
\newcommand\ainf{\mbox{$a_{\rm inf}$}}
\newcommand\abg{\mbox{$a_{\rm bg}$}}
\newcommand\xpbg{\mbox{$x_{\rm p}^{\rm bg}$}}
\newcommand\xpinf{\mbox{$x_{\rm p}^{\rm inf}$}}
\newcommand\xmasbg{\mbox{$x_{\rm mas}^{\rm bg}$}}
\newcommand\xmasinf{\mbox{$x_{\rm mas}^{\rm inf}$}}
\newcommand\thetainf{\mbox{$\theta^{\rm inf}$}}
\newcommand\thetabg{\mbox{$\theta^{\rm bg}$}}
\newcommand\xHbg{\mbox{$1/{\rm H}_{\rm bg}$}}
\newcommand\xHinf{\mbox{$1/{\rm H}_{\rm inf}$}}

\section{Introduction}
\label{introduction}
In this Chapter we discuss in more detail the initial conditions
required for inflation. In particular we examine what scales of
non-linear perturbations can be smoothed out by inflation.

%This Chapter is in a slightly different area
%to the rest of the thesis. However it does deal with inflation and
%inhomgeneties which are both central to the theme of the thesis.

% Although this work deals with inhomogeneities,
%they are not necessarily perturbative. So this work is in a slightly
%different area of Cosmology from the rest of the thesis and for that
%reason has been included in an Chapter instead of as an extra
%Chapter.

Inflation is the foremost idea for explaining the large scale
homogeneity and near flatness of the universe, which are the two main
unexplained observational features in the standard hot
big-bang model.  The large scale homogeneity or horizon problem
amounts to the fact that under hot big-bang evolution, due to a
decelerating scale factor ${\ddot a} < 0$, sufficiently separated
regions of the present day observable universe would never have been in
causal contact.  Nevertheless, it appears the
universe we observe looks very much the same, and in particular very
smooth, in all directions.  This fact is best seen in the 
%cosmic microwave
%background radiation (
 CMB which has  temperature fluctuations of
only one part in $10^5$ when measured from any direction in the sky
\cite{COBE}.

The inflation solution to this horizon problem is to picture the
universe during an early epoch to undergo an accelerated expansion,
${\ddot a} > 0$.  Such expansion can take an initially small causally
connected patch and enlarge it to a size that comfortably encompasses
our present day observed universe, thereby solving the horizon problem.
To realize inflation, the equation of state within the inflationary
patch must be of a very special form, possessing negative pressure.
The potential energy of a scalar field has an equation of state that
satisfies this requirement. This fact has been a key link towards a
dynamical realization of inflation and thereby has further motivated
the inflation solution.

As inflation is meant to solve the horizon problem, it is important
that it does not require acausal initial conditions. 
In particular, 
the picture of inflation considered here
is for the universe to emerge from an initial singularity
and then 
enter into a hot big-bang radiation dominated evolution.  At some
time $t_i$ after the initial singularity, the conditions appropriate
for inflation should occur within a small patch that is contained
within the causal horizon at that time.  
Chaotic inflation \cite{ci,linde}  does not fall into this picture
as there inflation 
is thought to start at the Planck epoch with homogeneity assumed on
the Planck scale.
To realize the picture of a
``local'' inflation, two requirements must be satisfied.  {}First a
physically sensible embedding must be demonstrated of the inflating
patch immersed within a non-inflating 
background \cite{embedd,Vachaspati:2000dy,Trodden:1999wc}.  
%CGCG
By embedding we mean matching the inflationary space time with the
background space time at the boundary of the patch.
Second, it must be
shown that
for $t> t_i$, the patch is dynamically stable to sustain
inflation \cite{piran181,kb,kb90,gpprl,gold43,other,Goldwirth:1992rj}.
{}For example, large fluctuations, which conceivably could
enter the inflating patch at a maximum rate limited only by causality,
should not destroy the inflationary conditions within the patch.

Recently, a convenient methodology has been developed
in \cite{Vachaspati:2000dy,Trodden:1999wc} for analyzing
the embedding problem, based
on the null Raychaudhuri equation \cite{ray,Wald:1984rg}.  In Sec. \ref{embedding_conditions}
the flat spacetime ($\Omega=1$)
formulation in \cite{Vachaspati:2000dy} 
is generalized to arbitrary spacetime curvature.
Then, an alternative solution from \cite{Vachaspati:2000dy}
to the embedding problem will be identified, which is especially
attractive for an 
inflating patch with an open geometry.
In Sec. \ref{dynamic_conditions}, initial conditions
for scalar field dynamics are presented, 
which are consistent with causality and our
embedding solution and which evolve into
successful supercooled or warm inflationary regimes.
%CG:
In Sec. \ref{dynamical_effects_on_embedding} we combine the results in
Sec. \ref{embedding_conditions}  and Sec. \ref{dynamic_conditions} to evaluate
the effect of inhomogeneities on the embedding problem.
Finally, Sec. \ref{conclusion} presents our conclusions.

\section{Embedding conditions}
\label{embedding_conditions}
The embedding of an inflationary
%CG
patch within a background space time generally is
regarded as unacceptable if there are any negative energy
regions. This requirement often is referred to as
the weak energy condition.  
The null Raychaudhuri equation is a useful diagnostic
tool for determining the validity of this condition.  This
equation determines the evolution
of the divergence $\theta \equiv \nabla^{\alpha}N_{\alpha}$
of the null ray vector $N^\alpha$ 
in a spacetime
%CG
with an arbitrary, and {\em not necessarily homogeneous},
energy density distribution.  
In order for the weak energy
condition to be valid, this equation
implies that for a null geodesic the condition 
%CG
\cite{Vachaspati:2000dy,Wald:1984rg}
\begin{equation}
\frac{d \theta}{ds} \leq 0
\label{dtheta}
\end{equation}
must be satisfied,
where $s$ is an affine parameter along the
null geodesic.  

{}For application of Eq.\ (\ref{dtheta}) to the inflation
embedding problem, the concept of anti-trapped
and normal regions should be understood.
%CG
Consider a sphere centered on a comoving observer.
If the space time were not expanding, photons emitted from the surface
of the sphere, radially towards the observer, would converge at
the observer. However, the expansion tends to work against the bundle
of rays converging to a point. If the expansion is rapid enough, the
bundle of rays will have diverging trajectories and then the spherical
surface from which the rays originated is
said to be an {\em anti-trapped\/} surface \cite{Wald:1984rg}.  
The spherical surface 
with a radius $\xmas$ is known as
the {\em minimally anti-trapped surface} (MAS) if any sphere with a
larger radius is anti-trapped.
{}For inwardly directed null rays, the divergence, $\theta$, will
be positive in anti-trapped regions of space.
On the other hand, 
if $\theta$ is negative for
inwardly directed null rays and positive for outwardly directed null
rays, the region is called {\it normal}.

These definitions are convenient  when
analyzing the weak energy condition based on Eq.\ (\ref{dtheta}).
{}For example, one can
%CG
immediately conclude \cite{Vachaspati:2000dy} that if an outer normal region
bounds an inner anti-trapped region in a not necessarily homogeneous,
spherically symmetric space time, the
weak energy condition would be violated, thus implying negative energy
is required at the boundary of such a configuration.

{}For the inflation problem, the inner
region is modeled as the putative inflation
patch (INF) and it is immersed within a outer background (BG)
expanding spacetime region.
%CG
As a first approximation, both the INF and BG regions are
characterized by FRW metrics of the form
\begin{equation}
ds^2 = -dt^2 + a^2 \left( \frac{dr^2}{1-K r^2} + r^2 d\Omega^2 \right),
\label{frw}   
\end{equation}
where $K=-1$ for an open geometry, 0 for a flat geometry and 1 for a
closed geometry. 
%CG
The effect of inhomogeneities in the BG will be considered in
Sec. \ref{dynamical_effects_on_embedding}. The INF patch should be homogeneous.
In what follows, $\theta$ and $\xmas$ will be computed for
a spacetime characterized by the metric Eq.\ (\ref{frw}).
The results below are applicable for both the
INF and BG regions, given that the appropriate parameters are used
in Eq.\ (\ref{frw}) for the two different spacetime regions.

Proceeding with this calculation,
the determinant of the metric Eq.\ (\ref{frw}) is
\begin{equation}
  \label{eq:determinant}
  g = |\mbox{det}(g_{\mu\nu})| = a ^ 6 r ^ 4 (1 -
K r^2)^{-1}\sin^2\psi ,
\end{equation}
where $\psi$ is the polar angle.
An incoming light ray (null vector) is given by
\begin{equation}
  \label{eq:null vector}
  N^\alpha
 = \frac{1}{a}\left[\delta^{\alpha}_{0} - a ^ {-1} (1 - K r^2)^
  {1/2}\delta^\alpha_1\right] ,
\end{equation}
with the divergence of the null rays being
\begin{equation}
  \label{eq:divergence}
  \theta
 = \frac{1}{\sqrt{g}} (\sqrt{g}N^\alpha)_{,\alpha}.
\end{equation}
Substituting equations (\ref{eq:determinant}) and (\ref{eq:null vector})
into (\ref{eq:divergence}) gives
\begin{equation}
  \label{eq:soln1}
  \theta
= \frac{2}{a} \left( {\rm H} - \frac{(1 - K r^2)^{1/2}}{ar} \right)
%\label{thetaio}
\end{equation}
with ${\rm H} \equiv {\dot a}/a$ the Hubble parameter.
The above recovers equation (10) 
of \cite{Vachaspati:2000dy} for $K = 0$.

The physical distance is given by
\begin{eqnarray}
  x &=& a \int dr (1-K r^2)^{-1/2} \\ &=& a\, \mbox{arcsinh(r)}, \quad
    \mbox{for $K = -1$}
  \label{eq:distance}
\end{eqnarray}
The MAS comoving distance, $\rmas$, must
give a zero divergence and so from Eq.\ (\ref{eq:soln1})
with, for example, $K=-1$
\begin{equation}
  \label{eq:condition}
  {\rm H} - \frac{{(1 + \rmas^2)}^{1/2}}{a\rmas} = 0,
\end{equation}
from which we get, using equation (\ref{eq:distance}),
\begin{equation}
  \label{eq:open_cond}
  \xmas = a\, \mbox{arcsinh}\left[({\rm H}^2a^2 - 1)^{-1/2}\right].
\end{equation}
{}From the Friedman equation:
\begin{equation}
  \label{eq:curv-density}
  {\rm H}^2 a^2 = \frac{K}{\Omega - 1}
\end{equation}
Substituting (\ref{eq:curv-density}) into (\ref{eq:open_cond})
gives the solution for $\xmas$ as a function of $\Omega < 1$.
A similar equation can be derived for the closed case $(1 < \Omega <
2)$ where an upper limit is needed on $\Omega$ to ensure 
that $1/{\rm H}$
is smaller than the radius of the Universe.  
Therefore, the general MAS size is given by
\begin{equation}
  \label{mascond}
  \xmas(t) = \frac{1}{\rm H}\left\{
\begin{array}{ll}
\frac{1}{(1-\Omega)^{1/2}} \mbox{arcsinh} \left(\sqrt{\frac{1 -
      \Omega}{\Omega}} \right), & 0 < \Omega < 1 \\ 1, & \Omega = 1 \\
      \frac{1}{(\Omega - 1)^{1/2}} \mbox{arcsin}
      \left(\sqrt{\frac{\Omega - 1}{\Omega}} \right), & 1< \Omega < 2 %CG(28/08)
\end{array}
\right.
\end{equation}
\begin{figure}[htbp]
   \begin{center}
         \includegraphics[angle=-90]{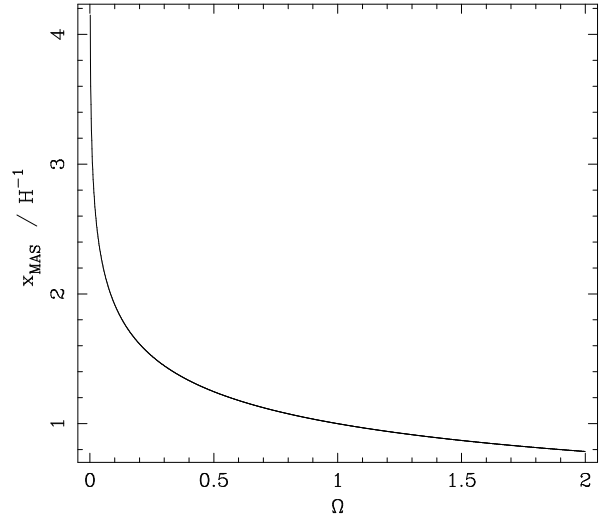}
     \caption{A plot of MAS size as function of the density.}
     
%     \caption{}
     \label{fig:xmas,Omega}
   \end{center}
\end{figure}
A plot of equation (\ref{mascond}) is given in Fig.
\ref{fig:xmas,Omega}. As can be seen, 
the size of the MAS becomes arbitrarily large relative
to the Hubble radius as $\Omega \rightarrow 0$.

The results for the $\xmas$ derived above are valid for either the
INF or BG region. In \cite{Vachaspati:2000dy}, the case $\Omega=1$ was
treated, and our calculation above is consistent with them in this
limit.  Before proceeding with our analysis for general $\Omega$, let
us review the results of \cite{Vachaspati:2000dy} for $\Omega=1$.  The
background spacetime in their work is pictured, as for us, as evolving
in a hot big bang regime.
{}For $\Omega=1$ and standard forms of matter,
e.g. radiation, non-relativitic matter, etc.,
the Hubble horizon sets the
the scale on which causal processes can take place.
For definiteness, if  we take the background causal horizon size as
$\xpbg = 1/\Hbg$, then by Eq.\ (\ref{mascond}) for $\Omega=1$, $\xpbg
=\xmasbg$.
\begin{figure}[htbp]
  \begin{center}
    \includegraphics[width=\hsize]{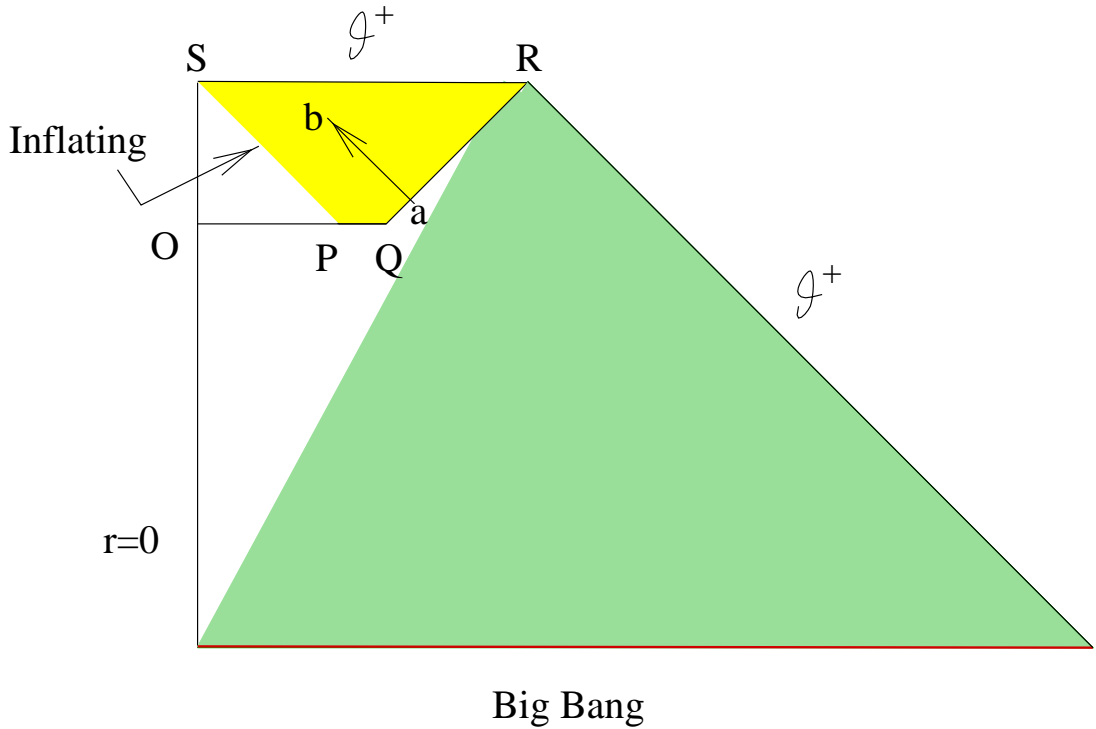}
    \caption[A Penrose diagram for local inflation.]{A Penrose diagram for local inflation (adapted from
      \protect \cite{Vachaspati:2000dy} ). The arrow (ab) denotes a
      radially directed null-geodesic going from the normal background
      space time to the anti-trapped space time inside the patch.
      Shading represents anti-trapped regions.}
    \label{fig:penrose}
  \end{center}
\end{figure}
Fig. \ref{fig:penrose} shows a Penrose diagram for an embedding that
violates the weak energy condition.  Here, at the beginning of
inflation, $\xpinf$ is represented by line OQ and $\xmasinf$ is
represented by line OP.  An inflating patch is pictured to develop
within some region inside the background (polygon OQRS in Fig.
\ref{fig:penrose}).  In the inflationary patch, they assume a
different Hubble parameter $\Hinf$ and assume the size of this patch
must be $\xpinf > 1/\Hinf$.  This assumption can be justified by
requiring the patch to be stable against perturbations and will be
discussed further in the next section.  By Eq.\ (\ref{mascond}) it
also implies $\xmasinf = 1/\Hinf$ for $\Omega=1$.  If the patch is to
be set up by causal processes then we would expect it to be smaller
than the background Hubble horizon, $\xpinf < \xHbg$.  These
conditions combined imply $\xmasbg = \xHbg > \xpinf > \xmasinf$.  As
such the region between $\xmasbg$ and $\xpinf$ is normal with respect
to the BG-region and the region between $\xpinf$ and $\xmasinf$ is
anti-trapped with respect to the inflating patch. Thus an in-going
null ray will go from a negative divergence region to a positive
divergence region.  Based on Eq.\ (\ref{dtheta}) this leads to a
violation of the weak energy condition.  Due to this fact, in
\cite{Vachaspati:2000dy} they conclude that the only way to avoid this
violation is to have $\xpinf > \xHbg$ which may be impossible or at
least very difficult to come about through causal processes.

%CG
Although this analysis assumed a homogeneous background, the general
result of \cite{Vachaspati:2000dy} was that if the inflationary patch
contained a MAS then it must be larger than the background MAS. One of
the main aims of this article is to further examine the implications
this has for a causal embedding of an inflationary patch.

Our first observation is that there is an embedding consistent with
both causality and the weak energy condition, in particular
\begin{equation}
\xpbg , \xmasinf > \xpinf\,.
\label{embed2}
\end{equation}
Since $\xpinf$ in Eq.\ (\ref{embed2}) is smaller than $\xpbg$,
it implies consistency with causality.
A region which is smaller than its MAS will have no anti-trapped surfaces.
%{}Furthermore, note that when the MAS of a region exceeds the
%size of the region, effectively that region has no MAS.
  Thus
in our proposed embedding Eq.\ (\ref{embed2}), the INF-BG boundary
is between two normal regions. 
As such, provided                                                           
$\thetainf]_{\rm boundary}$ is sufficiently smaller than $\thetabg]_{\rm
boundary}$,  the weak energy condition is satisfied.
One interesting feature of the embedding Eq. (\ref{embed2})
is that for $\Omegainf < 1$ it is acceptable for
$\xpinf > 1/{\rm H}_{\rm inf}$.

As inflation proceeds, $\Omegainf \rightarrow 1$ and so $\xmasinf
\rightarrow \xHinf$. One of the main features of inflation is that 
modes of a matter perturbation  become larger than the Hubble radius
as the Universe expands.
It follows that eventually $\xpinf > \xHinf, \xmasinf$ will need to
occur.
This will not
cause  violation of the weak energy condition provided that $\xpinf >
\xmasbg$ at that time.

{}Fig. \ref{fig:embed} shows a Penrose diagram of the proposed
embedding.
At point T in the figure, $\xpinf=\xmasbg$ and at point P,
$\xpinf=\xmasinf$.
\begin{figure}[htbp]
  \begin{center}
    \includegraphics[width=\hsize]{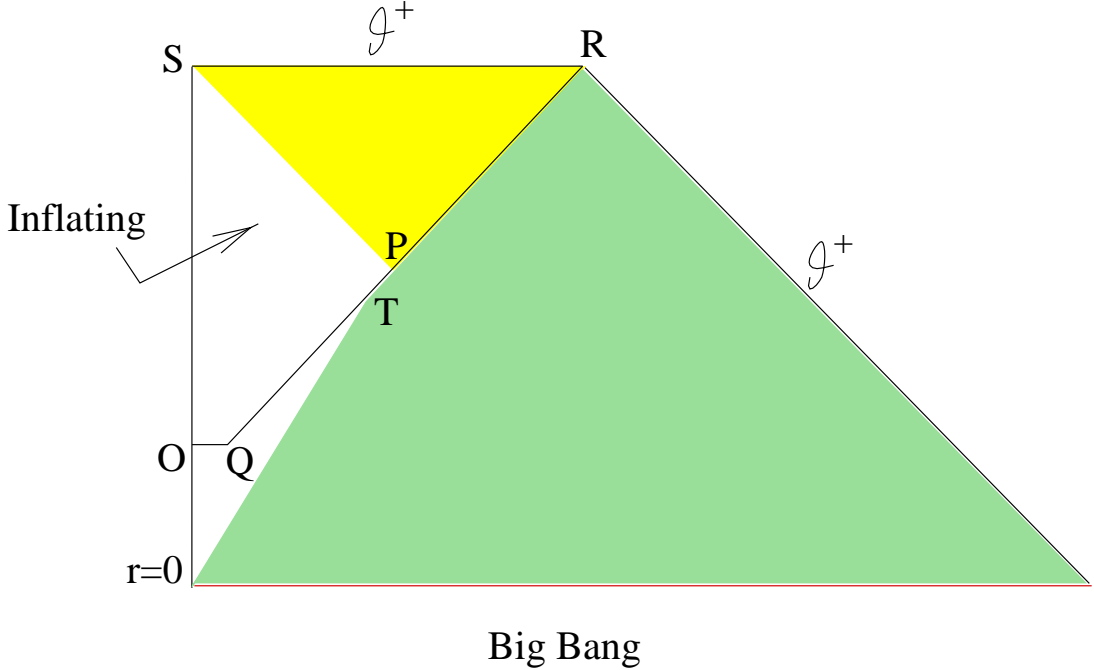}
    \caption[A successful embedding.]{A Penrose diagram for local inflation with an embedding
      that does not violate the weak energy condition and is not
      acausal.}
    \label{fig:embed}
  \end{center}
\end{figure}
 As can be seen the anti-trapped region develops in the
patch only after the patch has become greater than the background MAS.

There is another embedding which does not violate the weak energy
condition, in this case for $\Omegabg > 1$, but 
it proves not to be as useful,
\begin{equation}
  \label{eq:closed_bg}
  \xmasinf, \xmasbg < \xpinf < \xHbg\,.
\end{equation}
This solution is restricted by the upper limit on $\Omegabg$
that gives a lower limit (from Eq.\ \ref{mascond}) of $\xmasbg \approx
0.8/ \Hbg$ for $\Omega=2$. 
As the causal 
scale $\xpbg$ 
can be only of the order of $\xHbg$, such a large inflationary
patch may still be acausal.

Additional information about embedding constraints can be
obtained through the Israel junction conditions \cite{Israel:1966rt}.
In \cite{bkt,Sakai:1994fu}
conditions were derived for the  embedding of one FRW spacetime
within another. The Israel junction conditions require the extrinsic
curvature on the background side of the boundary to be
smaller than the extrinsic curvature on the patch side of the 
boundary if the boundary surface energy density is to be positive. 
It can then be shown \cite{bkt,Sakai:1994fu} that if the 
energy density of the background $\rhobg$
is greater than that of the patch $\rhoinf$, the Israel junction
conditions can be satisfied for an inflationary patch of arbitrary size 
and geometry with any
background
geometry. They also show for $\rhobg < \rhoinf$, an embedding
consistent with the Israel junction conditions and $\xpinf < \xHbg$
requires
the background geometry to be closed with the geometry
of the inflationary patch arbitrary.

\section{Dynamic Conditions}
\label{dynamic_conditions}
An immediate concern with the proposed embedding Eq.\ (\ref{embed2})
regards the size of the inflationary patch.  As far as the embedding
conditions are concerned, the patch size simply must lie below the
$\xmas$ line in Fig. 1.  In particular, for $\Omegainf > 1$ the
inflationary patch size must be 
$< 1/\Hinf$.  
However, for
$\Omegainf < 1$, 
the inflationary patch size can be 
$ >
1/\Hinf$.  The primary question which remains is what the minimal
initial inflation 
patch
size can be in order to be dynamically stable for
inflation to commence.  {}For example, if the initial patch is too
small, then large fluctuations initially outside the patch could enter
inside at a rate sufficiently fast to destroy the inflationary
conditions.  In this section, the dynamic conditions necessary
for inflation will be examined for both the scalar field
and a background radiation component.

\subsection{Scalar field}  
\label{scalar_field}
{}For scalar field driven inflation, the general dynamical conditions
necessary for a finite spatial patch to initiate and sustain inflation
have been considered in \cite{piran181,kb,kb90,gpprl,gold43,other} 
and a comprehensive review has
been given in \cite{Goldwirth:1992rj}. 
Here the requirements addressed in these
works, primarily the review \cite{Goldwirth:1992rj}, 
will be systematically examined
with emphasis on determining their implications for the
minimal initial inflationary patch size.  Our considerations 
of scalar field dynamics will generalize those in the 
above stated works, which focused on supercooled inflationary
dynamics \cite{siold,ni,ci}, 
in that here warm inflation dynamics \cite{wi} also will be
treated.

The classical evolution equation for the scalar inflaton
field has the general form
\begin{eqnarray}
& &{\ddot \phi} ({\bf x}, t)  + [3{\rm H}+\Gamma] {\dot \phi}({\bf x},t) 
\nonumber\\
& - &a^{-2} \nabla^2 \phi ({\bf x},t) + 
\frac{\delta V(\phi)}{\delta \phi({\bf x},t)} =0 ,
\label{phieom}
\end{eqnarray}
where $\Gamma$ is a dissipative coefficient, which represents
the effective interaction of the inflaton with other fields.
Here and below $\phi({\bf x},t)$ is a classical real number
which contains the ensemble average of the quantum and/or thermal
fluctuations.
In supercooled inflation, it is assumed the inflaton is isolated, in 
which case $\Gamma=0$. Thus no radiation is produced during inflation
and the universe inflates in a supercooled state.  On the other hand,
in warm inflation the inflaton interacts with
other fields, and the $\Gamma {\dot \phi}$ term is the simplest
representation of their effect on the evolution of the inflaton,
with generally $\Gamma > {\rm H}$.  Due to these interactions,
radiation is dissipated from the inflaton
system into the universe throughout the inflation period.
In general, $\Gamma$ can vary as a function of the inflaton field mode
to which it is associated, but here
$\Gamma$ is treated as a constant.

There are two types of potential $V(\phi)$ in Eq.\ (\ref{phieom})
that typically are treated in inflationary cosmology, a purely concave
potential of the generic form $V \sim m^2 \phi^2 + \lambda \phi^4$ 
and a double well potential such as $V \sim (\phi^2 - m^2)^2$.
Our discussion to follow will focus of the former type of potential
and at the end we will comment on the case of
the double well potential.  {}Furthermore, for most of our discussion
it will be adequate to consider the simplest case
of a quadratic potential, 
$V=\frac{1}{2} m^2 \phi^2$.
In this case,
going to Fourier space where we put the universe in a box
\begin{equation}
\phi({\bf x}, t) = \sum_{\bf k} \phi_{\bf k}(t) 
e^{i{\bf k} \cdot {\bf x}} ,
\end{equation}
Eq.\ (\ref{phieom}) becomes,
\begin{equation}
{\ddot \phi}_{\bf k} + [3{\rm H}+\Gamma] {\dot \phi}_{\bf k}(t)
+a^{-2} k^2 \phi_{\bf k}(t) + m^2 \phi_{\bf k}(t) =0 ,
\label{phikeom}
\end{equation}
where $k^2 \equiv |{\bf k}|^2$ with ${\bf k}$ the comoving
wave-vector and ${\bf k}/a(t)$ the corresponding physical wave-vector
at time $t$.

A necessary condition for inflation is that the zero mode
$\phi_{{\bf k} =0}$, must have a sufficiently large and
long-sustained amplitude
so that the potential energy $V=\frac{1}{2}m^2 \phi_0^2$ dominates the equation
of state of the universe, thereby driving inflation.
This leads to the familiar
slow-roll conditions, which require the curvature of the potential
to be sufficiently flat so that Eq.\ (\ref{phikeom}) 
for the zero mode becomes
first order in time
\begin{equation}
{\dot \phi}_0 = -\frac{dV(\phi)/d\phi_0}{3{\rm H}+\Gamma}.
\label{phik1eom}
\end{equation}

The dynamic initial condition problem is that the above requirements
should not be too special and in particular should not
violate causality.  The most acute initial condition
problem discussed in \cite{Goldwirth:1992rj} and related works is that generally
the initial inflaton field configuration will be very inhomogeneous,
thus considerable contribution of gradient energy ($(\nabla \phi)^2$)
should be present.  In its own right, the gradient term
for comoving mode ${\bf k}$ has energy density 
$\rho_{\nabla} = ({\bf k}^2/2a^2) \phi_{\bf k}^2$
and equation of state $p_{\nabla} = -\rho_{\nabla}/3$.
{}From the scale factor equation
\begin{equation}
\frac{\ddot a}{a} = -\frac{4 \pi G}{3}(\rho + 3p) ,
\label{scalefac}
\end{equation}
the effect of the gradient energy vanishes on the right hand
side. As such, any additional contribution from vacuum
energy still should drive inflation.  However,
excited modes with $k/a > 3{\rm H}+\Gamma$ not only will possess
gradient energy, but based on Eq.\ (\ref{phikeom})
are under-damped. As such, they also have a kinetic energy contribution
$\rho_{\dot \phi} \sim {\dot \phi}^2_{\bf k}$, which has
an equation of state $p_{\dot \phi} = \rho_{\dot \phi}$.
If the kinetic energy components of these modes dominates
the energy density in the universe, then from 
Eq.\ (\ref{scalefac}) inflation will cease to occur.

{}For the case of supercooled inflation, since $\Gamma=0$,
the simplest way to avoid
this problem is to require
that modes with $k/a \gtrsim {\rm H}$ initially
should not be excited. In other words, the inflaton field
initially should be smooth up to physical scales larger than
$\sim 1/{\rm H}$.  However, under general conditions, the causal
size of the pre-inflation patch  also will be of order
the Hubble radius $1/{\rm H}$. As such
this homogeneity requirement on the initial inflaton field
impinges on
being acausal,
since this condition
essentially requires initially smooth conditions up
to the causal scale.

To treat excited modes with $k/a > {\rm H}$
in the supercooled inflation case, the initial
condition dynamics are much more
complicated.  A simple analytical method applied to this
situation is the effective density approximation presented in
\cite{piran181,gold43} and reviewed 
in Sec. 7.2 of \cite{Goldwirth:1992rj}.  In this approach,
the inhomogeneities of the inflaton are treated in determining its
evolution, but the effect of these inhomogeneities on the
metric is only treated in the Friedmann equation through
homogeneous terms that represent the effective gradient and
kinetic energy densities as
\begin{eqnarray}
\left(\frac{\dot a}{a}\right)^2 \equiv {\rm H}^2 = 
 \frac{8\pi G}{3}
\left[\frac{1}{2} {\dot \phi}_0^2 + \frac{1}{2} m^2 \phi_0^2 \right.
+&& \nonumber\\ \left. \frac{1}{2} \sum_{\bf k}[{\dot \phi}_{\bf k}^2 + 
(\frac{{\bf k}^2}{a^2} +m^2) \phi_{\bf k}^2] +\rho_r\right]
-\frac{K}{a^2}.&& 
\end{eqnarray}
{}For initial field configurations with sufficiently excited
$k/a > {\rm H}$ modes so that their
gradient energies dominate the equation of state of the
universe, the effective density approximation indicates
the Friedmann and inflaton evolution equations are highly coupled.
In particular, the Hubble parameter will be dominated by the
gradient energy term and this will act back on the inflaton
evolution through the $3{\rm H} {\dot \phi}$ term.  
{}Furthermore, the initially large gradient terms also
can in turn induce large kinetic energy in the modes. The outcome
found in \cite{piran181,gold43,Goldwirth:1992rj}
(see also \cite{kb90}) 
for this case is that the universe expansion
is non-inflationary.  On the other hand, for the excited
$k/a > {\rm H}$ modes with smaller amplitudes, their evolution
will be oscillatory and once again the universe expansion is
non-inflationary.  {}For either
of these two possibilities,
this approximation method finds that after an initial period
of detaining the universe in a non-inflationary
regime, the effect of these higher modes becomes
negligible. At this point the remaining vacuum energy dominates and
eventually inflation proceeds.

Numerical simulations \cite{gpprl,gold43}
which exactly treat the effects on the
metric due to the inhomogeneities in the inflaton field,
in fact, do not support this final conclusion of the effective
density approximation.  Instead, the simulations
find that once modes with $k/a > {\rm H}$ are sufficiently
excited, inflation generally does not occur or
is highly suppressed.  Although the final conclusion
of the effective density approximation was incorrect, it at least
indicated that there is nontrivial interplay between the
inflaton and metric evolution equations 
once significant short distance inhomogeneities
are present in the inflaton field. 

In contrast to the supercooled case, once $\Gamma > {\rm H}$, the
effective density approximation indicates that for modes
with $\Gamma > k/a > {\rm H}$, there is almost no
feedback between the inflaton and Friedmann equations.
The evolution equation for these inflaton modes is
\begin{equation}
\Gamma \dot \phi_{\bf k}(t) = \left[\frac{{\bf k}^2}{a^2}+
m^2 \right] \phi_{\bf k}(t),
\end{equation}
and in particular 
is
independent of ${\rm H}$.  This in turn implies all these modes
are over-damped. As such, they only contribute gradient energy
to the equation of state of the universe, which alone is
ineffective in preventing inflation.  Thus up to the predictions
of the effective density approximation, in contrast to the
supercooled case,  in this warm inflation case modes with
$k/a < \Gamma$ show little coupling between the inflaton and
Friedmann equations. 
It is evident that dynamics with a $\Gamma {\dot \phi}$
damping term should have qualitative differences for the
initial condition problem compared to the case with $\Gamma=0$.
In particular, 
%CG
%the most enticing suggestion is that 
modes
of physical wavelength smaller that the Hubble radius
$1/{\rm H}$ but larger than $1/\Gamma$ could still be substantially
excited without destroying entrance into inflation.
This implies inflationary patch sizes smaller than the Hubble
radius may be dynamically stable in sustaining
inflation and based on the results of Sec. \ref{embedding_conditions}, 
they also
provide consistent spacetime embeddings, especially for 
$\Omegainf < 1$. 
{}Furthermore, studies of warm 
inflation \cite{wi,Berera:1997fm,taylor2000},
including the first principles quantum field theory
model \cite{wifp}, generally find
that to obtain adequate inflationary e-folds,
$N_e \stackrel{>}{\sim} 60$, it requires $\Gamma \approx N_em^2/{\rm H}$
with $m \gtrsim {\rm H}$ so that
$\Gamma \gtrsim N_e {\rm H}$ and thus $1/\Gamma \ll 1/{\rm H}$.
As such, for warm inflationary conditions this simple
analysis suggests that the smoothness requirement on the
initial inflationary patch is at scales much smaller than the
Hubble radius $1/{\rm H}$, which therefore imposes no
violation of causality.

These considerations can be extended to the more complicated situation
where the inflaton evolution equation has nonlinear terms,
such as the $\phi^4$ interaction term.  {}For supercooled inflation,
the effect of such interactions has been considered from the
perspective of mode mixing for some special cases in \cite{kb,kb90}
and through computer simulations of scalar fields in one spatial
dimension in \cite{gpprl,gold43}.  Up to the scope of these works,
their analysis concludes that nonlinear interactions do not present
any additional complications to the initial condition problem.
{}For the warm inflation case, this conclusion can be stated
in more general terms.  In particular, the dissipative coefficient 
$\Gamma$  completely damps the evolution, thus suppressing mode mixing,
of all modes with $k/a < \Gamma$.  As such, initially excited
modes with $k/a < \Gamma$ will evolve independently within
the time scales relevant to the initial period when inflation begins.
The only effect of excited modes with $k/a < \Gamma$ is to
contribute gradient energy to the equation of state of the universe,
and this alone is ineffective in suppressing inflation.
Therefore, provided the inflaton field also initially
maintains some vacuum energy, no aspect of the inflaton's dynamics
in the early period acts to circumvent entrance into the inflation 
phase.  

Although the above discussion addressed the 
main impediment in scalar field dynamics that
can prevent inflation,
there are a few smaller concerns worth
mentioning here.
One detail treated in \cite{Goldwirth:1992rj} is
inflaton initial conditions with large kinetic energy.
{}For $\Gamma=0$, it is shown in \cite{Goldwirth:1992rj} that this problem is 
self-correcting, and so should not pose a major barrier
in entering into the inflation phase.  {}For $\Gamma > {\rm H}$, the severity
of this problem diminishes further since this damping
term is an additional effect that helps suppress the
initial kinetic energy.  Another secondary detail is that above
we only treated concave potentials.  {}For double well potentials,
$V \sim (\phi^2 - m^2)^2$, as in the case of new 
inflation \cite{ni},
inflation requires the field to be well localized 
at the top of the potential hill $\phi =0$ and almost at
rest.  These requirements are necessary, otherwise as found
in various studies \cite{niearly}, the field can easily break up
into several small domains with randomly varying signs
of the amplitude.  {}For the supercooled case, new inflation,
the basic conclusion about the initial
conditions  required for inflation has been that they
are not very robust \cite{niearly,niic,Goldwirth:1992rj}.  The inclusion
of a dissipative term $\Gamma > {\rm H}$ will not necessarily improve this
situation.  On the one hand, such a term helps
considerably to damp kinetic energy, thus allowing
the inflaton to remain at the top of the hill longer
and drive inflation.  On the other hand, if such a damping term
arises too early before inflation is to begin,  since it freezes
the evolution of excited inflaton modes, it could prevent
the initial inflaton field configuration 
from equilibriating to $\phi=0$.

{}Finally in the above discussion $\Gamma \ne 0$ generally
represented warm inflation dynamics, although even for supercooled
inflation, if radiation is present at the onset of inflation,
such as in the new \cite{ni} and thermal \cite{ti}
inflation pictures, the effective evolution of the
inflaton could have a damping term of the form $\Gamma {\dot \phi}$.
Since our focus is on the initial phase
at the onset of inflation, it is possible
the effect of such damping terms also may
be applicable to the initial condition problem
in some supercooled inflation models.

\subsection{Radiation component}
\label{radiation_component}
So far we only considered the inflaton field system, but
in addition the universe could possess some background component of
radiation energy density, $\rhorbg$.  To realize inflation, minimally
it requires the vacuum energy density $\rho_{\rm v}$
to dominate in the patch, $\rhorbg < \rho_{\rm v}$. 
If initially $\rhorbg$
is larger than $\rho_{\rm v}$, expansion of the universe will
red-shift the radiation. Thus provided the vacuum energy sustains itself
during this early period, inflation eventually will start.
This is the standard new and warm inflation pictures.  

One could imagine that in some small patch inside a larger 
causally created spatial region,
the inflaton field configuration is reasonably smooth and is sustaining a 
sizable vacuum energy of magnitude
\begin{equation} 
\rho_{\rm v} \approx \rhorbg.
\label{rvapprr} 
\end{equation} 
As the universe expands $\rhorbg$ decreases and provided
$\rho_{\rm v}$ sustains itself, inflation will begin in the patch.
However at this moment,  there could be 
large energy fluctuations in the radiation
bath or other fields that are just outside the putative inflationary
patch.  One would like to understand how probable it is for such a
situation to curtail the inflation that had started in the patch.

To answer this question based solely on causality considerations,
one should consider the case of a perturbation
starting on the patch boundary and moving across the patch at the speed
of light, and make the extreme assumption that as the perturbation
overruns 
regions of the inflation patch, those regions convert back
to being non-inflating.  The question is what minimal initial
inflationary patch size is needed so that its expansion
under inflation is faster than its contraction due to the
impinging perturbations.  {}For the case of flat geometry $\Omega=1$,
this question was addressed in \cite{Goldwirth:1992rj} 
and they concluded that
the patch size should be at least 3 times the Hubble radius in
order for inflation to succeed in enlarging the
patch.  
{}For a general $\Omega$,
we can address this  question
by evaluating 
the maximum distance a perturbation can travel 
in the patch between the initial
$t_i$ and final
$t_f$ time of inflation
\begin{equation}
%  \label{eq:distance travelled}
  x_{\rm pert} = a(t) \int_{t_i}^{t_f} a^{-1} \, dt .
\label{xpert}
\end{equation}
Note that since $a$ is growing rapidly, there is negligible error
in taking $t_f \rightarrow \infty$ which implies
$x_{\rm pert}$ essentially is the event horizon.
We have examined Eq.\ (\ref{xpert}) for a variety of supercooled
and warm inflation models and generally find
$x_{\rm pert} \approx \xmas$.  

Thus in the most ideal case, an
inflationary patch smaller than the event horizon can not
be stable.  Here an important point of syntax should be noted,
that the generic size here is the event horizon and not the Hubble radius,
although for the flat case, $\Omega=1$, both are 
of the same order.
The above is the ideal bound based on causality.  However,
realistic dynamics also must be considered.

One case which 
corresponds to external perturbations entering
into an initially inflating patch is the case treated
in Subsect. IIIA of the mixing of high wavenumber modes.  
{}For the inflaton field interacting with itself or with other fields,
we argued above that if a dissipative term of the form $\Gamma {\dot \phi}$
is present in the inflaton effective evolution equation, then mode
mixing up to wave-numbers $k/a < \Gamma$ will not
occur within the time scale relevant to the initial condition
problem.  

In terms of the radiation bath, suppose a small vacuum 
dominated inflationary  regime
emerges, which is immersed inside a larger 
region containing radiation and gradient energy density.
Since the inflationary patch has negative pressure, the surrounding
radiation will flow into it.  The degree to which
the pressure differences are significant to this
process depends on the magnitudes
of the radiation, gradient and vacuum energies in both
regions as well as detailed dynamical 
considerations\cite{bkt,bubble}\footnote{The dynamics involves viscosity effects
at the interface between the inflation and background regions.
This viscosity is unrelated to the dissipative coefficient
$\Gamma$ in Sec. \ref{scalar_field}, which represents
the damping of the scalar field amplitude within the inflationary
patch.}.
Nevertheless, suppose the radiation tries to become uniform over
the entire region, including the patch.  
Since the patch started to inflate, it meant that initially
$\rho_{\rm v} \gtrsim \rhorbg$.  As time commences,
provided $\rho_{\rm v}$ remains constant, since expansion of the background
will decrease $\rhorbg$,  it means the amount of radiation
that flows into the inflationary patch will be less than $\rho_{\rm v}$.
Thus irrespective of whether the inflaton dynamics is supercooled
or warm, the patch should sustain a type of warm inflation due to
this influx of radiation from the regions that surround it. 

A thorough understanding of the initial
condition dynamics in presence of a background radiation component
is complicated.  To our knowledge, no quantitative analysis
has been done along these lines.  The review \cite{Goldwirth:1992rj}
only illustrated the problem with the example 
quoted above Eq.\ (\ref{xpert})
based on causality considerations but offered no dynamical 
examples with respect to a background radiation component.
Here we have offered one
scenario where the background radiation component  
should not prevent inflation from occurring.  In particular,
this example demonstrates that causality bounds 
on the rate at which radiation or other energy fluctuations
enter the patch are not the only aspect of this problem.
In addition, the magnitude of the entering
radiation must be adequately large to overwhelm the
vacuum energy.  We have offered arguments above that indicate this
latter requirement is not generic.  
 
%CG
\section{Dynamical effects on embedding}
\label{dynamical_effects_on_embedding}
As discussed in Sec. \ref{embedding_conditions}, a causally favourable
embedding requires the inflationary patch to start smaller than the
background MAS and then grow larger than it. A minimum condition for
this to happen is that the particle horizon of the
background eventually becomes greater than the background MAS.

For $p=\omega \rho$, the particle horizon is given by \cite{KT}
\begin{equation}
  \label{eq:particle_horizon}
  d(t) = \frac{1}{H} \int_0^1 \frac{dx}{\left[x^2(1-\Omega(t)) + \Omega(t) 
         x^{1-3w} \right]^{1/2}}.
\end{equation}
Comparing this equation with Eq.~(\ref{mascond}) it can be seen
that the particle horizon will only become greater than the MAS if
$\omega < 1/3$. This condition was also noted in \cite{DC} from a
slightly different perspective. There the patch was taken to start
larger than the background MAS.

It follows that our proposed embedding will not work for a pure
radiation background which has $\omega=1/3$. However, in general the
background will consist of radiation and an inhomogeneous scalar
field.  As discussed in Sec. \ref{dynamic_conditions}, the gradient
energy of the scalar field has $\omega = -1/3$ and the potential
energy has $\omega=-1$ while the kinetic energy has $\omega=1$. It
follows that if the background is gradient dominated our causal
embedding scheme will be viable, whereas our scheme fails if the
kinetic term dominates.  However, the kinetic energy of the scalar
field will be suppressed if the dissipative coefficient
$\Gamma$ of Sec. \ref{dynamic_conditions} is sufficiently large.
This is already required inside the putative inflation patch
in order to realize inflation.  Thus, it is not
unreasonable to expect $\Gamma$ in the background region
to be of the same magnitude. 

\section{Conclusion}
\label{conclusion}
This Chapter has investigated the initial condition
problem of inflation from the perspective both
of spacetime embedding and inflaton dynamics.  Our study
has highlighted two attributes of this problem which
have not been addressed in other works.  {}First, from
the perspective of spacetime embedding, we have observed that
the global geometry can play an important role in determining
the size of the initial inflationary patch that is consistent with
the weak energy condition.  Second, from the perspective of inflaton dynamics,
we have noted that a $\Gamma {\dot \phi}$
damping term could alleviate several
problems which traditionally have led to large scale homogeneity 
%CG
requirements before inflation.  

The purpose of this Chapter was to note for both these attributes, their
salient features with respect to the initial condition problem.
In the wake of this, several details emerge that must be understood.
Below, we will review the main result we found for both
attributes and then discuss the questions that must be addressed
in future work.  

{}For a causally generated patch a successful embedding can be achieved
if the patch does not contain an anti-trapped region.  We have shown
that the MAS size can be arbitrarily larger than the Hubble scale
provided $\Omega$ is made small enough.  So if the patch Hubble
horizon is taken as the minimum stable patch size, then the patch does
not have to contain an anti-trapped region if $\Omegainf < 1$.
This generalizes the analysis of \cite{Vachaspati:2000dy} which was
only for  $\Omega=1$. 
%An analysis based on the Israel junction
%conditions supports this conclusion. 
However, without the effects of damping, it
appears that the event horizon, not the Hubble horizon, 
is the minimum stable patch size. {}For
the de Sitter case one can see analytically that the event horizon is
equal to the MAS size regardless of $\Omega$ and numerical calculations
suggest the same is true for power law inflation. However, radiation
damping of perturbations could stabilize a patch smaller than the event
horizon. In this case an open geometry for the patch
would allow the patch not to contain an anti-trapped region and thus
allow a causal embedding in an expanding background which does not
violate the weak energy condition. 
Eventually the patch should develop a MAS within it, but by then it
could have expanded to be larger than the background MAS.

{}For the dynamics problem, with respect to the scalar field
the new consideration was the effect of a $\Gamma {\dot \phi}$
damping term.  We found that such a term could suppress
many of the effects from initial inhomogeneities of the inflaton,
which in studies traditionally done without this term
lead to important impediments to
entering the inflation regime.  It appears evident that
inclusion of such a damping term will
lead to qualitative differences in the
initial condition problem.  The most interesting
outcome is initial inflationary patches
smaller than the Hubble radius $1/{\rm H}$ may be able to inflate.

This Chapter examined the consequences of damping terms but did
not delve into their fundamental origin. 
{}For the cosmological setting,
such damping terms are typically associated with
systems involving a scalar field interacting
with fields of a radiation bath.  In this case, 
such damping terms have been
found in first principles calculations
for certain warm inflation models \cite{wifp},
although more work is needed along these lines. 
It is worth noting here that in the early stages of certain
supercooled inflaton scenarios where radiation is present, 
in particular new \cite{ni} and
thermal \cite{ti} inflation, a careful examination of the
dynamics may reveal damping terms similar to this.
Since the initial stages are the crucial period for the
initial condition problem, if further study supports the
importance of such damping terms, it may be useful
to better understand damping effects also in such
scenarios.

Since the most suggestive situation for the damping terms is
where in addition a radiation component is present in the universe,
in Subsect. IIIB we also studied the effects of this 
component on the initial condition problem.
Specifically we studied the case most suggestive
for the initial stages of new and warm inflation, 
where a small inflation patch is submerged inside a larger
radiation dominated spatial region.  Our main
observation has been that the minimal condition for
the patch to inflate is that its vacuum energy density must be larger than
the background radiation energy density.  Provided the vacuum
energy sustains itself, since expansion of the universe
will dilute the radiation energy density, it is not clear-cut
that the external radiation energy can act with sufficient magnitude
to impede inflation inside the patch.

%CG
We have shown that the gradient terms, due to the inhomogeneous scalar
field in the background space time, make it possible for the patch
boundary to overtake the background MAS. However the kinetic energy of
the scalar field must not be dominant for this to happen.  This can be
ensured by including a damping term in the scalar field equation.

%\end{appendix}

\chapter{Conclusions}

\label{ch:conc}

\def \etal {{\it et al.\ }}

In this Chapter the results of the thesis are summarised and
current and possible extensions of the work are discussed.

 In Chapter \ref{ch:adent} adiabatic and entropy perturbations from
multi-field models were studied \cite{gordonadent}.  A general perturbation was
decomposed into components parallel (adiabatic) and perpendicular
(entropy) to the background trajectory. We derived the evolution
equations for the entropy and adiabatic perturbations. From these
equations it was evident that on large scales the evolution of the
adiabatic perturbation is driven by the entropy perturbation, whereas
the entropy is not driven by the adiabatic perturbation. We found that
inflationary potentials that have a small effective entropy mass can
generate non-negligible entropy perturbations. Also, it was shown that
it is only for a non-straight background trajectory that the entropy
will source the adiabatic component on large scales.

We also analysed how correlations in the adiabatic and entropy
components can arise in multi-field models. We found that when the
adiabatic component is driven by the entropy component, it becomes
correlated with it. The magnitude of this correlation depends on the
relative magnitude of the homogeneous and inhomogeneous solutions of
the adiabatic evolution equation. The magnitude of the isocurvature
perturbation, in the radiation era, depends on the magnitude of the
entropy perturbation at the end of inflation and the details of how
the transition from the inflation to radiation era takes place. But,
on large scales, it is independent of the adiabatic
perturbation. 

We analysed the non-interacting double inflation model using our
decomposition.
% and recovered the results of Langlois \cite{Langlois}, who
%used the usual field decomposition. 
Our approach was useful in
identifying the reason for the features of the resulting power spectra
during the radiation era.

The question of whether preheating could affect the spectrum of
perturbations on scales relevant to structure formation was addressed
in Chapter \ref{preheating} \cite{gordonadent,gordonsting}.
  A range of models that might have this
property were examined. A potential which has efficient preheating and
does not suppress the entropy perturbation during inflation was shown
to be necessary for preheating to affect large scale structure.

We found that when the mass of the entropy field is larger than the Hubble
parameter during inflation, the entropy perturbation will be
exponentially suppressed and so preheating will not be able to effect
large scale structure. We also found that our new entropy evolution
equation was much more numerically robust, in tracking this
suppression, when compared to using the usual field perturbation
equations and constructing the entropy perturbation algebraically.

The formalism discussed in Chapter \ref{ch:adent} (which originally
appeared in \cite{gordonadent}) has subsequently been extended by
Bartolo \etal \cite{bartolo}. They derived generic slow roll
expressions for the adiabatic and entropy perturbation. One 
question that we believe still needs to be resolved is how the entropy
perturbation is transferred from the inflation to the radiation
era. When there are couplings between the scalar fields, how these
couplings influence the transfer is still not properly understood.

In Chapter \ref{ch:cmb} we studied the effect of including correlated
adiabatic and entropy perturbations on parameter estimation using the
CMB \cite{gordoncmb}. We proposed a generic form for the initial power spectra which
should encompass a broad range of two-field inflation models. We then
modified the program CMBFAST and showed how considerable
savings in computation can be achieved by using library spectra.

We found that the current CMB data was consistent with a correlated
CDM entropy perturbation of as high as twice the magnitude of the
adiabatic perturbation. We also found a degeneracy between the
primordial slope and the amount of correlated CDM entropy
perturbation. 

Subsequently Bartolo, \etal \cite{bartolo2} have used the
formalism we developed in \cite{gordonadent} to derive consistency
relations between the scalar and tensor slopes for two field
models. These consistency relations could provide a detectable
signature for two-field inflation models in the CMB.

In Chapter \ref{ch:branes} we studied the the effect of having an
extra spatial dimension on density perturbations in the brane world
scenario \cite{gordonbranes}.  We analysed super-Hubble scalar perturbations using a
covariant approach, and found that they evolve differently from
general relativity due to bulk effects.  The metric perturbation is no
longer constant on super-Hubble scales during inflation and high
energy radiation.
% , which means
% that COBE limits on the inflationary potential are different from
% general relativity and may be sensitive to the form of the potential.

We examined how the bulk Weyl curvature can be treated as an
additional fluid on the brane with possibly non-adiabatic
perturbations. We showed that on large scales the density perturbations on the
brane are described by a closed system of equations, while on small
scales there is no equation describing the evolution of the
anisotropic pressure of the projected Weyl curvature on the brane and so
the full five dimensional problem has to be studied.  
% However it
% becomes difficult to apply the covariant approach for these
% scales. Instead, one can use the metric-based bulk perturbation
% formalism (see e.g.\ \cite{brane,lmw}).
Subsequently Langlois {\it et al.} \cite{LMSW} have shown how even
the SW effect cannot be computed on the brane without solving the 5D
equations. 

In Chapter \ref{ic} we examined how non-linear inhomogeneities
effected the start of inflation \cite{gordonic}. We extended previous
work to show that even for a non-flat geometry it was not possible to
initiate inflation if there were perturbations on a scale smaller than
the Hubble horizon.

We then examined the effect of including a damping term due to decay
of the inflaton into radiation during inflation as in the `warm
inflation' model. We found this damping term set a new minimum wave
length for the non-linear perturbations that could be smoothed by
inflation. This minimum wave-length could be smaller than the Hubble
distance and so this allows a causal mechanism to be responsible for
the initial configuration of the inflationary patch.

%Overall, we hope this thesis has highlighted the potential importance
%of non-adiabatic perturbations both in understanding the early
%universe and explaining current cosmological observations.

\bibliographystyle{plain}

\begin{thebibliography}{99}
%\begin{references}
%\bibitem{erase}

\bibitem{gordonadent}
C. Gordon, D. Wands, B. A. Bassett and R. Maartens,
%``Adiabatic and entropy perturbations from inflation,''
Phys.\ Rev.\ D {\bf 63} 023506 (2001). 


\bibitem{gordonsting}
B. A. Bassett, C. Gordon, R. Maartens and D. I. Kaiser,
%``Restoring the sting to metric preheating,''
Phys.\ Rev.\ D {\bf 61} 061302 (2000).


\bibitem{gordoncmb}
L. Amendola, C. Gordon, D. Wands and M. Sasaki,
% %``Correlated perturbations from inflation and the cosmic microwave  background,''
 astro-ph/0107089. 
% %%CITATION = ASTRO-PH 0107089;%%

\bibitem{gordonbranes}
C. Gordon and R. Maartens,
%``Density perturbations in the brane world,''
Phys.\ Rev.\ D {\bf 63} 044022 (2001). 
%%CITATION = HEP-TH 0009010;%%

\bibitem{gordonic}
A. Berera and C. Gordon,
%``Inflationary initial conditions consistent with causality,''
Phys.\ Rev.\ D {\bf 63}  063505 (2001).


% \end{thebibliography}

%\def\bibname{References for Chapter 1}
%\begin{thebibliography}{99}

%\bibitem{gordon_art}
% L. Amendola, C. Gordon, D. Wands and M. Sasaki,
% %``Correlated perturbations from inflation and the cosmic microwave  background,''
% astro-ph/0107089;
% %%CITATION = ASTRO-PH 0107089;%%
% C. Gordon and R. Maartens,
% %``Density perturbations in the brane world,''
% Phys.\ Rev.\ D {\bf 63}  044022 (2001)
% [hep-th/0009010];
% %%CITATION = HEP-TH 0009010;%%
% C. Gordon, D. Wands, B. A. Bassett and R. Maartens,
% %``Adiabatic and entropy perturbations from inflation,'' 
% Phys.\ Rev.\ D {\bf 63}  023506 (2001) [astro-ph/0009131];
% %%CITATION = ASTRO-PH 0009131;%%
% B. A. Bassett, C. Gordon, R. Maartens, and D. I. Kaiser,
% %``Restoring the sting to metric preheating,''
% Phys.\ Rev.\  D {\bf 61}  061302 (2000) [hep-ph/9909482]. 

\bibitem{mnf}
C. Gordon, {IEEE} Transactions on Geoscience and Remote Sensing {\bf
  38} 608
(2000).

\bibitem{KT}
E. W. Kolb and M. S. Turner,
{\em The Early Universe,}
   Addison-Wesley (1990); M. S. Longair, {\em Galaxy Formation},
   Springer (1998).

\bibitem{LiLy00} A. R. Liddle and D. H. Lyth,
{\em Cosmological inflation and large-scale structure,}
%\it
  Cambridge University Press (2000).

\bibitem{COBE}
C. L. Bennett {\it et al.},
%``4-Year COBE DMR Cosmic Microwave Background Observations: Maps and Basic Results,''
Astrophys.\ J.\  {\bf 464} L1 (1996). 
%%CITATION = ASTRO-PH 9601067;%%

\bibitem{SME}
W. Stoeger {\it et al.}, Astrophys.\ J.\  {\bf 443} L1 (1995). 

\bibitem{MFB} V. F. Mukhanov, H. A. Feldman and R. H. Brandenberger,
%``Theory of cosmological perturbations. Part 1. Classical perturbations. Part 2. Quantum theory of perturbations. Part 3. Extensions,''
Phys.\ Rept.\ {\bf 215} 203 (1992). 
%%CITATION = PRPLC,215,203;%%

\bibitem{net}
C. B. Netterfield {\it et al.},
%``A measurement by BOOMERANG of multiple peaks in the angular power
%spectrum of the cosmic microwave background,'' 
astro-ph/0104460. 
%%CITATION = ASTRO-PH 0104460;%%

\bibitem{halverson}
N. W. Halverson {\it et al.},
%``DASI First Results: A Measurement of the Cosmic Microwave Background Angular Power Spectrum,''
astro-ph/0104489. 
%%CITATION = ASTRO-PH 0104489;%%

\bibitem{lee}
A. T. Lee {\it et al.},
%``A High Spatial Resolution Analysis of the MAXIMA-1 Cosmic Microwave
%Background Anisotropy Data,'' 
astro-ph/0104459. 
%%CITATION = ASTRO-PH 0104459;%%



%\bibitem{newcmb}
% C. B. Netterfield {\it et al.},
% %``A measurement by BOOMERANG of multiple peaks in the angular power
% %spectrum of the cosmic microwave background,'' 
% astro-ph/0104460;
% %%CITATION = ASTRO-PH 0104460;%%
% N. W. Halverson {\it et al.},
% %``DASI First Results: A Measurement of the Cosmic Microwave Background Angular Power Spectrum,''
% astro-ph/0104489;
% %%CITATION = ASTRO-PH 0104489;%%
% A. T. Lee {\it et al.},
% %``A High Spatial Resolution Analysis of the MAXIMA-1 Cosmic Microwave
% %Background Anisotropy Data,'' 
% astro-ph/0104459. 
% %%CITATION = ASTRO-PH 0104459;%%




\bibitem{linde}
A. D. Linde,
{\em Particle Physics And Inflationary Cosmology,}
 Harwood (1990).

\bibitem{mathieu}
N. W. Mac Lachlan, {\em Theory and applications of the Mathieu
  functions,} Dover (1961).

\bibitem{KLS1}
L. Kofman, A. D. Linde and A. A. Starobinsky,
%``Reheating after inflation,''
Phys.\ Rev.\ Lett.\  {\bf 73} 3195 (1994). 


\bibitem{KLS2}
L. Kofman, A. D. Linde and A. A. Starobinsky,
%``Towards the theory of reheating after inflation,''
Phys.\ Rev.\ D {\bf 56} 3258 (1997). 



\bibitem{ADD}
N.~Arkani-Hamed, S.~Dimopoulos and G.~R.~Dvali,
%``The hierarchy problem and new dimensions at a millimeter,''
Phys.\ Lett.\ B {\bf 429} 263 (1998). 


\bibitem{RS1} L. Randall and R. Sundrum, Phys. Rev. Lett. {\bf 83} 
4690 (1999). 

\bibitem{RS2}
L. J. Randall and R. Sundrum,
%``Out of this world supersymmetry breaking,''
Nucl.\ Phys.\ B {\bf 557} 79 (1999). 


\bibitem{roy}
R. Maartens,
%``Geometry and dynamics of the brane-world,''
gr-qc/0101059. 
%%CITATION = GR-QC 0101059;%%

%\end{thebibliography}


%\def\bibname{References for Chapter 2 
%(Adiabatic and entropy pertrubations from inflation) 
%References
%}
%\begin{thebibliography}{99}


\bibitem{GBW} J. Garc\'\i a-Bellido and D. Wands,
        %``Constraints from inflation on scalar - tensor gravity theories,''
        Phys.\ Rev.\  D {\bf 52}  6739 (1995);
        J. Garc\'\i a-Bellido and D. Wands,
        %``Metric perturbations in two-field inflation,''
        Phys.\ Rev.\  D {\bf 53}  5437 (1996).
        

\bibitem{Shinji}
S. Tsujikawa and H. Yajima, hep-ph/0007351. 

\bibitem{HK}
H. Kodama and T. Hamazaki, Prog.\ Theor.\ Phys.\  {\bf 96}  949 (1996);
T. Hamazaki and H. Kodama, Prog.\ Theor.\ Phys.\  {\bf 96}  1123 (1996). 

\bibitem{Hu}
W. Hu, Phys.\ Rev.\  D {\bf 59}  021301 (1999). 


\bibitem{WMLL} D. Wands, K. A. Malik, D. H. Lyth and A. R. Liddle,
        %``A new approach to the evolution of cosmological
        %perturbations on large scales,''
        Phys.\ Rev.\ D {\bf 62}  043527 (2000). 

\bibitem{Bassett:1999mt}
B. A. Bassett, F. Tamburini, D. I. Kaiser, and R. Maartens,
%``Metric preheating and limitations of linearized gravity. II,''
Nucl.\ Phys.\  B {\bf 561}  188 (1999). 

\bibitem{BKM1} B. Bassett, D. Kaiser, and R. Maartens, Phys. Lett. 
B {\bf 455}  84 (1999. 

%\cite{Finelli:1999bu}
\bibitem{Finelli:1999bu}
F. Finelli and R. Brandenberger,
%``Parametric amplification of gravitational fluctuations during  reheating,''
Phys.\ Rev.\ Lett.\  {\bf 82}  1362 (1999). 
%%CITATION = HEP-PH 9809490;%%


\bibitem{SS} M. Sasaki and E. D. Stewart,
        %``A General analytic formula for the spectral index of the
        %density perturbations produced during inflation,''
        Prog.\ Theor.\ Phys.\  {\bf 95}  71 (1996). 

\bibitem{Star}
A. A. Starobinsky, Pis'ma Zh.\'{E}ksp. Teor. Fiz. {\bf 42}  45 (1985)
[JETP Lett. {\bf 42}  51 (1985)]. 

\bibitem{Salopek} 
D. S. Salopek, Phys.\ Rev.\  D {\bf 52}  5563 (1995);
T. T. Nakamura and E. D. Stewart, Phys.\ Lett.\  B {\bf 381}  413 (1996). 

\bibitem{SasTan}
M. Sasaki and T. Tanaka, Prog.\ Theor.\ Phys.\  {\bf 99}  763 (1998). 

\bibitem{LR}
D. H. Lyth and A. Riotto, Phys. Rep. {\bf 314}  1 (1999). 

\bibitem{Polarski:1992dq}
D. Polarski and A. A. Starobinsky,
%``Spectra of perturbations produced by double inflation with
%an intermediate matter dominated stage,''
Nucl.\ Phys.\  B {\bf 385}  623 (1992). 

\bibitem{PolSta94}
D. Polarski and A. A. Starobinsky,
%``Isocurvature perturbations in multiple inflationary models,''
Phys.\ Rev.\  D {\bf 50}  6123 (1994).

%\bibitem{Kofman:1994rk}
%L. Kofman, A. Linde and A. A. Starobinsky,
%%``Reheating after inflation,''
%Phys.\ Rev.\ Lett.\  {\bf 73}  3195 (1994) [hep-th/9405187]. 

\bibitem{SY}
A. A. Starobinsky and J. Yokoyama,
        %``Density fluctuations in Brans-Dicke inflation,''
        gr-qc/9502002. 

\bibitem{MukSte} 
V. F. Mukhanov and P. J. Steinhardt, Phys. Lett. B {\bf
         422}  52 (1998).

\bibitem{julien} 
J. Lesgourgues,
%``Features in the primordial power spectrum of double D-term inflation,''
Nucl.\ Phys.\ B {\bf 582}  593 (2000).
[%%CITATION = HEP-PH 9911447;%%

\bibitem{many}
A. D. Linde, Phys. Lett. B {\bf 158}  375 (1985); 
D. Seckel and M. S. Turner, Phys. Rev. D {\bf 32}  3178 (1985);
L. A. Kofman, Phys. Lett. B {\bf 173}  400 (1986); 
A. D. Linde and L. A. Kofman, Nucl. Phys. B {\bf 282}  555 (1987); 
D. S. Salopek, Phys. Rev. D {\bf 45}  1139 (1992);
A. D. Linde and V. F. Mukhanov, Phys. Rev. D {\bf 56}  535 (1997). 
M. Bucher and Y. Zhu, Phys. Rev. D {\bf 55}  7415 (1997);
P. J. E. Peebles, Astrophys. J. {\bf 510}  523 and 531
(1999);
A. R. Liddle and A. Mazumdar, Phys.\ Rev.\  D {\bf 61}  123507 (2000). 

\bibitem{baryon}
H. Kodama and M. Sasaki, Int. J. Mod. Phys. A {\bf 1}  265 (1986);
P. J. E. Peebles, Astrophys. J. {\bf 315}  L73 (1987); T. Chiba, N. 
Sugiyama, and Y. Suto, Astrophys. J. {\bf 429}  427 (1994). 

\bibitem{P}
H. Kodama and M. Sasaki,
% Int. J. Mod. Phys. A{\bf 1}  265 (1986);
Int. J. Mod. Phys. A{\bf 2}  491 (1987);
%H. Kodama and M. Sasaki, Int. J. Mod. Phys. A {\bf 2}  491 (1987); 
J. R. Bond and G. Efstathiou, Mon. Not. R. Astron. Soc. {\bf 218} 
103 (1986). 

\bibitem{BMT}
M. Bucher, K. Moodley and N. Turok,
%``The general primordial cosmic perturbation,''
Phys.\ Rev.\ D {\bf 62} 083508 (2000). 
%%CITATION = ASTRO-PH 9904231;%%

\bibitem{Langlois} 
D. Langlois,
        %``Correlated adiabatic and isocurvature perturbations from
        %double inflation,''
        Phys.\ Rev.\  D {\bf 59}  123512 (1999). 

\bibitem{BMT2}
M. A. Bucher, K. Moodley, and N. G. Turok, astro-ph/0007360. 

\bibitem{Langlois2}
D. Langlois and A. Riazuelo, Phys.\ Rev.\  D {\bf 62}  043504
(2000). 

\bibitem{isocmb}
J. R. Bond and G. Efstathiou, Mon. Not. R. Astron. Soc. {\bf 222} 
33 (1987); 
W. Hu, E. Bunn, and N. Sugiyama, Astrophys. J. {\bf
447}  L59 (1995);
R. Stompor, A. J. Banday, and
K. M. Gorski, Astrophys. J. {\bf 463}  8 (1996);
T. Kanazawa, M. Kawasaki, N. Sugiyama, and T. Yanagida, Prog. 
Theor. Phys. {\bf 102}  71 (1999);
K. Enqvist and H. Kurki-Suonio, Phys. Rev. D {\bf 61}  043002 (2000);
E. Pierpaoli, J. Garcia-Bellido, and S. 
Borgani, JHEP {\bf 10}  015 (1999);
K. Enqvist, H. Kurki-Suonio, and J. Valiviita, 
%``Limits on Isocurvature Fluctuations from Boomerang and MAXIMA,''
Phys.\ Rev.\ D {\bf 62} (2000) 103003.


%K. Enqvist, H. Kurki-Suonio and J. Valiviita,
%``Open and closed CDM isocurvature models contrasted with the CMB data,''
%astro-ph/0108422. 
%%CITATION = ASTRO-PH 0108422;%%

\bibitem{enqvistBM}
K. Enqvist, H. Kurki-Suonio and J. Valiviita,
%``Limits on Isocurvature Fluctuations from Boomerang and MAXIMA,''
Phys.\ Rev.\ D {\bf 62}  103003 (2000). 


\bibitem{Karim}
K. A. Malik and D. Wands, Phys.\ Rev.\  D {\bf 59}  123501 (1999).


\bibitem{Bardeen} J. M. Bardeen, Phys. Rev. D {\bf 22}  1882 (1980). 

\bibitem{Hwang}
J. Hwang,
%``Evolution of scalar field cosmological perturbations,''
Astrophys.\ J.\  {\bf 427}  542 (1994). 

\bibitem{Ellis}
M. Bruni, G. F. R. Ellis and P. K. S. Dunsby, Class. Quantum Grav. {\bf
9}  921 (1992). 

\bibitem{Durrer}
R. Durrer, Fundamentals of Cosmic Physics {\bf 15}  209 (1994).


%\bibitem{MFB} V. F. Mukhanov, H. A. Feldman and R. H. Brandenberger,
%        Phys. Rep. {\bf 215}  203 (1992). 

\bibitem{KS} 
H. Kodama and M. Sasaki, Prog. Theor. Phys. Suppl. {\bf
        78}  1 (1984). 

\bibitem{SM} 
M. Sasaki, Prog. Theor. Phys. {\bf 76}  1036 (1986);
V. F. Mukhanov, Zh.\'{E}ksp. Teor. Fiz. {\bf 94}  1 (1988)
        [Sov. Phys. JETP {\bf 68}  1297 (1988)]. 

\bibitem{TN} A. Taruya and Y. Nambu, Phys. Lett. B {\bf 428}  37
(1998). 

\bibitem{Lukash}
V. Lukash, Zh. Eksp. Teor. Fiz. {\bf 79}  1601 (1980) [Sov. Phys. JETP
{\bf 52}  807 (1980)]. 

\bibitem{Lyth85}
D. H. Lyth, Phys. Rev. D {\bf 31}  1792 (1985). 


\bibitem{BST} 
J. M. Bardeen, P. J. Steinhardt and M. S. Turner,
        Phys.\ Rev.\  D {\bf 28}  679 (1983). 

\bibitem{MarSch} 
J. Martin and D. J. Schwarz,
        Phys.\ Rev.\  D {\bf 57}  3302 (1998).
        

\bibitem{LL93}
A. R. Liddle and D. H. Lyth,
%``The Cold dark matter density perturbation,''
Phys.\ Rep.\  {\bf 231}  1 (1993).


\bibitem{FB}
F. Finelli and R. H. Brandenberger,
%``Parametric amplification of metric fluctuations during reheating in two  field models,''
Phys.\ Rev.\ D {\bf 62} (2000) 083502.
%%CITATION = HEP-PH 0003172;%%


\bibitem{StewartWalker}
J. M. Stewart and M. Walker, Proc. R. Soc. Lond. A {\bf 341}  49 (1974). 


\bibitem{mm}
A. Mazumdar and L. E. Mendes, Phys. Rev. D {\bf 60}  103513 (1999). 
%hep-ph/9902274. 

\bibitem{BV}
B. A. Bassett and F. Viniegra, Phys. Rev. D {\bf 62}  043507 (2000). 


\bibitem{Jedamzik:2000um}
K. Jedamzik and G. Sigl,
%``On metric preheating,''
Phys.\ Rev.\  D {\bf 61}  023519 (2000);
%^\bibitem{Ivanov:2000hz}
P. Ivanov,
%``On generation of metric perturbations during preheating,''
Phys.\ Rev.\ D {\bf 61}  023505 (2000)/

\bibitem{LLMW}
A. R. Liddle, D. H. Lyth, K. A. Malik and D. Wands,
%``Super-horizon perturbations and preheating,''
Phys.\ Rev.\ D {\bf 61}  103509 (2000).


%\bibitem{Bassett:2000ta}
%B. A. Bassett, C. Gordon, R. Maartens, and D. I. Kaiser,
%``Restoring the sting to metric preheating,''
%Phys.\ Rev.\  D {\bf 61}  061302 (2000) [hep-ph/9909482]. 

\bibitem{BD}
N. D. Birrell and P. C. W. Davies, {\em Quantum fields in curved
space}  Cambridge University Press (1982). 


\bibitem{ZBS}
J. P. Zibin, R. H. Brandenberger and D. Scott,
%``Backreaction and the parametric resonance of cosmological  fluctuations,''
Phys.\ Rev.\ D {\bf 63} 043511 (2001). 
%%CITATION = HEP-PH 0007219;%%



\bibitem{TBV}
S. Tsujikawa, B. A. Bassett, and F. Viniegra, JHEP {\bf 08}  019
(2000).


%\end{thebibliography}

%\def\bibname{References for Chapter 3 
%Preheating
%}
%\begin{thebibliography}{99}


%\bibitem{KLS} L. Kofman, A. Linde, and A. Starobinsky, Phys. Rev. D
%{\bf 56}  3258 (1997). 

%\bibitem{met} 
%B. A. Bassett, D. I. Kaiser, and R. Maartens,
%Phys. Lett. B {\bf 455}  84 (1999); 
%B. A. Bassett, F. Tamburini,
%D. I. Kaiser, and R. Maartens, Nucl. Phys. B {\bf 561}  188 (1999). 


%\bibitem{BV}
%B. A. Bassett and F. Viniegra, hep-ph/9909353. 

%\bibitem{FB} F. Finelli and R. H. Brandenberger, Phys. Rev. Lett. {\bf
%82}  1362 (1999). 

%\cite{JS}
%\bibitem{JS}
%K. Jedamzik and G. Sigl,
%``On metric preheating,''
%Phys.\ Rev.\  D {\bf 61}  023519 (2000). 
%[hep-ph/9906287]. 
%%CITATION = HEP-PH 9906287;%%
%\href{http://www. slac. stanford. edu/spires/find/hep/www?eprint=HEP-PH/9906287}{SPIRES}

%\cite{Ivanov:2000hz}
%\bibitem{I}
%P. Ivanov,
%``On generation of metric perturbations during preheating,''
%Phys.\ Rev.\  D {\bf 61} 023505 (2000). 
%[astro-ph/9906415]. 
%%CITATION = ASTRO-PH 9906415;%%
%\href{http://www. slac. stanford. edu/spires/find/hep/www?eprint=ASTRO-PH/9906415}{SPIRES}

%\bibitem{nonlin} 
%M. Parry and R. Easther, hep-ph/9903550
%and hep-ph/9910441;  
%A. Sornborger and M. Parry,  Phys. Rev. Lett. {\bf 83}  666 (1999)

%\bibitem{j} 
%J. Baacke, K. Heitmann and C. Paetzold,
%Phys. Rev. D {\bf 55}  2320 (1997); 
%B. A. Bassett and S. Liberati,
%Phys. Rev. D{\bf 58}  021302 (1998); 
%B. A. Bassett and F. Tamburini, Phys. Rev. Lett. {\bf 81}  2630 (1998). 

%\bibitem{subir} 
%G. G. Ross, Phys. Lett. B {\bf 171}  46 (1986);
%G. G. Ross, and S. Sarkar, Nucl. Phys. B {\bf 461} 597 (1996). 

%\bibitem{KLS94}L. Kofman, A. Linde, and A. Starobinsky, Phys. Rev. 
%Lett. {\bf 73}  3195 (1994). 

%\bibitem{GPR} 
%B. R. Greene, T. Prokopec, and T. G. Roos, Phys. Rev. D
%{\bf 56}  6484 (1997). 

%\bibitem{j2} 
%J. Baacke, K. Heitmann, and C. Paetzold, Phys. Rev. 
%D {\bf 55}  7815  (1997). 

% \bibitem{hyb} 
% J. Garcia-Bellido and A. Linde, Phys. Rev. D {\bf 57} 
% 6075 (1998); 
% M. Bastero-Gil, S. F. King, and J. Sanderson,
% Phys.\ Rev.\ D {\bf 60}  103517 (1999).



%\bibitem{GB} 
%J. Garcia-Bellido, Phys. Lett. B {\bf 418}  252 (1998). 

% \bibitem{doppler}  
% A. Lewin and A. Albrecht, 
% %``Can inflationary models of cosmic perturbations evade the secondary  oscillation test?,''
% Phys.\ Rev.\ D {\bf 64}  023514 (2001).
% %D. Langlois, Phys. Rev. D {\bf 59}  123512 (1999). 

\bibitem{nongauss} 
H. M\"uller and C. Schmid, gr-qc/9412022. 

\bibitem{pbvm} 
B.~A.~Bassett, G.~Pollifrone, S.~Tsujikawa and F.~Viniegra,
%``Preheating as cosmic magnetic dynamo,''
Phys.\ Rev.\ D {\bf 63} 103515 (2001). 

\bibitem{mag} 
M. S. Turner and L. W. Widrow, Phys. Rev. D {\bf 37}  2743
(1988). 

%\end{thebibliography}

%\def\bibname{References for Chapter 5
%(Effects on the Cosmic Microwave Background)
% References
%}
%\begin{thebibliography}{99}


\bibitem{tegmark}
M. Tegmark, M. Zaldarriaga and A. J. Hamilton,
%``Towards a refined cosmic concordance model: joint 11-parameter constraints from CMB and large-scale structure,''
Phys.\ Rev.\ D {\bf 63} (2001) 043007.
%%CITATION = ASTRO-PH 0008167;%%

\bibitem{adiablikli}
P. de  Bernardis {\it et al.}, 
%``Multiple Peaks in the Angular Power Spectrum of the Cosmic
%Microwave Background: Significance and Consequences for Cosmology''
astro-ph/0105296;
R. Stompor {\it et al.},
%``Cosmological implications of the MAXIMA-I high resolution Cosmic
%Microwave Background anisotropy measurement''
astro-ph/0105062;
C. Pryke {\it et al.},
%``Cosmological Parameter Extraction from the First Season of
%Observations with DASI,'' 
astro-ph/0104490;
%%CITATION = ASTRO-PH 0104490;%%
X. Wang {\it et al.},
%``Is cosmology consistent?''
astro-ph/0105091. 

%\bibitem{multiinf}
%A. D. Linde, {\em Phys. Lett. } B {\bf  158} (1985) 375; 
%L. A. Kofman, {\em Phys. Lett. } B {\bf  173} (1985) 400; 
%A. D. Linde and L. A. Kofman, { Nucl. Phys. } B {\bf  282}   555 (1987). 
%D. Polarski and A. A. Starobinsky,
%``Isocurvature perturbations in multiple inflationary models,''
%Phys.\ Rev.\ D {\bf 50}  6123 (1994)
%[astro-ph/9404061]
%%CITATION = ASTRO-PH 9404061;%%
% A. A. Starobinsky and J. Yokoyama
% %, {\em Density fluctuations in Brans-Dicke inflation} 
% {\tt gr-qc/9502002}; 
%J. Garc\'\i a-Bellido and D. Wands, {\em Phys. Rev. } D {\bf  52} (1995) 
%6739; 
%M. Sasaki and E. D. Stewart, 
% {\em Prog. Theor. Phys. } {\bf 95} (1996) 71;
%J. Garc\'\i a-Bellido and D. Wands, { Phys. Rev. } D {\bf  53}  5437  
%(1996); 
%A. D. Linde and V. Mukhanov, Phys. Rev. D {\bf 56}   535 (1997);
%M. Sasaki and T. Tanaka, 
%Prog.\ Theor.\ Phys.\ {\textbf{99}}  763 (1998) [gr-qc/9801017]. 

%\bibitem{LangloisI}
%D. Langlois,
%``Correlated adiabatic and isocurvature perturbations from double
%inflation,'' 
%Phys.\ Rev.\ D {\bf 59}  123512 (1999). 
%[astro-ph/9906080]. 
%%CITATION = PHRVA,D59,123512;%%

%\bibitem{GWBM}
%C. Gordon, D. Wands, B. A. Bassett and R. Maartens,
%``Adiabatic and entropy perturbations from inflation,''
%Phys.\ Rev.\ D {\bf 63}  023506 (2001)
%[astro-ph/0009131]. 
%%CITATION = ASTRO-PH 0009131;%%

%\bibitem{Hwang1}
%J. Hwang and H. Noh,
%``Cosmological perturbations with multiple scalar fields,''
%Phys.\ Lett.\ B {\bf 495}  277 (2000)
%[astro-ph/0009268]. 
%%CITATION = ASTRO-PH 0009268;%%

\bibitem{HwangNoh}
J. Hwang and H. Noh,
%``Cosmological perturbations with multiple scalar fields,''
Phys.\ Lett.\ B {\bf 495} 277 (2000). 
%%CITATION = ASTRO-PH 0009268;%%


\bibitem{bartolo}
N. Bartolo, S. Matarrese and A. Riotto,
%``Oscillations during inflation and cosmological density perturbations,''
astro-ph/0106022. 
%%CITATION = ASTRO-PH 0106022;%%

%\bibitem{efstathiou} 
%G. Efstathiou and J. R. Bond, Mon.\ Not.\ Roy.\ Astron.\ Soc.\  {\bf
%  218}  103 (1986);
%H. Kodama and M. Sasaki, Int. J. Mod. Phys. A{\bf 1}  265 (1986);
%A{\bf 2}  491 (1987). 

%\bibitem{bmt1}
%M. Bucher, K. Moodley and N. Turok,
%``The general primordial cosmic perturbation,''
%Phys.\ Rev.\ D {\bf 62}  083508 (2000);
%M. Bucher, K. Moodley and N. Turok,
%``Characterising the primordial cosmic perturbations using MAP and
%PLANCK,'' 
%astro-ph/0007360;
%M. Bucher, K. Moodley and N. Turok,
%``Constraining Isocurvature Perturbations with CMB Polarization'',
%astro-ph/0012141. 

%\bibitem{indep}
%  R. Stompor, A. J. Banday and K. M. Gorski, 
% { Ap. J. } {\bf 463}  8 (1996);
% E. Pierpaoli, J. Garcia-Bellido and S. Borgani,
% %``Microwave background anisotropies and large scale structure
% % constraints  on isocurvature modes in a two-field model of inflation,'' 
% JHEP {\bf 9910}  015 (1999);
% M. Kawasaki and F. Takahashi,
% %``Adiabatic and isocurvature fluctuations of Affleck-Dine field in
% %D-term  inflation model,'' 
% hep-ph/0105134. 

%\bibitem{enqvistBM}
% K. Enqvist, H. Kurki-Suonio and J. Valiviita,
% %``Limits on Isocurvature Fluctuations from Boomerang and MAXIMA,''
% Phys.\ Rev.\ D {\bf 62}  103003 (2000). 

%\bibitem{peebles} P. J. E. Peebles, Astrophys.\ J.\ {\bf 510}  531 (1999). 

%\bibitem{langlois2}
%D. Langlois and A. Riazuelo,
%``Correlated mixtures of adiabatic and isocurvature cosmological
%perturbations,'' 
%Phys.\ Rev.\ D {\bf 62}  043504 (2000). 
%%CITATION = PHRVA,D62,043504;%%

\bibitem{trotta}
R. Trotta, A. Riazuelo and R. Durrer,
%``Reproducing Cosmic Microwave Background anisotropies with mixed
%isocurvature perturbations,'' 
astro-ph/0104017. 
%%CITATION = ASTRO-PH 0104017;%%

%\bibitem{WMLL1}
%D. Wands, K. A. Malik, D. H. Lyth and A. R. Liddle,
%``A new approach to the evolution of cosmological perturbations on
%large scales,'' 
%Phys.\ Rev.\ D {\bf 62}  043527 (2000)
%[astro-ph/0003278]. 
%%CITATION = ASTRO-PH 0003278;%%

\bibitem{bon}
J. R. Bond, A. H. Jaffe and L. Knox, Ap. J. {\bf 533} (2000). 

\bibitem{sel}
U. Seljak and M. Zaldarriaga, Ap. J. {\bf 469} (1996). 

%\end{thebibliography}

%\def\bibname{References for Chapter 4
%(Brane world perturbations) 
%References
%}
%\begin{thebibliography}{99}



%\bibitem{rs}
%L. Randall and R. Sundrum, Phys. Rev. Lett. {\bf 83}  4690 (1999). 

\bibitem{mwbh}
R. Maartens, D. Wands, B. A. Bassett, and I. P. C. Heard, Phys. Rev. 
D {\bf 62}  041301R (2000). 

\bibitem{cll}
E. J. Copeland, A. R. Liddle and J. E. Lidsey,
%``Steep inflation: Ending braneworld inflation by gravitational particle  production,''
Phys.\ Rev.\ D {\bf 64} 023509 (2001). 
%%CITATION = ASTRO-PH 0006421;%%

%\bibitem{wmll}
%D. Wands, K. A. Malik, A. R. Liddle, and D. H. Lyth, Phys. Rev. D
%{\bf 62}  043527 (2000). % (astro-ph/0003278). 

\bibitem{m}
R. Maartens, Phys. Rev. D {\bf 62}  084023 (2000). 

\bibitem{LMSW}
D.~Langlois, R.~Maartens, M.~Sasaki and D.~Wands,
%``Large-scale cosmological perturbations on the brane,''
Phys.\ Rev.\ D {\bf 63} 084009  (2001). 
%%CITATION = HEP-TH 0012044;%%


\bibitem{pert} 
H. Kodama, A. Ishibashi, and O. Seto, Phys. Rev. D
{\bf 62}  064022 (2000); 
D. Langlois,
%``Brane cosmological perturbations,''
Phys.\ Rev.\ D {\bf 62} 126012 (2000); 
D.~Langlois,
%``Evolution of cosmological perturbations in a brane-universe,''
Phys.\ Rev.\ Lett.\  {\bf 86} 2212 (2001); 
K. Koyama and J. Soda,
%``Bulk gravitational field and cosmological perturbations on the brane,''
hep-th/0108003;
%%CITATION = HEP-TH 0108003;%%
S. Mukohyama,
%``Perturbation of junction condition and doubly gauge-invariant  variables,''
Class.\ Quant.\ Grav.\  {\bf 17} 4777 (2000). 
%%CITATION = HEP-TH 0006146;%%

\bibitem{bd}
C. van de Bruck, M. Dorca, R. H. Brandenberger and A. Lukas,
%``Cosmological perturbations in brane-world theories: Formalism,''
Phys.\ Rev.\ D {\bf 62} 123515 (2000); 
%%CITATION = HEP-TH 0005032;%%
C. van de Bruck, M. Dorca, C. J. Martins and M. Parry,
%``Cosmological consequences of the brane/bulk interaction,''
Phys.\ Lett.\ B {\bf 495} 183 (2000). 
%%CITATION = HEP-TH 0009056;%%

\bibitem{lmw}
D. Langlois, R. Maartens, and D. Wands, Phys. Lett. B {\bf 489} 
259 (2000). % (hep-th/0006007). 

\bibitem{BM}
J.~D.~Barrow and R.~Maartens,
%``Kaluza-Klein anisotropy in the CMB,''
gr-qc/0108073.
%%CITATION = GR-QC 0108073;%%


\bibitem{mw}
K. Maeda and D. Wands,
%``Dilaton-gravity on the brane,''
Phys.\ Rev.\ D {\bf 62} 124009 (2000); 
%%CITATION = HEP-TH 0008188;%%
A. Mennim and R. A. Battye,
%``Cosmological expansion on a dilatonic brane-world,''
Class.\ Quant.\ Grav.\  {\bf 18} 2171 (2001). 
%%CITATION = HEP-TH 0008192;%%

\bibitem{msm}
S. Mukohyama, T. Shiromizu, and K. Maeda, Phys. Rev. D {\bf 61} 
024028 (2000). 

\bibitem{bdel}
P. Binetruy, C. Deffayet, U. Ellwanger, and D. Langlois, Phys. 
Lett. B {\bf 477}  285 (2000). 

\bibitem{sms}
T. Shiromizu, K. Maeda, and M. Sasaki, Phys. Rev. D {\bf 62} 
024012 (2000). 

\bibitem{eu}
J. Cline, C. Grojean, and G. Servant, Phys. Rev. Lett. {\bf 83} 
4245 (1999); 
L. E. Mendes and A. R. Liddle,
%``Initial conditions for hybrid inflation,''
Phys.\ Rev.\ D {\bf 62}  103511 (2000);
%%CITATION = ASTRO-PH 0006020;%%
A. Mazumdar,
%``Interesting consequences of brane cosmology,''
Phys.\ Rev.\ D {\bf 64} (2001) 027304;
A. Mazumdar,
%``Post-inflationary brane cosmology,''
Nucl.\ Phys.\ B {\bf 597} (2001) 561.


\bibitem{ehb}
G. F. R. Ellis, J. Hwang and M. Bruni, Phys. Rev. D {\bf 40} 
1819 (1989). 

\bibitem{gs}
J. Garriga and M. Sasaki, Phys. Rev. D{\bf 62}  043523 (2000);
A. Chamblin, A. Karch and A. Nayeri,
%``Thermal equilibration of brane-worlds,''
Phys.\ Lett.\ B {\bf 509} 163 (2001).
%%CITATION = HEP-TH 0007060;%%


%\end{thebibliography}


%\def\bibname{References for Appendix A}
%\begin{thebibliography}{99}

%\newcommand{\wwwspires}{http://www. slac. stanford. edu/spires/find/hep/www}
%\cite{Bennett:1996ce}
%\bibitem{Bennett:1996ce}
%C. L. Bennett {\it et al.} 
%``4-Year COBE DMR Cosmic Microwave Background Observations: Maps and Basic Results,''
%Astrophys.\ J.\  {\bf 464}  L1 (1996). 
%[astro-ph/9601067]. 
%%CITATION = ASTRO-PH 9601067;%%
%\href{\wwwspires?eprint=ASTRO-PH/9601067}{SPIRES}

\bibitem{ci} A. Linde, Phys. Lett. {\bf 129B}  177 (1983). 

%\cite{Linde:1990nc}
%\bibitem{Linde:1990nc}
%A. Linde,
%``Particle Physics And Inflationary Cosmology,''
%{\it  Chur, Switzerland: Harwood (1990)  
%(Contemporary concepts in physics, 5)}. 

\bibitem{embedd} 
K. Sato, H. Kodama, M. Sasaki, and K. Maeda,
Phys. Lett B {\bf 108}  103 (1982); 
E. Fahri and A. Guth, Phys. Lett. B {\bf 183}  149 (1987);
S. Blau, E. Guendelman and A. Guth, Phys. Rev. D {\bf 35}  1747 (1987); 
E. Fahri, A. Guth, and J. Guven, Nucl. Phys. B {\bf 339}  417 (1990). 

%\newcommand{\wwwspires}{http://www. slac. stanford. edu/spires/find/hep/www}
%\cite{Vachaspati:2000dy}
\bibitem{Vachaspati:2000dy} 
T. Vachaspati and M. Trodden,
%``Causality and cosmic inflation,''
Phys.\ Rev.\ D {\bf 61}  023502 (2000). 
%[gr-qc/9811037]. 
%%CITATION = GR-QC 9811037;%%
%\href{\wwwspires?eprint=GR-QC/9811037}{SPIRES}

\newcommand{\wwwspires}{http://www. slac. stanford. edu/spires/find/hep/www}
%\cite{Trodden:1999wc}
\bibitem{Trodden:1999wc} 
M. Trodden and T. Vachaspati,
%``What is the homogeneity of our universe telling us?,''
Mod.\ Phys.\ Lett.\ {\bf A14}  1661 (1999). 
%[gr-qc/9905091]. 
%%CITATION = GR-QC 9905091;%%
%\href{\wwwspires?eprint=GR-QC/9905091}{SPIRES}

%\newcommand{\wwwspires}{http://www. slac. stanford. edu/spires/find/hep/www}
%\cite{Starkman:1999pg}
%\bibitem{Starkman:1999pg} G. Starkman, M. Trodden and T. Vachaspati,
%``Observation of cosmic acceleration and determining the fate of the
%universe,'' 
%Phys.\ Rev.\ Lett.\ {\bf 83}  1510 (1999). 
%[astro-ph/9901405]. 
%%CITATION = ASTRO-PH 9901405;%%
%\href{\wwwspires?eprint=ASTRO-PH/9901405}{SPIRES}

\bibitem{piran181} 
T. Piran, Phys. Lett. B {\bf 181}  238 (1986). 

\bibitem{kb} 
J. H. Kung and R. Brandenberger, Phys. Rev. D {\bf 40} 
2532 (1989). 

\bibitem{kb90} 
J. H. Kung and R. Brandenberger,
Phys. Rev. D {\bf 42}  1008 (1990). 

\bibitem{gpprl} 
D. S. Goldwirth and T. Piran, Phys. Rev. Lett. 
{\bf 64}  2852 (1990). 
 
\bibitem{gold43} 
D. S. Goldwirth, Phys. Rev. D {\bf 43}  3204 (1991). 

\bibitem{other} 
H. Kurki-Suonio, P. Laguna, and R. A. Matzner,
Phys. Rev. D {\bf 48}  3611 (1993);
E. Calzetta and M. Sakellariadou, Phys. Rev. D {\bf 45}  2802 (1992);
T. Chiba, K. Nakao, and T. Nakamura, Phys. Rev. D {\bf 49}  3886 (1994);
O. Iguchi and H. Ishihra, Phys. Rev. D {\bf 56}  3216 (1997). 

%\newcommand{\wwwspires}{http://www. slac. stanford. edu/spires/find/hep/www}
%\cite{Goldwirth:1992rj}
\bibitem{Goldwirth:1992rj}
D. S. Goldwirth and T. Piran,
%``Initial conditions for inflation,''
Phys.\ Rept.\  {\bf 214}  223 (1992). 
%%CITATION = PRPLC,214,223;%%
%\href{\wwwspires?j=PRPLC\%2c214\%2c223}{SPIRES}

\bibitem{ray} A. K. Raychaudhuri, Phys. Rev. {\bf 98}  1123 (1955). 

%\newcommand{\wwwspires}{http://www. slac. stanford. edu/spires/find/hep/www}
%\cite{Wald:1984rg}
\bibitem{Wald:1984rg}
R. M. Wald,
{\em General Relativity,}
  University of Chicago Press (1984).

%\cite{Israel:1966rt}
\bibitem{Israel:1966rt}
W. Israel,
%``Singular Hypersurfaces And Thin Shells In General Relativity,''
Nuovo Cim.\  B {\bf 44}  1 (1966). 
%%CITATION = NUCIA,B44S10,1;%%

\bibitem{bkt} V. A. Berezin, V. A. Kuzmin and I. I. Tkachev,
Phys. Rev. D {\bf 36}  2919 (1987). 

%\cite{Sakai:1994fu}
\bibitem{Sakai:1994fu}
N. Sakai and K. Maeda,
%``Junction conditions of Friedmann-Robertson-Walker space-times,''
Phys.\ Rev.\  D {\bf 50}  5425 (1994). 
%[gr-qc/9311024]. 
%%CITATION = GR-QC 9311024;%%

\bibitem{siold} 
A. H. Guth, Phys. Rev D {\bf 23}  347 (1981);
E. Gliner and I. G. Dymnikova, Sov. 
Astron. Lett. {\bf 1}  93 (1975);
R. Brout, F. Englert and E. Gunzig, Ann. 
Phys. {\bf 115}  78 (1978); 
R. Brout, F. Englert and P. Spindel,
Phys. Rev. Lett. {\bf 43}  417 (1979);
A. A. Starobinsky, Phys. Lett. B {\bf 91}  154 (1980);
L. Z. Fang, Phys. Lett. B {\bf 95}  154 (1980);
E. Kolb and S. Wolfram, Astrophys. J. {\bf 239}  428 (1980);
K. Sato, Phys. Lett. B {\bf 99}  66 (1981); {\it ibid}  Mon. Not. R. 
Astron. Soc. {\bf 195}  467 (1981); 
D. Kazanas, Astrophys. J. {\bf 241}  L59 (1980). 

\bibitem{ni} 
A. Albrecht and P. J. Steinhardt, Phys. Rev. Lett. 
{\bf 48}  1220 (1982); A. Linde, Phys. Lett. {\bf 108B}  389 (1982). 

\bibitem{wi} 
A. Berera,  Phys. Rev. Lett. {\bf 75}  3218 (1995);
A. Berera, Phys. Rev. D {\bf 54}  2519 (1996). 

%\newcommand{\wwwspires}{http://www. slac. stanford. edu/spires/find/hep/www}
%\cite{Berera:1997fm}
\bibitem{Berera:1997fm}
A. Berera,
%``Interpolating the stage of exponential expansion in the early
%universe:  A possible alternative with no reheating,'' 
Phys.\ Rev.\  D {\bf 55}  3346 (1997). 
%[hep-ph/9612239]. 
%%CITATION = HEP-PH 9612239;%%
%\href{\wwwspires?eprint=HEP-PH/9612239}{SPIRES}

\bibitem{taylor2000} 
A. N. Taylor and A. Berera, Phys. Rev. 
D {\bf 62}  083517 (2000). 

\bibitem{wifp} 
A. Berera, M. Gleiser and R. O. Ramos,
Phys. Rev. Lett. {\bf 83} (1999) 264; 
A. Berera, Nucl. Phys. B {\bf 585}  666 (2000). 

\bibitem{niearly} 
A. D. Linde, Phys. Lett. B {\bf 132}  317 (1983);
A. D. Linde, Phys. Lett. B {\bf 162}  281 (1985);
G. F. Mazenko, W. G. Unruh, and R. M. Wald, Phys. Rev. D {\bf 31} 
273 (1985). 

\bibitem{niic} 
D. S. Goldwirth, Phys. Lett. B {\bf 243}  41 (1990). 

\bibitem{ti} 
D. H. Lyth and E. D. Stewart, Phys. Rev. Lett. 
{\bf 75}  201 (1995). 

\bibitem{bubble} 
P. J. Steinhardt, Phys. Rev. D {\bf 25}  2074 (1982). 

%\bibitem{KT}
%E. W. Kolb and M. S. Turner,
%``The Early Universe,''
%{\it  Redwood City, USA: Addison-Wesley (1990)  (Frontiers in physics, 69)}. 

\bibitem{DC}
D. H. Coule,
%``Quantum creation and inflationary universes: A critical
%appraisal,''
Phys.\ Rev.\ D {\bf 62} 124010 (2000). 
%[gr-qc/0007037].


%\def\bibname{References for Chapter 6
%(Conclusions)
 %References
%}

%\begin{thebibliography}{99}
%\bibitem{gordonadent1}
%C. Gordon, D. Wands, B. A. Bassett and R. Maartens,
%``Adiabatic and entropy perturbations from inflation,'' 
%Phys.\ Rev.\ D {\bf 63}  023506 (2001) [astro-ph/0009131];

%\bibitem{bartolo1}
%N. Bartolo, S. Matarrese and A. Riotto,
%``Oscillations during inflation and cosmological density perturbations,''
%astro-ph/0106022. 
%%CITATION = ASTRO-PH 0106022;%%

\bibitem{bartolo2}
N. Bartolo, S. Matarrese and A. Riotto,
%``Adiabatic and isocurvature perturbations from inflation: Power spectra  and consistency relations,''
astro-ph/0107502. 
%%CITATION = ASTRO-PH 0107502;%%

% \bibitem{brane}
% D. Langlois,
% %``Brane cosmological perturbations,''
% Phys.\ Rev.\ D {\bf 62} 126012 (2000); 
% %[hep-th/0005025];
% %%CITATION = HEP-TH 0005025;%%
% %D. Langlois, R. Maartens and D. Wands,
% %``Gravitational waves from inflation on the brane,''
% %Phys.\ Lett.\ B {\bf 489} (2000) 259
% %[hep-th/0006007]; 
% H. A. Bridgman, K. A. Malik and D. Wands,
% %``Cosmological perturbations in the bulk and on the brane,''
% astro-ph/0107245; K. Koyama and J. Soda,
% %``Bulk gravitational field and cosmological perturbations on the brane,''
% hep-th/0108003. 
% %%CITATION = HEP-TH 0108003;%%

%\end{thebibliography}


\end{thebibliography}

% \def\bibname{References for the Abstract}

%\end{references}

%%%%%%%%%%%Double spacing%%%%%%%%%%%%%%%%%%%
%\end{double}

\end{document}